\documentclass{aa}
\usepackage{graphicx}
\usepackage{txfonts}
\usepackage{float}
\usepackage{color}
\definecolor{darkblue}{rgb}{0.00,0.00,0.25}
\definecolor{aipbrown}{rgb}{0.43,0.27,0.16}
\usepackage[colorlinks=true,urlcolor=blue,citecolor=darkblue]{hyperref}
\usepackage{xspace}
\newcommand{\DATA}{{\tt DATA}\xspace}
\newcommand{\DQ}{{\tt DQ}\xspace}
\newcommand{\STAT}{{\tt STAT}\xspace}
\newcommand{\kms}{\,km\,s$^{-1}$\xspace}
\newcommand{\eg}{e.\,g.\xspace}
\newcommand{\ie}{i.\,e.\xspace}
\newcommand{\oi}{[\ion{O}{i}]\xspace}
\newcommand{\cgsflux}{\,erg\,s$^{-1}$\,cm$^{-2}$\,\AA$^{-1}$\xspace}
\newcommand{\sinc}{\ensuremath \mathrm{sinc}}
\newcommand{\case}[1]{{\bf [#1]}}
\newcommand{\spp}{$s_\mathrm{p}^2$\xspace}

\begin{document}
\title{The Data Processing Pipeline for the MUSE Instrument}
\author{Peter M.\ Weilbacher\inst{1}
        \and
        Ralf Palsa\inst{2}
        \and
        Ole Streicher\inst{1}
        \and
        Roland Bacon\inst{3}
        \and
        Tanya Urrutia\inst{1}
        \and
        Lutz Wisotzki\inst{1}
        \and
        Simon Conseil\inst{3,4}
        \and
        Bernd Husemann\inst{5}
        \and
        Aur\'elien Jarno\inst{3}
        \and
        Andreas Kelz\inst{1}
        \and
        Arlette P\'econtal-Rousset\inst{3}
        \and
        Johan Richard\inst{3}
        \and
        Martin M.\ Roth\inst{1}
        \and
        Fernando Selman\inst{6}
        \and
        Jo\"el Vernet\inst{2}
       }
  \institute{Leibniz-Institut f\"ur Astrophysik Potsdam (AIP),
             An der Sternwarte 16, 14482 Potsdam, Germany\\
             \email{pweilbacher@aip.de}
         \and
             ESO, European Southern Observatory, Karl-Schwarzschild-Str.\ 2,
             85748 Garching bei M\"unchen, Germany
         \and
             Univ Lyon, Univ Lyon1, Ens de Lyon, CNRS, Centre de Recherche
             Astrophysique de Lyon UMR5574, F-69230, Saint-Genis-Laval, France
         \and
             Gemini Observatory / NSF’s OIR Lab, Casilla 603, La Serena, Chile
         \and
             Max-Planck-Institut f\"ur Astronomie, K\"onigstuhl 17, 69117
             Heidelberg, Germany
         \and
             European Southern Observatory, Ave.~Alonso de C\'ordova 3107,
             Vitacura, Santiago, Chile
            }
   \date{Received 29 February 2020 / accepted 13 June 2020}
\authorrunning{Weilbacher et al.}
\abstract{
  Processing of raw data from modern astronomical instruments is nowadays often
  carried out using dedicated software, so-called ``pipelines'' which are
  largely run in automated operation.
  In this paper we describe the data reduction pipeline of the Multi Unit
  Spectroscopic Explorer (MUSE) integral field spectrograph operated at ESO's
  Paranal observatory. This spectrograph is a complex machine: it records data
  of 1152 separate spatial elements on detectors in its 24 integral field units.
  Efficiently handling such data requires sophisticated software, a high degree
  of automation and parallelization.
  We describe the algorithms of all processing steps that operate on
  calibrations and science data in detail, and explain how the raw science
  data gets transformed into calibrated datacubes.
  We finally check the quality of selected procedures and output data
  products, and demonstrate that the pipeline provides datacubes ready for
  scientific analysis.
}
\maketitle

\section{Introduction}
MUSE \citep[the {\bf M}ulti-{\bf U}nit {\bf S}pectroscopic {\bf E}xplorer;][]
{Bacon+10,2014Msngr.157...13B} is a large-field, medium resolution
integral field spectrograph operated at the European Southern Observatory (ESO)
Very Large Telescope (VLT) since October 2014.

\subsection{Historical background}
MUSE was developed as one of the 2nd generation instruments at ESO's Paranal
observatory.
At its inception, the landscape of optical integral field spectrographs was led
by a few instruments at 4\,m-class telescopes, like SAURON \citep{BCM+01} and
PMAS \citep{2005PASP..117..620R}, as well as fiber-based units like VIMOS
\citep{2003SPIE.4841.1670L} and
GMOS \citep{2002PASP..114..892A} on 8\,m telescopes.
While the Euro3D network \citep{2004AN....325...83W} had been created to
develop software for such instruments \citep{2006NewAR..50..252R}, data
reduction remained cumbersome \citep[\eg,][]{2005ApJ...628L.139M}. Depending on
observatory and instrument in question, only basic procedures were widely
available. While there were exceptions \citep[][among others]{2003A&A...408..455W,ZGS+05},
researchers often struggled to subtract the sky and to combine multiple
exposures, and many homegrown procedures were developed to do that and produce
datacubes ready for analysis. The results were frequently suboptimal
\citep[\eg,][]{2005MNRAS.359..895V}.

In this environment, the specifications for MUSE were developed by ESO based
mostly on the experience with the FLAMES/GIRAFFE spectrograph
\citep{2002Msngr.110....1P}. Of the performance requirements, six were relevant
for the development of the pipeline. They included
(i) the capability to reconstruct images with a precision of better than 1/4
pixel,
(ii) a flux calibration accurate to $\pm$20\%,
(iii) the ability to support offsets,
(iv) sky subtraction to better than 5\% sky intensity outside the emission
lines, with a goal of 2\%,
(v) the capability to combine up to 20 dithered exposures with a S/N of at
least 90\% of the theoretical value, and
(vi) a wavelength calibration accuracy of better than 1/20th of a resolution
element.
Overall, the goal was to deliver software that would generate data cubes ready
for scientific use, with the best possible S/N to detect faint sources, with
only minimal user interaction.
More details of the history of 3D spectroscopic instruments, their properties,
and the development of MUSE can be found in \citet{BM17_3DSpec}.

\subsection{Instrument properties}
In its wide-field mode (WFM), the instrument samples the sky at approximately
$0\farcs2\times0\farcs2$ spatial elements and in wavelength bins of about
1.25\,\AA\,pixel$^{-1}$ at a spectral resolution of $R\sim3000$ over a
$1\arcmin\times1\arcmin$ field of view with a wavelength coverage of at least
4800\dots9300\,\AA\ (nominal) and 4650\dots9300\,\AA\ (extended mode). Since
2017, MUSE can operate with adaptive optics (AO) support \citep{SLA+12}. In
WFM, it is operated in a seeing-enhancing mode, correcting the ground layer
only \citep[see][for a first science result]{2018MNRAS.480.1689K}, the pixel
scale and wavelength range stay the same.

A high-order laser tomography AO correction \citep{2018SPIE10703E..1GO} has
been available in the so-called Narrow Field Mode (NFM) since 2018 \citep[][are
first science publications]{2019A&A...621L...5K,2019Icar..331...69I}. In this
mode, the scale is changed to 25\,mas\,pixel$^{-1}$ to better sample the
AO-corrected point spread function (PSF). The resulting field is then $8\times$
smaller (about $7\farcs5\times7\farcs5$). The wavelength range is the same as
in nominal mode. In all these cases, the data is recorded on a fixed-format
array of 24 CCDs (Charge Coupled Devices), each of which is read out to deliver
raw images of $4224\times4240$ pixels in size.
We summarize the instrument modes in Table~\ref{tab:modes} and show a
sketch of its operation in Fig.~\ref{fig:instlayout}.

For an instrument with such complexity, combined with the size of the raw data
(about 800\,MB uncompressed), a dedicated processing environment is a necessity.
This pipeline was therefore planned early-on during the instrument development
to be an essential part of MUSE. While some design choices of the
instrument were mainly driven by high-redshift Universe science cases, many
other observations were already envisioned before the start of MUSE
observations. By now MUSE actually evolved into a general purpose instrument,
as documented by recent publications on the topics of (exo-)planets
\citep{2018Icar..302..426I,2019NatAs...3..749H}, Galactic targets
\citep{WeilbacherM42,2015MNRAS.450.1057M}, and resolved stellar populations
\citep{2016A&A...588A.148H,2018MNRAS.473.5591K}, via nearby galaxies
\citep{2015MNRAS.452....2K,2015A&A...576L...3M} and galaxy clusters
\citep{2015MNRAS.446L..16R,2018MNRAS.477...18P} to high-redshift Lyman-$\alpha$
emitters \citep{2015A&A...575A..75B,2016A&A...587A..98W,2018Natur.562..229W},
to name only a few science applications. In contrast to some other instruments
with a dominating scientific application
\citep[see][]{2015AN....336..324S,2018MNRAS.481.2299S}, the MUSE pipeline thus
cannot be concerned with astrophysical analysis tasks. Its role is confined to
the transformation from the raw CCD-based data to fully calibrated data cubes.
Depending on the science case in question, other tools were then created to
handle the data cubes
\citep[MUSE Python Data Analysis Framework, MPDAF,][]{2016ascl.soft11003B,MPDAF_1710.03554},
to extract stellar \citep[PampelMUSE,][]{KWR13} or galaxy
\citep[TDOSE,][]{2019A&A...628A..91S} spectra, or to detect emission lines
sources \citep[LSDCat,][]{2017A&A...602A.111H}, among others. Describing these
external tools is not the purpose of this paper.

\subsection{Paper structure}
The data processing of MUSE at different stages of implementation was
previously described in \citet{WGR+09}, \citet{SWB+11}, \citet{WSU+12}, and
\citet{2014ASPC..485..451W}.  These papers still reflect much of the final
pipeline software, and explain some of the design choices in more detail.
The present paper aims to first describe the {\em science} processing steps on
a high level (Sect.~\ref{sec:sciproc}) to let the user get an idea of the steps
involved in the aforementioned transformation. Afterwards, in
Sect.~\ref{sec:fullproc}, we give a detailed description of all steps involved
in generating master calibrations and science products. Some algorithms that
are used in multiple steps are then presented in Sect.~\ref{sec:algo} while
Sect.~\ref{sec:impl} briefly discusses key parameters of the implementation.
Sect.~\ref{sec:qual} investigates the data quality delivered by the pipeline
processing.  We conclude with a brief outlook in Sect.~\ref{sec:concl}.

This paper is based on MUSE pipeline version 2.8.3 as publicly released in
June 2020\footnote{\url{https://www.eso.org/sci/software/pipelines/muse/muse-pipe-recipes.html}}.
v2.8 was the first version to support all modes of the instrument, in particular the
NFM.\footnote{Milestones of earlier versions were v1.0 in Dec.~2014 to support
  the first observing runs with only seeing-limited WFM, and v2.2 which first
  supported WFM AO data in Oct.~2017.}
Where applicable, we note in which version a new feature was introduced.

\section{Science processing overview}\label{sec:sciproc}
\begin{figure}
\includegraphics[trim={56 133 56 55},clip,width=\linewidth]{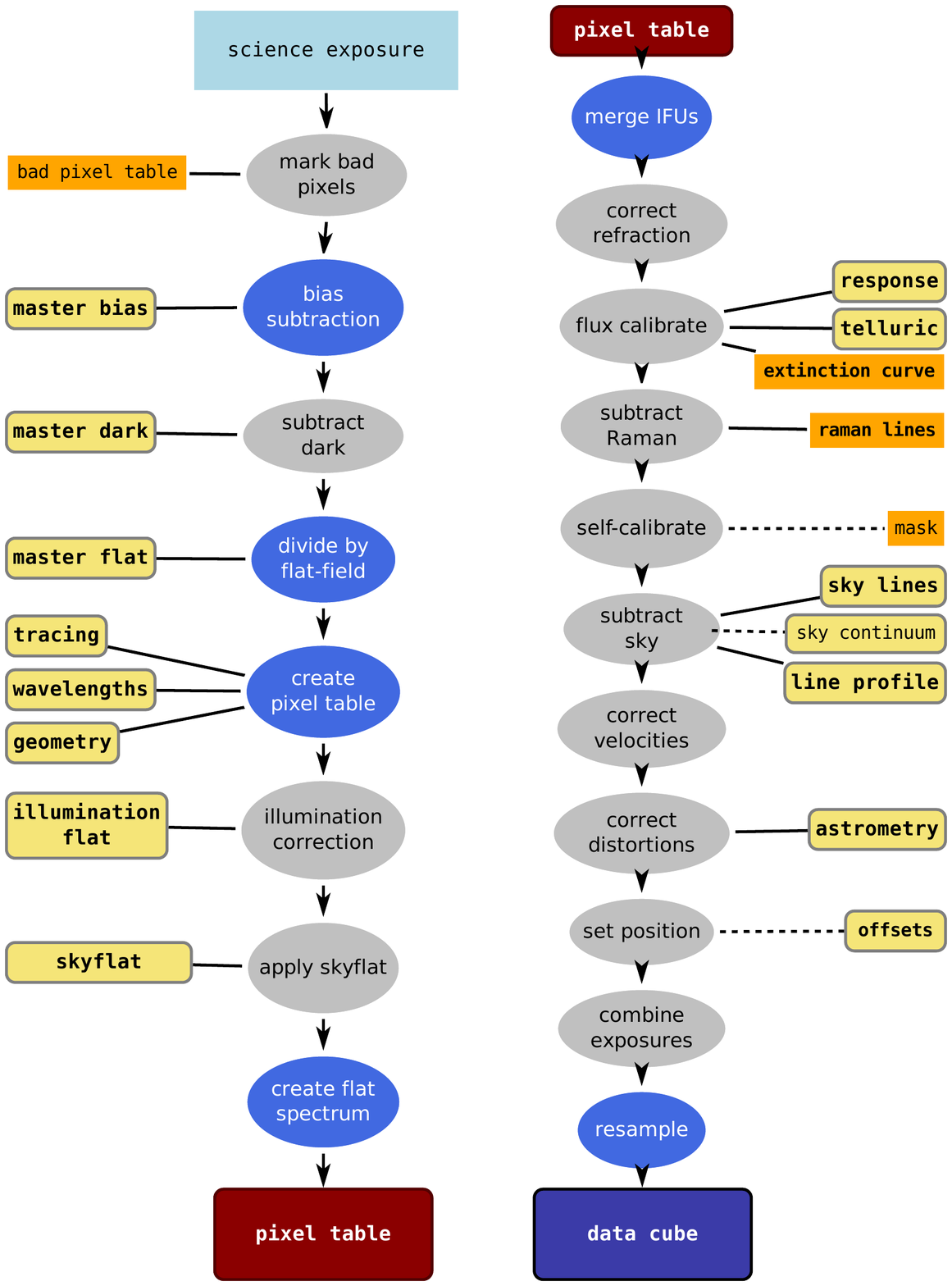}
\caption{{\bf Left}: {\em Basic processing} from raw science data to the
         intermediate pixel table.
         {\bf Right}: {\em Post-processing} from pixel table to the final
         datacube.
         Optional steps are marked grey, mandatory ones in blue. Manually
         created input files have an orange background, calibrations are
         highlighted. Inputs that are needed are connected with a solid line,
         dotted lines signify inputs that are not required.}
\label{fig:sciproc}
\end{figure}

The main processing steps to calibrate the data and transform it from the
image-based format of the raw data via a table-based intermediate format during
processing to the output cube are visualized in Fig.~\ref{fig:sciproc}.  The
computations are split into two parts, the {\em basic processing} -- this
calibrates and corrects data on the basis of single CCDs -- and the {\em
post-processing} -- carrying out on-sky calibrations and construction of the
final datacube. The intermediate data, the {\em pixel tables}, are the files
that connect both processing levels.

In this section, we only briefly mention the processing steps, and point to
later sections where they are described in more detail. In a few cases a
processing step is not connected to a calibration file, and hence not described
further. Then we describe this step here in greater depth.

\subsection{Raw data}\label{sec:raw}
\begin{figure}
\includegraphics[width=\linewidth]{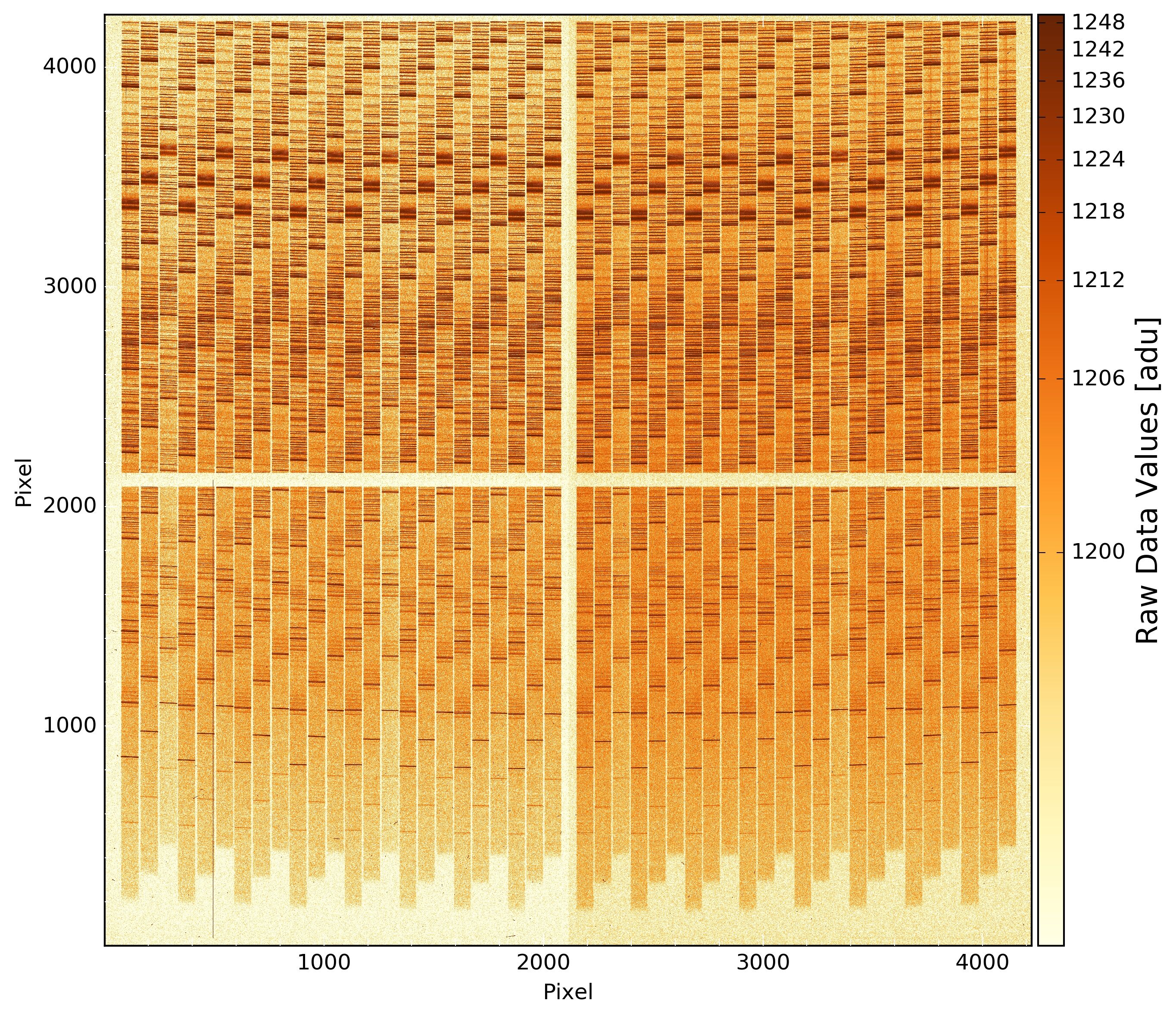}
\caption{Raw science data of one of the 24 CCDs, displayed in
         negative arcsinh scaling. This shows the data of IFU 10 of the exposure
         started at 2014-06-26T01:24:23 (during MUSE Science Verification). The
         48 slices of the IFU are the stripes oriented almost vertically, that
         appear dark in this representation. The blue end of the MUSE wavelength
         range is located at the bottom, the red limit near the top, the step-pattern
         is created by the geometry of the image slicer. Since this
         was a 600\,s exposure, the sky emission and continuum dominate over the
         relatively faint object signal in this part of the cube.  The overscan
         regions of the CCDs are creating the cross in the center of the image,
         the prescan regions are the empty borders.  This exposure was taken in
         nominal mode (WFM-NOAO-N), the 2nd-order blocking filter removed the
         blue light, so that the bottom part of the image appears empty.}
\label{fig:raw}
\end{figure}

The MUSE raw data comes in multi-extension FITS (Flexible Image Transport System) files, where each of the 24
CCD images is stored in one extension. Each of the images is $4224\times4240$\,pixels
in size, and stored as unsigned integers of 16\,bit. The CCD is read-out on four ports,
so that the CCD has four regions of equal size, called {\em quadrants}. These
quadrants have a data section of $2048\times2056$ pixels, and pre- and overscan
regions of 32 pixels in width. The images are accessible in the FITS files via
extension names, formed by the IFU number prefixed by CHAN, for example, {\tt
CHAN01} for the first IFU. A typical raw science image of {\tt CHAN10} is
displayed in Fig.~\ref{fig:raw}.

Several additional FITS extensions may be present for on-sky data, depending
on the instrument mode used for an exposure. These are concerned with ambient
conditions, performance of the auto-guiding system of the VLT, the slow-guiding
system of the MUSE instrument, and the atmospheric turbulence parameters used
by the AO system.
These extra extensions are not used by the MUSE pipeline.

\subsection{Basic science processing}\label{sec:scibasic}
The first step of the science processing is done with the module {\tt
muse\_scibasic}. The inputs to this recipe are one or more raw science
exposures, optionally one corresponding illumination flat-field exposure, and a
number of calibration files. Outputs are pre-reduced pixel tables.

Processing is internally performed on each individual CCD,
so that all of the following is done 24 times. The raw CCD image is read in
(\DATA image) from the raw data (the corresponding {\tt CHANnn} extension), and
two images of the same size are added, one for the data quality (\DQ, see
Sect.~\ref{sec:badpix}), one for the variance (\STAT, see
Sect.~\ref{sec:errorprop}). Both new images are empty at the start.  Next, if the
optional bad pixel table was given, those pixels are marked in the \DQ image.
In any case, saturated pixels are detected and marked, if the data
allow such determination to be made (raw values zero or above 65500).

Next, the overscan regions of the MUSE CCDs are analyzed to determine
corrective factor and slopes to apply to the bias (see Sect.~\ref{sec:ovsc} for
more details about this step), before the master bias image of the
corresponding CCD is subtracted.  Then, the CCD gain is used to transform the
input data from analog-to-digital units ({\tt adu}) to electrons (internally
called {\tt count}). The gain value is taken from the FITS headers of the raw
exposure. If the optional master dark was given, the pipeline subtracts the
dark current image given by that calibration file. An optional step in the
CCD-level calibration is to detect and mark cosmic rays \citep[using the DCR
algorithm,][]{2004PASP..116..148P}.  However, this is usually not necessary
at this stage (as explained in detail in Sect.~\ref{sec:cosmics}).
The science image is then divided by the master
lamp flat-field image provided to the processing routine. The last step in the
direct CCD-level processing propagates the relative IFU flux level from the
twilight sky cube, if this optional calibration was given as input calibration
file.

The mandatory input calibrations, trace table, wavelength calibration table, and
geometry table (their content and purpose are explained in
Sects.~\ref{sec:tracing}, \ref{sec:wavecal}, and \ref{sec:geo}),
are used to assign coordinates -- two spatial components in
pseudo pixel units relative to the MUSE field of view, and one wavelength
component -- to each CCD-based pixel. Thereby, a {\em pixel table} is created
for each input science (on-sky) exposure.

All these steps are also applied in the same way to the optional raw
illumination flat-field exposure if one was supplied.\footnote{This illumination
  flat-field is a lamp-flat exposure that is taken by the observatory at least
  once per hour, or if the ambient temperature changes significantly.}
The following steps are exclusively applied to exposures taken on-sky at night.

The wavelength zeropoint is corrected using sky emission lines, if applicable
(see Sect.~\ref{sec:applwave}).
Afterwards, the pixel table is usually cropped to the useful wavelength range,
depending on the mode used for the observations. The useful wavelength is
defined by the range for which the MUSE field of view is fully sampled. It
extends from 4750\dots9350\,\AA\ for the nominal and 4600\dots9350\,\AA\ for
the extended mode.\footnote{The exact ranges are slightly different for the AO
  modes, see Table \ref{tab:modes}. For those, the region affected by the NaD
  narrow-band blocking filter is also specially marked at this stage.}

If the optional raw illumination flat-field exposure was given as input, it is
then used to correct the relative illumination between all slices of one IFU.
For this, the data of each slice is multiplied by the normalized median flux
(over the wavelength range 6500\dots7500\,\AA, to use the highest S/N data in
the middle of the wavelength range) of that slice in that special flat-field
exposure. Since the illumination of the image slicer changes with time and
temperature during the night, this correction removes these achromatic
variations of the illumination and thereby significantly improves flux
uniformity across the field.

The last step in the basic science processing interpolates the master twilight
sky cube (Sect.~\ref{sec:twilight}, if it was given as input) to the coordinate
of each pixel in the pixel table.  Spatially, the nearest neighbor is taken, in
wavelength a linear interpolation between adjacent planes is carried out. The
data values in the pixel table are then divided by the interpolated twilight
sky correction.

At this stage, the pre-reduced pixel table for each on-sky exposure is saved to
disk, in separate files for each IFU, including the corresponding averaged lamp
flat spectrum in one of the file extensions.

The module discussed in this section, {\tt muse\_scibasic}, is also used to
process other, non-science on-sky exposures taken for calibration purposes.
Specifically, standard star fields, sky fields, and astrometric exposures of
globular clusters are handled by {\tt muse\_scibasic}, but then further
processed by specialized routines.

\subsection{Science post-processing}\label{sec:scipost}
The input to the post-processing are the pre-reduced pixel tables,
the main output is the fully reduced data cube.
The first step is to merge the pixel tables from all IFUs into a common table.
This step has to take into account the relative efficiency of each IFU as
measured from twilight sky exposures. It is applied as scaling factor relative
to the first channel.
When merging the (science) data, all lamp flat spectra of the IFUs are averaged
as well. Since removing the large-scale flat-field spectrum from the (science)
data is desirable, without re-introducing the small-scale variations
corrected for by flat-fielding, this
mean lamp flat spectrum is smoothed over scales larger than any small-scale
features like telluric absorption or interference filter fringes. The on-sky
data is then divided by this spectrum.\footnote{The correction for the lamp
  flat-field spectrum is done since v2.0 of the pipeline, since v2.6 the
  smoothing is applied.}

Then, the merged pixel table is put through several corrections. The
atmospheric refraction is corrected (for WFM data, the NFM uses an optical
corrector) relative to a reference wavelength (see Sect.~\ref{sec:dar}).
In case a response curve is available, the flux calibration is carried out
next.  It converts the pixel table data (and variance) columns into flux units.
This uses an atmospheric extinction curve that has to be passed as input table.
If a telluric correction spectrum was provided, this is applied as well
(Sect.~\ref{sec:fluxcal}.)

For exposures taken with AO in WFM, a correction of atmospheric emission lines
caused by Raman scattering of the laser light can be carried out (see
Sect.~\ref{sec:raman}).
A per-slice self-calibration can be run next to improve background
uniformity across the field of view of MUSE (explained in detail in
Sect.~\ref{sec:autocal}).

Typically, sky subtraction is carried out next. This step has multiple ways of
deriving the sky contribution which also depends on the user input and the type
of field being processed (sky subtraction is not needed for all science cases
and can be skipped). In case of a filled science field, an offset sky field has
to be used to characterize the sky background (see Sect.~\ref{sec:skydecom} and
\ref{sec:skysub}), sky lines and continuum are then needed as inputs. The
procedure is the same for a largely empty science field, just that the sky
spectrum decomposition does not need extra inputs.
Since the sky lines change on short timescales, they usually have to be
re-fitted using a spectrum created from a region of the science exposure
devoid of objects.
(This is the default behavior, but one can choose to skip the refit.) The
continuum, however, only changes slowly and is subtracted directly.  In all
cases, the user usually has to tell the pipeline which spatial fraction of an
exposure is sky-dominated, so that the software can use that portion of the
data to reconstruct the sky spectrum.

The science data is then corrected for the motion of the telescope. This radial
velocity correction is done by default relative to the barycenter of the solar
system, but for special purposes heliocentric and geocentric corrections are
available. Algorithms from G.~Torres ({\tt bcv}) and the FLAMES pipeline
\citep{2002SPIE.4844..310M} and transformations from
\citet{1994A&A...282..663S} are used to compute the values.

The spatial coordinate correction is applied in two steps and makes use of the
astrometric calibration (Sect.~\ref{sec:astcal}). First, linear transformation
and rotation are applied and the spherical projection is carried out. This
transforms the pixel table spatial coordinates into native spherical
coordinates, following \citet{2002A&A...395.1077C}.
The second step is the spherical coordinate rotation onto the celestial
coordinates of the observed center of the field.  This step can be influenced
to improve both absolute and relative positioning of the exposure by supplying
coordinate offsets. Such offsets can be computed manually by the user or
automatically by correlating object detections in overlapping exposures
(Sect.~\ref{sec:offcalc}).

Once all these corrections and calibrations are applied on the pixel table
level, the data are ready to be combined over multiple exposures. This can be a
very simple concatenation of the individual pixel tables or involve relative
weighting. By default, only linear exposure time weighting is carried out, such
that twice as long exposures are weighted twice as strongly. Another
possibility is seeing-weighted exposure combination which is implemented to
primarily use the FWHM measured by the VLT auto-guiding system during each
(non-AO) exposure. More complex weighting schemes are possible, but require the
users to determine the weights themselves, depending on exposure content and
science topic.

\begin{figure}
\includegraphics[width=\linewidth]{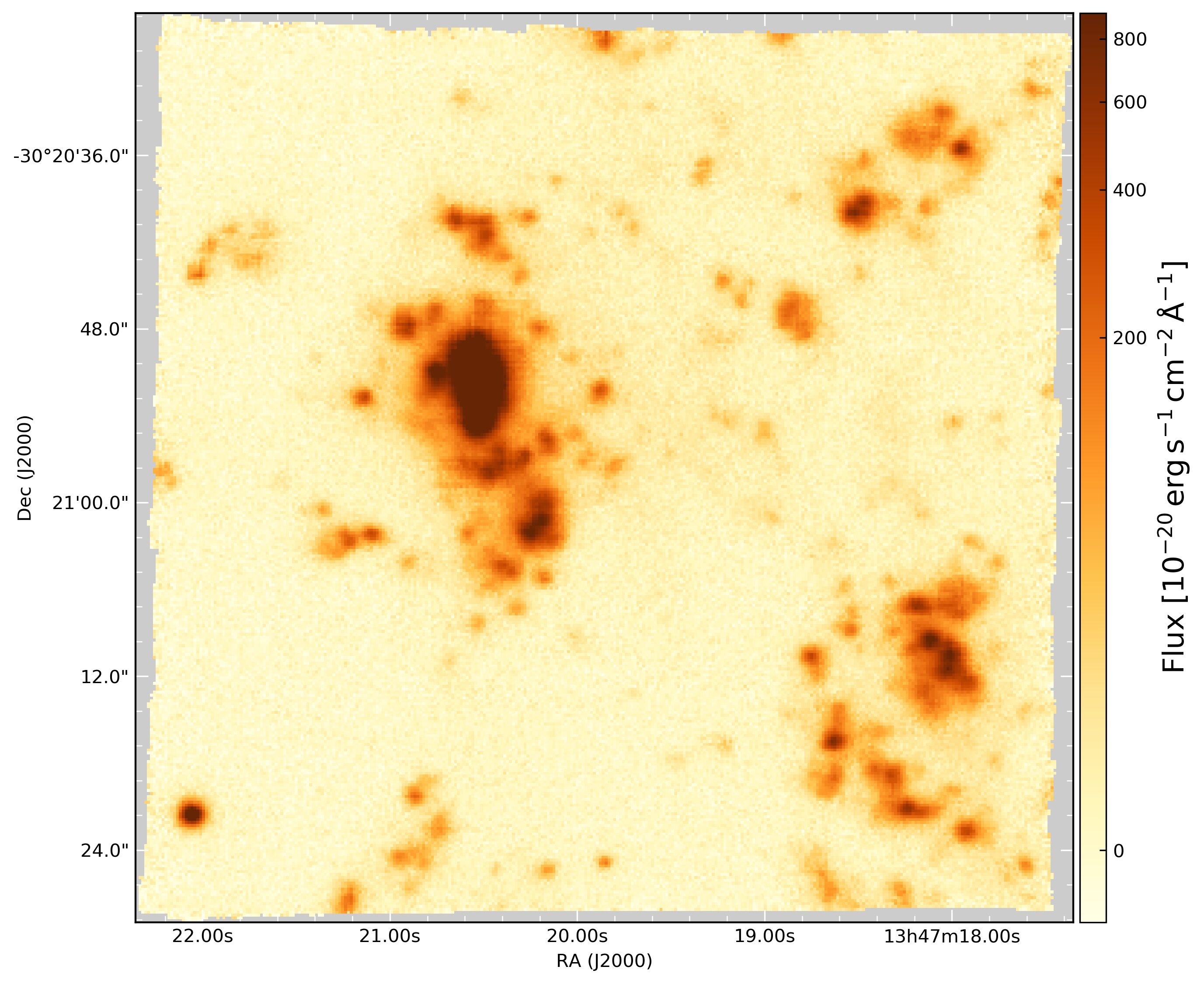}
\caption{Reduced science data. A combination of three science
         exposures taken on 2014-06-26 between 1:00 and 2:00 UTC, including the
         image displayed in Fig.~\ref{fig:raw}.
         This image shows a cut of the data cube at the wavelength of H$\alpha$
         (redshifted to 6653.6\,\AA), displayed in negative arcsinh scaling.
         Regions at the edge that were not covered by the MUSE data are
         displayed in light grey.}
\label{fig:reduced}
\end{figure}

Finally, the data of one or more exposures are resampled into a datacube. The
process is described in detail in Sect.~\ref{sec:cuberec}, by default it uses a
drizzle-like \citep{Drizzle09} algorithm to conserve the flux (\eg, of emission
lines over a spatially finite object).
This cube is normally computed such that north is up and east is left and the
blue end of the wavelength range is in the first plane of the cube, and so that
all data contained in the pixel table is encompassed by the grid of the output
cube. The example in Fig.~\ref{fig:reduced} shows a single wavelength plane from
a reduced cube \citep[this data of NGC\,5291\,N was published
by][]{2016A&A...585A..79F}.
Both the extent of the cube as well as the sampling of the output data can
be adjusted by the user for a given science project. A logarithmic wavelength
axis can be defined as well and the user can choose to have it saved in vacuum
instead of air wavelengths.
The cube is the base for the computation of reconstructed images of the field of
view (see Sect.~\ref{sec:imagerec}) which are integrated over a built-in
``white'' or any other filter function.

These post-processing routines for the science data are offered in the pipeline
by {\tt muse\_scipost}, the combination of multiple exposures is also available
separately as {\tt muse\_exp\_combine}. The module {\tt muse\_exp\_align}
(Sect.~\ref{sec:offcalc}) can be used to compute the offset corrections, as
mentioned above, between multiple (overlapping) exposures in an automated
fashion.

\section{Full processing details}\label{sec:fullproc}
Calibration of an instrument usually includes creation of master calibration
data which is then applied to subsequent calibration steps and the science data
itself. This is no different for MUSE, where the usual calibration exposures are
done during daytime, with the help of the calibration unit \citep{KBB+10,KBH+12}.
They are supplemented by on-sky calibrations done with the telescope during
twilight and during the course of the night. The full details of how these
specific calibrations are processed and then applied are provided in this
section. The purpose and frequency of the different calibrations are
further described in the ESO MUSE User Manual \citep{MUSE_User_Manual_v10.4}.
At the end of each subsection we also point out within which pipeline module
the described steps are implemented and how the significant parameters are
named.

\subsection{Bias level determination}\label{sec:bias}
The first step to remove the instrumental pattern from CCD exposures is always
to subtract the bias level. To this end, daily sequences of typically 11
bias frames with zero exposure time and closed shutter are recorded.
In case of MUSE, the CCDs are read out in four
quadrants and so the bias level already in the middle of the read-out image
exhibits four different values. On top of this quadrant pattern, the bias shows
horizontal and vertical gradients so that bias images have to be subtracted
from science and calibration data as 2D images. Finally, a variation of the
bias level with time means that before subtraction, the bias needs to be offset
to the actual bias determined from the other exposure, using values from the
overscan (Sect.~\ref{sec:ovsc}).

The bias images are also used to measure the read-out noise (RON) of each CCD.
This is computed on difference images ($B_1 - B_2$) of one whole bias sequence.
On each difference image, 400 boxes (100 in each CCD quadrant) of $9\times9$ pixels
are distributed within which the standard deviation of the pixel values is
recorded. The median of these standard deviations are taken as the
$\sigma_{B_1-B_2}$ value for each difference image, the average of all these
values is the $\overline{\sigma_{B_1-B_2}}$ for each CCD quadrant.
To estimate the error, the standard deviation of the standard deviations
of all boxes is taken. If the error of the $\overline{\sigma_{B_1-B_2}}$ is
found to be larger than 10\% of the $\overline{\sigma_{B_1-B_2}}$, the
procedure is repeated, with a differently scattered set of boxes. The
$\overline{\sigma_{B_1-B_2}} / \sqrt{2}$ is then used as initial variance of
the individual bias images (see Sect.~\ref{sec:errorprop}).
To compute a meaningful final read-out noise value and its error, the CCD
gain\footnote{This correct gain value has to be set up in the FITS header of the
  raw images so that this can happen in practice.}
$g$ is used to convert it to units of electrons
\citep[see][]{2006hca..book.....H}:
\begin{equation*}
\mathrm{RON} = \dfrac{g\ \overline{\sigma_{B_1-B_2}}}{\sqrt{2}}
\end{equation*}

The only other processing involved in creating the master bias image, is to
combine the individual images, using the algorithm described in
Sect.~\ref{sec:imcomb}. By default, a $3\sigma$ clipped average is computed.

Some CCD defects show up as columns of different values already on bias images.
To find them, column statistics of median and average absolute deviation are
used to set thresholds above the typical column level on the master bias image.
Normally, a $3\sigma$ level is used to find bright columns. High pixels within
these columns get flagged in the master image. Since finding dark pixels is not
possible on bias images, flagging of low-valued pixels is set to $30\sigma$, so
that this does not happen.

Application of the master bias image to a higher-level exposure involves the
overscan-correction described in Sect.~\ref{sec:ovsc}, after checking that the
same overscan handling was used for both. This is then followed by the
subtraction of the master bias image which includes propagation of any flagged
pixels to the resulting image.

The routine producing the master bias is available as {\tt muse\_bias} in the
pipeline.

\subsection{Dark current}
Estimating the dark current -- the electronic current that depends on exposure
time -- is a typical aspect of the characterization of CCDs,
so this procedure is available in the MUSE pipeline as well. Darks, long
exposures with the shutter remaining closed and external light sources switched
off, can also be used to search for hot pixels and bright columns on the
resulting images. Darks for MUSE are usually recorded as sequences of fives
frames of 30\,min once a month.

\begin{figure}
\includegraphics[trim={32 11 80 25},clip,width=\linewidth]{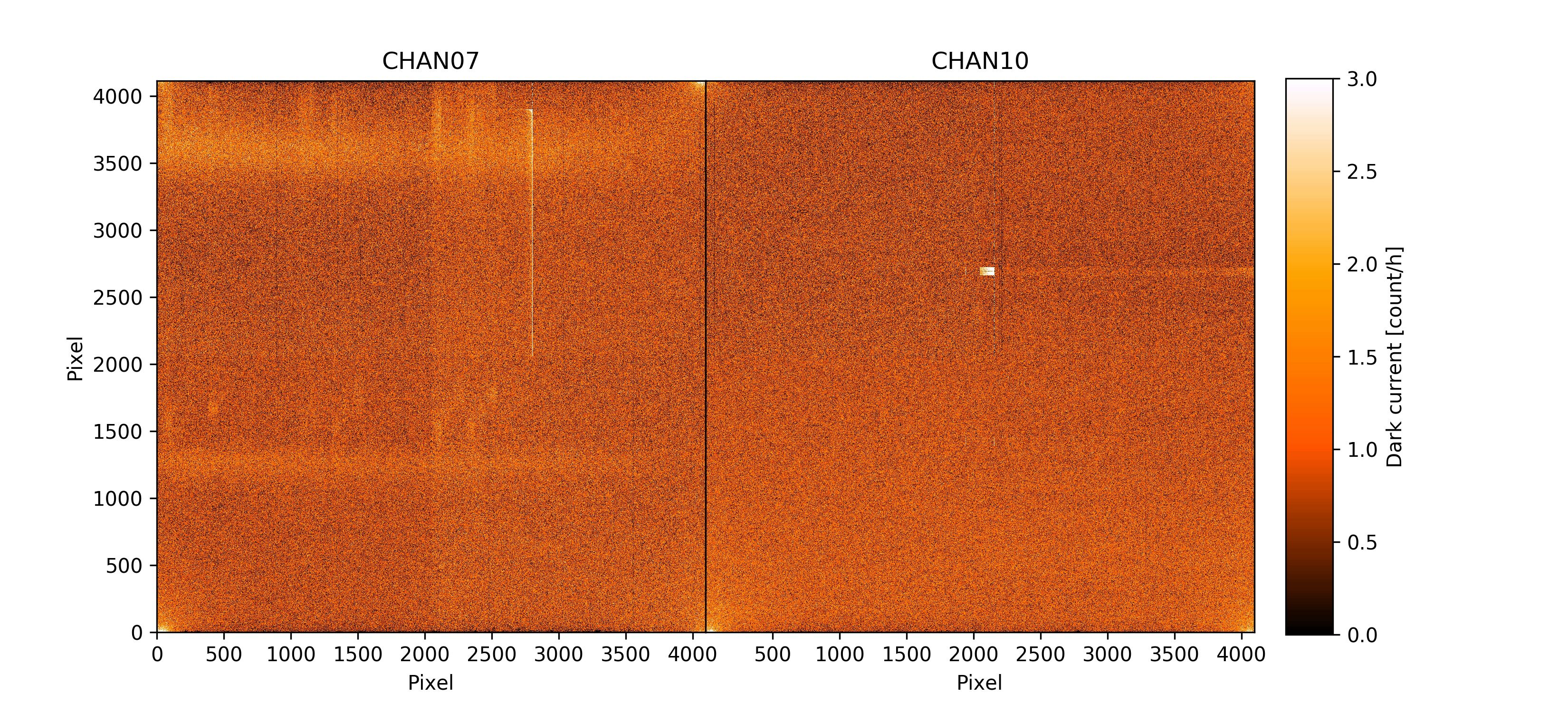}
\caption{Master dark images for two MUSE CCDs, both combined from 149 raw dark
         images taken between 2018-06-25T11:20:47 and 2018-08-21T06:56:51.
         The CCD of channel 7 shows the broad horizontal stripes while the CCD
         of channel 10 shows a noticeable block of hot pixels. The location of
         the hot corners is different for both CCDs, and while channel 10 shows
         a vertical gradient seen in most MUSE CCDs, this is less apparent for
         the CCD in channel 7.}
\label{fig:darks}
\end{figure}

Processing of dark exposures is as follows: from each of a sequence of darks,
the bias is subtracted using the procedures outlined in Sect.~\ref{sec:ovsc} and
\ref{sec:bias}. All exposures are then converted to units of electrons, scaled
to the exposure time of the first one, and combined using the routines described
in Sect.~\ref{sec:imcomb}, by default using the $\pm3\sigma$-clipped average.
If enough exposures were taken, these output images are free of cosmic rays.
The resulting images, one for each CCD, are then normalized to a 1 hour
exposure, so that the master dark images are in units of
e$^-$\,h$^{-1}$\,pixel$^{-1}$.
Hot pixels are then located using statistics of the absolute median deviation
above the median of each of the four data sections on the CCD.
Typically, a $5\sigma$ limit is used. Such hot pixels are marked in the \DQ
extension to be propagated to all following steps that use the master dark.
The master dark image that was thus created look as shown in Fig.~\ref{fig:darks}
for two example IFUs.

As an optional last step, a smooth model of the dark can be
created\footnote{This feature is available since v2.8.}.  Since the MUSE CCDs
have very low dark current (typically measured to be about
$1$\,e$^-$\,h$^{-1}$\,pixel$^{-1}$)
averaged across their data regions, a model is necessary to subtract
the dark current from science exposures at the pixel level, to avoid adding
additional noise.
The model consists of several steps: Some CCDs show evidence of low-level light
leaks. These can be modeled first, over horizontal regions 280-340 pixels high,
using two-dimensional polynomials of order 5 in both directions.
After subtracting these, one can refine the detection of bad pixels. To
represent the large scale dark current, a bilinear polynomial is used. The fit
for this ignores 500 pixels at the edge of the CCD, so that uneven borders and
the corners heated by the read-out ports do not contribute.
Finally, these read-out ports are modeled separately, in a region of 750 pixels
radius around each of the four corners. Here, a 5th order polynomial is used
again. To make the fit stable against small noise fluctuations, a 100 pixel
annulus is used to tie the corner model to the global fit.
The sum of all three polynomial models is the final dark model.
Searching for bad pixels can then be done again, using statistics of the
differences between the master dark and this smooth model.

If neither the master dark nor the dark model are used, one can transfer the
hot pixels found by this procedure to separate bad pixel tables (see
Sect.~\ref{sec:badpix}) instead.

The routine producing the master dark is available as {\tt muse\_dark} in
the pipeline. The sigma-clipping level to search for hot pixels is adjustable by
the parameter {\tt hotsigma}, the smooth modeling activated with {\tt model}.

\subsection{Flat-fielding}\label{sec:flat}
Flat-field correction in the MUSE pipeline has the same purpose as in classical
CCD imaging, to correct pixel-to-pixel sensitivity variations, and to
locate dark pixels.
The process is simple: from each of a sequence of exposures taken with the
continuum lamps switched on, the bias is subtracted using the procedures
outlined in Sect.~\ref{sec:ovsc} and \ref{sec:bias}. A dark image can
optionally be subtracted; this is typically not done, since the exposure times
are short. The main purpose of subtracting a dark here would be to propagate
its hot pixels map. The data is then converted from units of adu to electrons,
using the gain value provided with the raw data. In case different exposure
times were used, the images are then scaled relative to the first one. Then all
images are combined, using one of the algorithms described in
Sect.~\ref{sec:imcomb}, using a $3\sigma$-level clipped mean by default.  The
final master flat image is then normalized to 1 over the whole data section of
the CCD.

Once the slice tracing (Sect.~\ref{sec:tracing}) is done (in the
implementation, this is done in the same pipeline module as the flat-fielding)
and the slice edges on the CCD are known, the master flat-field can also be
used to search for dark pixels.
This is done for each CCD row, in the region between the edges of each slice.
The dark pixels are determined as outliers in the sigma-clipped statistics of
normally $5\times$ the absolute median deviation below the median. This
effectively marks all dark columns. Since for MUSE CCDs some of the electrons
lost in the dark columns appear as bright pixels on their borders, we also flag
pixels $5\sigma$ above the median.
Both limits can be adjusted.

In the case of flat-field exposures in nominal mode, the blue (lower) end of
each CCD image contains many pixels that are not significantly illuminated. Due
to noise, some of these are below the bias level and hence are negative in the
master flat-field image. These pixels are flagged as well, to exclude them from
further processing, in case the science data is not automatically clipped at
the blue end (the default).

Application of the master flat-field correction to any higher-level exposure
simply involves division by the master-flat image of the respective CCD. Pixel
flags are thereby propagated from the master flat image to the other image, the
pixel variances are propagated as well.

The routine producing the master flat-field is available as {\tt muse\_flat} in
the pipeline. The bad pixel detection threshold is set with the parameters {\tt
losigmabadpix} and {\tt hisigmabadpix}.

\subsection{Slice tracing}\label{sec:tracing}
The tracing algorithm is the part of the MUSE pipeline that determines the
location of those areas on the CCDs that are illuminated by the light from the
48 slices of each image slicer in each IFU.

This step uses the flat-field exposures to create well-illuminated images for most
of the wavelength range. These are prepared into a master-flat image for each
CCD, as described in Sect.~\ref{sec:flat}.
Tracing starts by extracting an averaged horizontal cut of the $\pm15$ CCD rows
around the center of the data area. By averaging the data from upper and lower
quadrants, the influence of a possible dark column in one of the quadrants is
reduced.
Pixels that fall below a fraction of 0.5 (by default) of the median value of
this cut determine the first-guess edges of each slice. The first slice (on the
left-hand of the CCD image) is the most critical one. If it is not
detected within the first
100 pixels of the cut or if it is narrower than 72.2 or wider than 82.2 pixels,
then the detection process is stopped and the tracing fails, otherwise the rest
of the slices are detected the same way. If more or less than 48 slices were
detected this way, or if some of the slices were too narrow, the process is
iteratively repeated with edge levels that are $1.2\times$ smaller, until a
proper number of slices is found. The horizontal centers (the mean position of
both edges) of these slices are subsequently used to trace curvature and tilt
of all slices on the image. For this, horizontal cuts are placed along a
sufficient number of rows along the whole vertical extent of every slice. The
number of cuts is determined by the number of rows $n_\mathrm{sum}$ that are
averaged with each such trace point.
(By default, $n_\mathrm{sum}=32$, so there are 128 trace points over the
vertical extent of each slice.)
Starting at the center of each cut, that is at the first-guess position of the
center of the slice, pixel values are compared to the median value across the
slice, in both directions. If a pixel falls below a given fraction (typically
0.5, as above), the slice edge is found and linearly interpolated to a
fractional pixel position of the given limit. This determines two initial slice
edge positions (left and right) and an implicit slice center (the average of the
two).
\begin{figure}
\centering
\includegraphics[width=\linewidth]{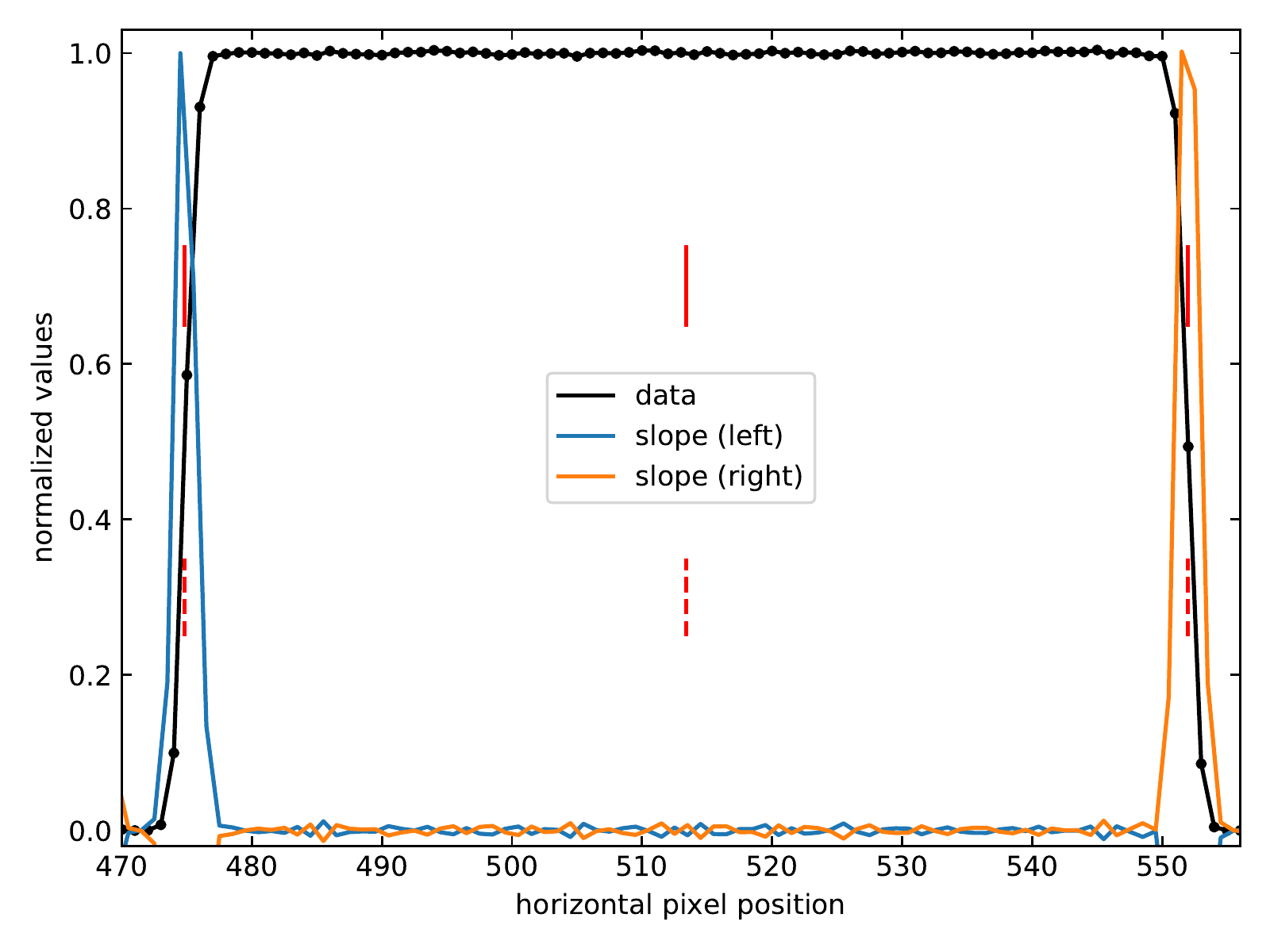}
\caption{Plot to illustrate the tracing procedure and the edge refinement. We
         show a cut through a MUSE slice at the CCD level, illuminated by a
         flat-field lamp. The data itself is displayed normalized to the constant
         region in between the edges (black). The slope of the data for the
         left edge (blue) and the right edge (orange) are displayed as well.
         The original trace position (vertical dashed red lines) and the refined
         edges (solid red lines) are shown as well. In this case the refinement
         shifted the positions by less than 0.1 pixels.}
\label{fig:trace}
\end{figure}
Since the slice edge can be robustly defined by the position where the flux
falls to 50\% of the flux inside the slice, both edge positions are refined.
This is done using the slope of all pixel values along the cut which contains a
peak at the 50\% position (see Fig.~\ref{fig:trace}). This peak is then fitted with a Gaussian to give a
very robust edge position, that is more accurate than a linearly interpolated
fractional edge.\footnote{An exception are the slices at the bottom-right corner
  in the MUSE field of view, where the field is affected by unfocused
  vignetting. The relevant slices are numbers 37 to 48 on the CCD in IFU 24, and
  in these, the Gaussian edge refinement is switched off for the affected edge.}

After determining the edges at all vertical positions, those with unlikely
results are filtered, using the range of expected slice widths (again, 72.2 to
82.2 pixels) as criterion. This effectively removes CCD columns where the
illuminated area of a slice is strongly affected by a dark column or some other
artifact.
Then, a polynomial (typically of 5th order) is iteratively fitted to the
remaining good trace points, using a $5\sigma$ limit for rejection. The final
tracing information then includes 3 polynomials for each slice, marking the
slice center and its left and right edge on the CCD.

In the pipeline, this routine computing the master trace table is part of the
{\tt muse\_flat} module, for efficiency reasons, and executed if the {\tt trace}
parameter is set to {\em true}. Parameters to change the edge detection fraction
({\tt edgefrac}), the number of lines over which to average vertically ({\tt
nsum}), and the polynomial order of the fitted solution ({\tt order}) can be
adjusted.

\subsection{Wavelength calibration}\label{sec:wavecal}
The wavelength calibration is essential for a spectrographic instrument. In the
MUSE pipeline, a dedicated module computes the wavelength solution for every
slice on the CCD.
This solution is a two-dimensional polynomial, with a ``horizontal'' order (2 by
default) describing the curvature of the arc lines in each slice on the CCD, and
a ``vertical'' order (6 by default) describing the dispersion relation with
wavelength.\footnote{This is similar to the {\sc fitcoords} task in the {\sc
  longslit} package of the IRAF environment.}

Because the MUSE spectrographs do not operate in vacuum, the wavelength
calibration is based on arc line wavelengths in standard air. However, if
convenient for the scientific analysis, the final cube can be constructed by
the pipeline in vacuum wavelengths at a later stage.

\subsubsection{Computing the wavelength solution}
This module expects to get a series of exposures, in which a number of images
exist for each of the three arc lamps built into MUSE (HgCd, Ne, and Xe). The
use of different lamps ensures a reasonable coverage of the wavelength range of
the instrument. Typically five exposures per lamp are used, to maximise the S/N
for fainter lines without saturating the brightest ones.
All images of the sequence are bias subtracted as discussed before. Optionally,
they can be dark corrected and flat-fielded as well, but these calibrations are
typically not used for the wavelength calibration. The units are then converted from
adu to electron, normally using the gain value given in the raw data.
Contrary to other modules, the exposures are not {\em all} combined. Instead
they are sorted into the sub-sequences of exposures illuminated by each of the
three arc lamps which are then combined, so that the following analysis is done
separately on three images. This ``lamp-wise'' handling has the advantage that
the images contain fewer blends of emission lines, and so the lines are easier
to identify and measure. The combination method used by default is the
$3\sigma$-clipped average (see Sect.~\ref{sec:imcomb}).

The actual analysis works separately for each slice.
From the reference list of arc lines, only the lines for the relevant lamp are
selected.
To detect the corresponding lines on the real exposure, a S/N spectrum is
created from the central CCD column of each slice, by dividing the \DATA image
by the square root of the \STAT image. This lets the detection work equally
well, if the arc exposures were flat-fielded or not. After subtracting a
median-smoothed version to remove any constant background, lines are detected
using $1\sigma$ peaks (by default) in terms of mean of the absolute median
deviation above the residual median of the full spectrum. The initial line center in
CCD pixels is then determined using Gaussian fits to each detection. Artifacts
that are not arc lines in these detections are filtered out, by rejecting
single-pixel peaks and those with FWHM outside the range $1.0\dots5.0$\,pixel, a
flux below 50\,e$^{-}$, and with an initial centering error $>1.25$\,pixel.
The detections then need to be associated with known arc lines. This is done
using an algorithm based on one-dimensional pattern matching
\citep{2008eic..work..191I,FORS_PipeMan}. This only assumes that the known lines
are part of the detected peaks and that the dispersion is locally linear,
inside a range of $1.19\dots1.31$\,\AA\,pixel$^{-1}$.
A tolerance of 10\% is assumed by default when associating distances measured in
the peaks with distances in the known arc lines. For WFM data, this process
typically detects between 100 and 150 peaks which are then associated with 90
to 120 known arc lines, all arc lamps taken together. For NFM, where the arc
lines do not reach the same illumination level due to the $64\times$ lower
surface brightness levels, 85--110 detections turn into 70--100 identified
lines.
The analysis on each of the per-lamp images continues by fitting each of the
identified lines with a 1D Gaussian in each CCD column, over the full width of
the slice as determined from the trace table, to determine the line center. To
reject deviant values among these fits, which might be due to hot pixels or dark
columns (not all of which are detectable on previous calibration exposures),
this near-horizontal sequence of the center of a line is iteratively fit with a
one-dimensional polynomial of the ``horizontal'' order. The iteration by default
uses a $2.5\sigma$ rejection.
After all individual lines and multiplets of all arc lamps have been treated in
this way, the final solution is computed by an iterative, two-dimensional
polynomial fit to all measured arc line centers and their respective reference
arc wavelengths.
This iteration uses a $3\sigma$ clipping level.
This fit is typically weighted by the errors of all line centroids added in
quadrature with the scatter of each line around the previous 1D fit.
The coefficients of the fit are then saved into a table.

In the pipeline, the calibration routine is implemented in {\tt muse\_wavecal}.
The parameters for the polynomial orders are {\tt xorder} and {\tt yorder}.
The line detection level can be tuned with the {\tt sigma} parameter, and the
pattern matching with {\tt dres} and {\tt tolerance}. The iteration $\sigma$
level for the individual line fits is called {\tt linesigma}, the one for the
final fit {\tt fitsigma}, and the final fit weighting scheme can be adapted
using {\tt fitweighting}.

\subsubsection{Applying the wavelength solution}\label{sec:applwave}
Applying the wavelength solution simply evaluates the 2D polynomial for the
slice in question at the respective CCD position. This provides high enough
accuracy for other daytime calibration exposures.

When applying this to night-time (science) data, however, it becomes important
to notice that those data are usually taken a few hours apart from the arc
exposures and the ambient temperature might have significantly changed in that
time.  This results in a non-negligible zero-point shift of the wavelength
of all features on the CCDs, of up to 1/10th of the MUSE spectral resolution or
more.

The procedure to correct this was already briefly mentioned in
Sect.~\ref{sec:scibasic}. After applying the wavelength solution to night-time
data, they are stored in one pixel table for each IFU.
From this table, a spectrum is created, simply by averaging all pixel table
values whose central wavelength fall inside a given bin.
By default, the bins are 0.1\,\AA\ wide which oversamples the MUSE spectral
resolution about $25\times$. In effect, this results in high S/N spectra of the
sky background, since one averages approximately 3600 spectra (all illuminated
CCD columns). This requires the objects in the cube to be faint compared to the
sky lines.
Since this is not always the case, the spectrum is reconstructed iteratively, so
that pixels more than $\pm15\sigma$ from the reconstructed value in each
wavelength bin are rejected once.  Here, the $\sigma$-level is used in terms of
standard deviation around the average value.
For cases with very bright science sources in the field, this does not remove
the sources well enough from the sky spectrum and iterative parameters may have
to be tuned to $2\sigma$ for the upper level and 10 iterations.
The spectrum is only reconstructed in regions of around the brightest sky lines
(by default, $\pm5$\,\AA\ around the \oi lines at 5577.339 and 6300.304\,\AA).
Any shift from the tabulated central wavelengths of these sky lines
\citep{NIST_ASD_2014} in the real data is then corrected by just adding or
subtracting the difference from the wavelength column of the pixel table.
Because the reconstructed spectrum contains data as well as variance
information, the fitted Gaussian centroids are computed together with error
estimates. The final shift is computed as the error-weighted mean centroid
offset of all lines given. Since the mentioned \oi lines are the brightest
lines in the MUSE wavelength range, and the only bright lines that are isolated
enough for this purpose, these are selected by default.
Only if the science data contains a similarly strong feature at the same
wavelength that covers much of an IFU, the user should select a different line.

The method described here is implemented in the MUSE pipeline in the {\tt
muse\_scibasic} module. The parameter {\tt skylines} gives the zeropoint
wavelengths of the sky emission lines to use, {\tt skyhalfwidth} determines the
extraction window for the fit, with {\tt skybinsize} one is able to tune the
binning, and {\tt skyreject} allows to set the rejection parameters for the
iterative spectrum resampling.

\subsection{Geometric calibration}\label{sec:geo}
One central, MUSE-specific, calibration is to determine where within the field
of view the 1152 slices of the instrument are located. This is measured in the
``geometric'' calibration. The ``astrometric'' calibration then goes one step
further and aims to remove global rotation and distortion of the whole MUSE
field.
The instrument layout underlying this procedure is displayed in
Fig.~\ref{fig:instlayout}.

\begin{figure*}
\centering
\includegraphics[width=0.8\linewidth]{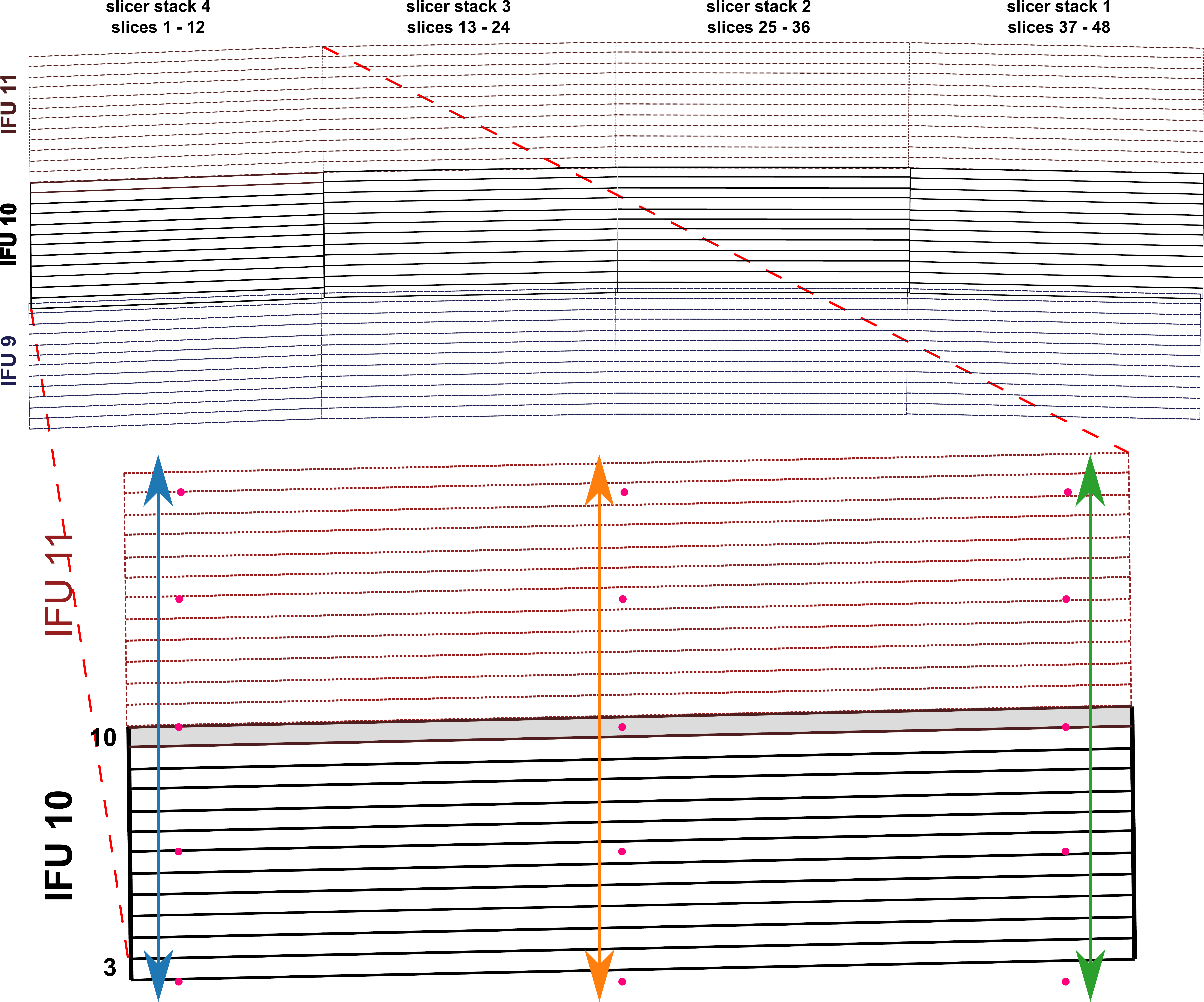}
\caption{Sketch of the geometry of selected IFUs.
         The four stacks of slices are marked in the top figure which shows
         three selected IFUs of MUSE. The upper part shows that IFUs 9 and 10
         partially overlap in projection to the VLT focal plane; these IFUs are also
         significantly offset horizontally.
         The lower part displays the approximate location of the pinholes (the pink
         dots) relative to the slices of the leftmost slicer stack in two of
         those IFUs. Slice 10 of IFU 10 is highlighted with a grey background.
         During the exposure sequence, the pinholes are moved vertically, the
         arrows represent the motion that resulted in the flux distribution
         depicted in Fig.~\ref{fig:geoseq}, where the curves are displayed in
         the same color for the same pinhole.
}
\label{fig:geosketch}
\end{figure*}

\begin{figure}
\includegraphics[width=0.9\linewidth]{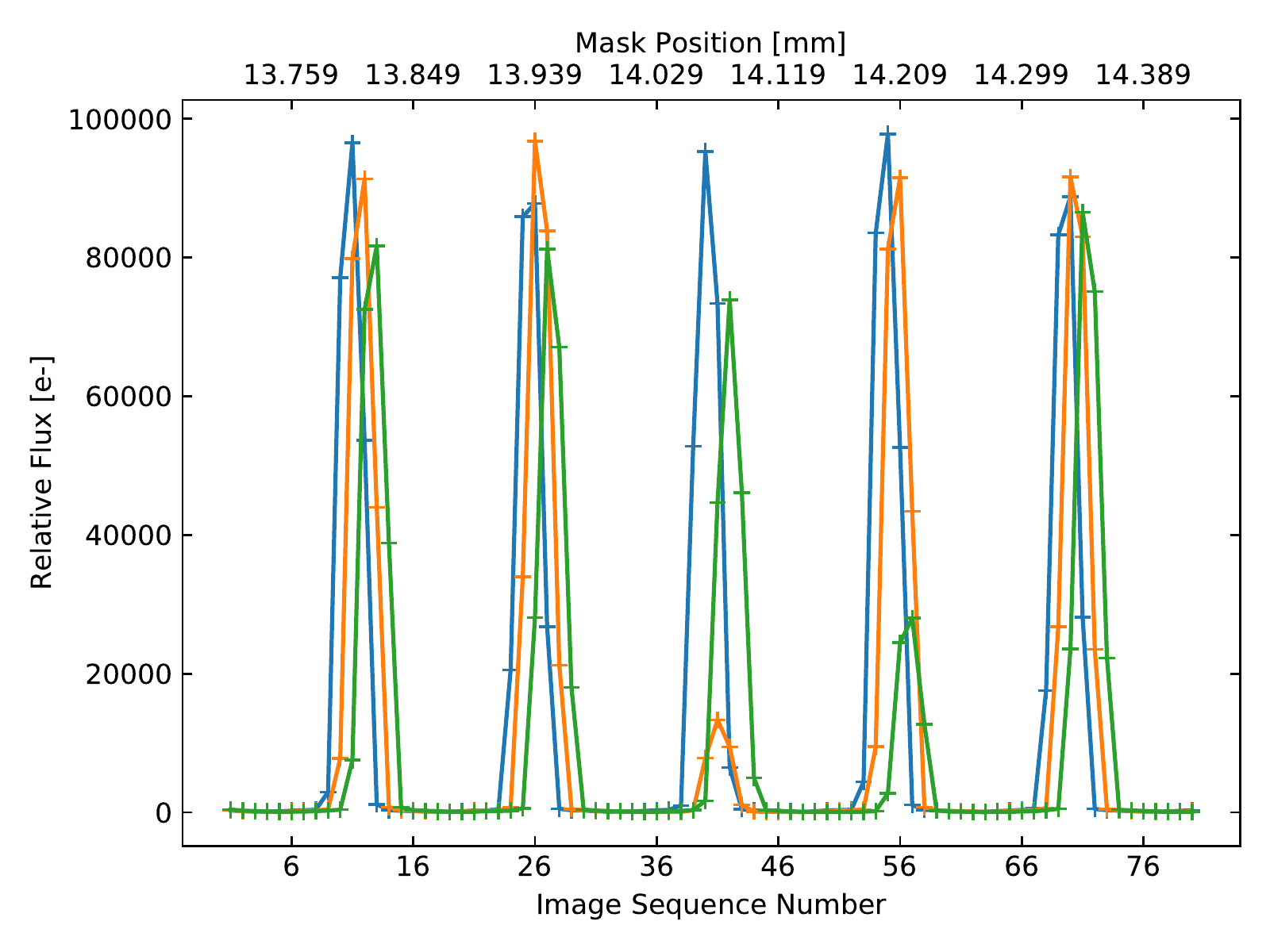}
\caption{Relative illumination of slice 10 of IFU 10 from the geometrical
         calibration sequence taken on 2015-12-03. The measurements are
         from the arc line \ion{Ne}{i}\,6383. The three colors represent the
         fluxes measured for the three different pinholes (see
         Fig.~\ref{fig:geosketch}) that illuminate the slice.  The slice is
         illuminated by the pinholes five times. Since the peaks occur at
         different positions for the different pinholes, one can already tell
         that the slice is tilted in the MUSE field of view, by 0.773\degr\ in
         this case.
         (Two of the pinholes are dirty, so that the orange peak near exposure
         40 and the green peak at exposure 56 reach lower flux levels.)
}
\label{fig:geoseq}
\end{figure}

To measure this geometry, and determine x and y positions, width, and angle for
each slice, a pinhole mask is used. This mask contains 684 holes,
distributed in 57 rows of 12 holes, with a horizontal distance of 2.9450\,mm and
a vertical distance of 0.6135\,mm between adjacent holes; the covered field is
therefore approximately equal to the $35\times35\,$\,mm$^2$ field that
corresponds to the MUSE field in the VLT focal plane\footnote{The focal plane
  scale for VLT UT4 with MUSE is 1\farcs705\,mm$^{-1}$.}.
The mask gaps are chosen so that a given slice in its place in the image slicer
is simultaneously illuminated by three pinholes, and every fifth slice is
illuminated in vertical direction.  A partial visualization of this setup is
shown in Fig.~\ref{fig:geosketch}.  This mask is then vertically moved across
the field, in 60-80 steps of 9\,$\mu$m, while being illuminated by the Ne and
HgCd arc lamps.\footnote{Contrary to the wavelength calibration procedure, both
  arc lamps simultaneously illuminate the same exposure.}
If the light from a given pinhole illuminates a slice in one MUSE IFU, the
corresponding location on the CCD is illuminated as well, in the form of a
bright spot, whose size is dominated by the instrumental PSF. The expected
position of the spot can be well determined from the mask layout together with
the trace table (Sect.~\ref{sec:tracing}) and the wavelength calibration table
(Sect.~\ref{sec:wavecal}). As the pinholes are moved and the pinhole disappears
from the position in the MUSE field that this slice records, this spot gets
less bright or completely disappears. The outcome of such a sequence is plotted
in Fig.~\ref{fig:geoseq} which shows the illumination (flux distribution) of
three pinholes observed by one slice through the sequence of 60-80 exposures.
Note that slices on the edge of the field of view are only illuminated two or
three times. Together with the known properties of the pinhole mask and the
instrument as well as other calibrations, the position, width, and tilt of each
slice in the field of view can be determined as follows.

The processing has two stages, the first separately handles the data from all
IFUs (in parallel), the second then derives a global solution using the data
from all IFUs. In the IFU-specific part, the raw data is first handled as other
raw data, so that it is bias-subtracted and trimmed, converted to units of
electrons, and optionally dark subtracted and flat-fielded. Next, all input
mask exposures of each IFU are then averaged, and this combined image, together
with the trace table and wavelength calibration as well as the line catalog are
used to detect the spots of the arc lines.
For this purpose, the line catalog of the wavelength calibration is taken, but
reduced to the 13 brightest
isolated lines of Ne, Hg, and Cd in the wavelength range 5085\dots8378\,\AA.
Based on tracing and wavelength calibration, rectangular windows are
constructed for each slice and arc line, over the full width of the slice on
the CCD, and $\pm$7 pixels in wavelength direction. In this window, a simple
threshold-based
source detection is run, by default using 2.2$\sigma$ in terms of absolute
median deviation above the median value. Only if exactly three spots are found,
the routine continues, otherwise artifacts may have been detected. Since the
detection is run for all arc lines, and the geometry is not wavelength
dependent, failed detection for a single line is not critical.
The flux of all spots is then measured at the detected position, in {\em all}
60-80 exposures. This is done using simple integration in a rectangular window
of $\pm$5 pixels in all directions, subtracting a background in an aperture of
$\pm$7 pixels.
The centroid and FWHM of each spot are measured as well, by using a 2D Gaussian
fit.\footnote{A simpler barycenter measurement together with a direct
  determination of the FWHM from the pixel values was initially used, but the
  resulting precision was not sufficient. Since the spots in the wavelengths
  used for this calibration are very compact, the Gaussian is a good
  representation.  The barycenter method can still be switched on by setting the
  {\tt centroid} parameter to {\tt barycenter}.}
The corresponding exposure properties (especially the mask position) is taken
from the FITS header of the exposure. The result of these measurements is that
for each of the three spots in each slice, for each arc line, and each exposure
we have a set of CCD and spatial mask positions as well as the flux. Altogether
these are a maximum of 149760 values, but in practice about 140000 of them are
valid and are used for the further analysis.

While the vertical distance of the pinholes is known to good enough
precision, the motion of the mask is not calibrated well enough, since
it includes an uncertainty about the angle of the mask inside the mask
wheel and the relative motion. The effective pinhole distance
therefore needs to be self-calibrated from the actual data.
To do this, the centroid of all flux peaks (visible in Fig.~\ref{fig:geoseq})
is measured in terms of the vertical position and the distance between all
peaks on the scale of the vertical position. The difference between all peaks
is then tracked and, after rejection of outliers with unphysical distances, the
average difference is the effective vertical pinhole distance, this is then
converted to a scale factor $f_\mathrm{dy}$. Its standard deviation indicates
an accuracy of 5\,$\mu$m or better, about 4\% of the slice height.

To start determining the slice position in the field of view, the
central flux peak of the exposure series is taken. From this, the
weighted mean CCD position in x and y is computed, taking into account
the flux of the spot in each exposure of the peak. The mask position
of this peak is recorded as well.

Using all three spots of a given slice, one can compute the scale $s$
of a slice, using the average distance $\left<\delta x\right>$ between the pairs of
pinholes of a slice:
\begin{eqnarray*}
\left<\delta x\right> &=& (\delta x_1 + \delta x_2) / 2\enspace\mathrm{pix}\\
                    s &=& 2.9450\,\mathrm{mm} / \left<\delta x\right> \mathrm{pix}
\end{eqnarray*}
The width $w$ of a slice in the field of view then follows from its
width $w_\mathrm{CCD}$ as measured on the CCD:
\begin{equation}
w = 1\farcs705\,\mathrm{mm}^{-1} / 0\farcs2\,\mathrm{pix}^{-1}\ s\ w_\mathrm{CCD}
\label{eq:geowidth}
\end{equation}
Here, $w_\mathrm{CCD}$ can be taken directly from the width of the slice as
determined by the trace function (Sect.~\ref{sec:tracing}) at the vertical
position of the detected spots.
A simple error estimate $\sigma_w$ is propagated from the standard deviation
of the two measured distances $\delta x_1$ and $\delta x_2$ within each slice.
If the width of a slice for a given arc line was determined to be outside
predefined limits (72.2 to 82.2 pixels), the measurement is discarded.

The angle $\varphi$ of each slice can by computed using the known distance of the
pinholes in the mask and the distance between the mask positions of the maxima
$p$ between two pinholes:
\begin{equation}
\varphi = \arctan(\Delta p / 2.9450\,\mathrm{mm})
\label{eq:geoangle}
\end{equation}
Since it contains three spot measurements, each slice has two independent angle
measurements. These are averaged to give one angle value per slice and arc
line.
Any angles above 5\degr\ are discarded at this stage since they are unphysical.
An error estimate $\sigma_\varphi$ is computed using the standard deviation of the
two individual measurements.

\begin{figure}
\includegraphics[width=\linewidth]{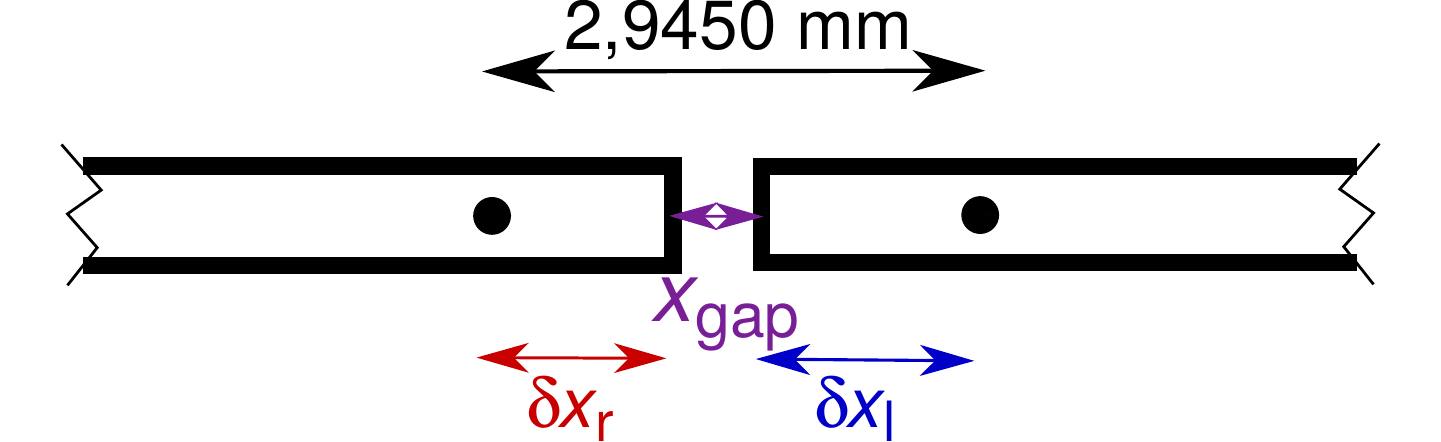}
\caption{Illustration of the computation of the gap between slices.}
\label{fig:geogap}
\end{figure}

The next step is to combine the measurements from all arc lines into one.
This is done by an error-weighted average of all measurements of $w$ and
$\varphi$, after rejecting outliers using sigma-clipping
in terms of average absolute deviation around the median.

Since any horizontal gap between the adjacent slices together with the width
$w$ of the involved slices determines the central position of these slices, we
compute the gap based on the positions of the spots within the involved slices.
For this, we take the distance $\delta x_\mathrm{l}$ of the left-most spot to
the left edge of the slice on the CCD and $\delta x_\mathrm{r}$ of the
right-most spot to the right edge of the slice, in CCD pixels.
Of these we again compute a sigma-clipped mean over all arc lines and then
compute the (central) gap in x-direction by subtracting the measured distances
from the total width of the slice as given by the mask design. With the scale
factors involved we get
\begin{equation*}
x_\mathrm{gap} = 2.9450\,\mathrm{mm}\ 1\farcs705\,\mathrm{mm}^{-1}
               / 0\farcs2\,\mathrm{pix}^{-1}\
               \left(1 - \dfrac{\delta x_\mathrm{l}}{\left<\delta x\right>}
                     - \dfrac{\delta x_\mathrm{r}}{\left<\delta x\right>}\right)
\end{equation*}
for the width of the gap in units of pixels.\footnote{In case this gap estimate
  is determined as negative, an average gap of 0.04\,pix is taken for an inner
  and 0.07\,pix for an outer gap. These values were determined as typical values
  for the gaps during instrument commissioning.}
This is illustrated in Fig.~\ref{fig:geogap}.  For the two central slices, the
initial x positions are then
\begin{equation}
x_\mathrm{init} = \pm(x_\mathrm{gap} / 2 + w / 2)
\end{equation}
with $w$ from Eq.~\ref{eq:geowidth}.  The error estimates of these are
propagated from the individual measurements.  The positions of the outer slices
are computed in a similar way, but then depend on the positions of the inner
slices.

This initial estimate of the horizontal position is already accurate for the
relative positions {\em within each IFU}, but needs to be refined to give a
correct global estimate. For this, the relative positions of central spots that
pass through the slices in adjacent IFUs (\eg, the top slice in IFU 10 and the
bottom slice in IFU 11, as shown in Fig.~\ref{fig:geosketch}) are compared.
If the slices were centered at the same horizontal position, the spots would
have the same position relative to the slice center. Any deviation
$x_\mathrm{diff}$ from this can therefore be corrected by shifting all slices
in all following IFUs. Since after this correction, the field of view is not
centered any more, the middle of the central gap of the top row of slices of
IFU 12 is taken as a horizontal reference point $x_\mathrm{ref}$, which is
shifted to zero position. The fully corrected horizontal position $x$ of the
slices is then:
\begin{equation}
x = x_\mathrm{init} - \left< x_\mathrm{diff}\right> - x_\mathrm{ref}
\label{eq:geox}
\end{equation}
where $\left< x_\mathrm{diff}\right>$ is the weighted average value, after
rejecting outliers, of the horizontal shifts determined for spots of all arc
lines.

While deriving the weighted mean width and angle for each slice, the weighted
mean mask position for the central flux peak is computed as well, which in turn
is converted into a sigma-clipped weighted mean position using the data from
all arc lines. This is now used to determine the last missing entry of the
geometrical calibration, namely the vertical center, $y$. After subtracting
an offset, to center the field at the top row of IFU 12, we loop through all
four slicer stacks (see Fig.~\ref{fig:geosketch}) of all IFUs. Within each
slicer stack, we go through all slices in vertical direction. Starting at the
central reference slice and going in both directions, we convert the central
mask position into a relative position. While doing this, we can detect if a
flux peak originates from a different pinhole in the mask, since we know the
approximate vertical size of the slices ($\sim 120\,\mu$m). This procedure
results in monotonically increasing vertical positions in units of mm
$y_\mathrm{mm}$ which are then converted to the final y-position in pixels
using
\begin{equation}
y = 1\farcs705\,\mathrm{mm}^{-1} / 288\ f_\mathrm{dy}\ y_\mathrm{mm}
\label{eq:geoy}
\end{equation}
where 288 is the number of vertical slices, representing the nominal number of pixels,
and $f_\mathrm{dy}$ is the scale factor that relates the nominal and effective
vertical pinhole distance of the mask. An error estimate is propagated from
the initial measurement of the mask position of the flux peak.
One last trick in the computation of $y$ concerns the top and bottom slices of
each IFU. Since by design, these may only partially be illuminated by light
coming from the relay-optics, their vertical position can appear to be offset,
so that they seem to overlap with adjacent slices from the same IFU. Since this
is unphysical, it is corrected by assuming that these edge slices will have the
same vertical distance as all the other (non-edge) slices in the respective
IFU.

For a few slices, the computed four coordinates may be different from the
surrounding slices. This occurs if artifacts like hot columns influence
detection or centroid computation of the spots. In early versions of the
pipeline (before v1.2) these few slices were corrected by hand in the resulting
table. Since then, a new procedure was implemented to automate this.
As the image slicer in each MUSE IFU is built of rigidly mounted stacks of
slices, one can take advantage of this, and demand that all properties should
change smoothly across the 12 vertical slices in each stack. By fitting a
linear relation to each property, with the number of each slice in vertical
direction as the abscissa used in the fit, one can easily find the outliers
among those 12 slices. The values for these are then set to the value of the
fit at that slice number. This non-iterative replacement was found to perform
best when used at a $1.5\sigma$ level.
Then less than 20\% of the slices are changed for each of the four properties,
and the resulting geometrical calibration is very smooth and no outliers are
left.

It should be mentioned that with this procedure, and using
Eq.~\ref{eq:geowidth}, \ref{eq:geox}, and \ref{eq:geoy} we arrive at a
geometrical calibration that is in units of pixels. However, these are really
``pseudo-pixels'' that are constructed such that each horizontal and vertical
element is defined to be 0\farcs2 on the sky. In reality, the sampling of MUSE
is slightly different and varies across the field. The astrometric calibration
(Sect.~\ref{sec:astcal}) is the step to correct this before producing the final
cube that will be analyzed for scientific purposes.
The geometric calibration changes very little over time. Unless instrument
interventions or earthquakes affect the relative positions, only a monthly or
quarterly recomputation is necessary. The geometric calibration does not need
to be run by normal users, but is typically provided with the science data by
ESO.

The pipeline module that carries out the procedure described in this section is
called {\tt muse\_geometry}. The parameter that determines the spot detection
sigma-level is called {\tt sigma} and
the sigma-clipping parameter for the final smoothing is called {\tt smooth}.

\subsection{Twilight sky-flat fielding}\label{sec:twilight}
The MUSE calibration unit \citep{KBH+12} provides an illumination that closely
but not perfectly resembles the illumination of objects on the sky. To remove
any residual gradients, the pipeline makes use of the bright sky background
exposures,
with the sequences started before sunset to reach high illumination levels in
short exposures.
They should very well resemble the illumination of the MUSE instrument with a
constant background.
These twilight skyflats are bias-subtracted, converted from adu to electrons,
optionally dark-subtracted, flat-fielded and then all exposures are combined,
using the chosen combination parameters (usually a $\sigma$-clipped average).
Using wavelength calibration, tracing solution and geometry table, each pixel
in these combined images is then assigned 3D coordinates (spatial and
wavelength), creating a pixel table for each IFU.  Since the red part of the
skyflats can be strongly affected by the spectral 2nd order depending on the
instrument mode, the pixel tables are cut in wavelength.
In AO modes, the light in the wavelength region of the NaD laser light are
blocked using notch filters, and pixels around this wavelength contain
only noise.  They are therefore excluded from further processing as well.
If the optional raw illumination flat-field exposure was given as input, it is
then used to correct the relative illumination between all slices of one IFU.
The sum of the values in these pixel tables are computed and saved, to later be
used for the relative scaling among all IFUs.
The pixel tables of all IFUs are then merged as already described in
Sect.~\ref{sec:scipost} for the science data. A cube is then resampled from the
merged data, using a sampling of 250\,\AA\,pixel$^{-1}$ in wavelength
direction, and a white-light image gets created. This skyflat cube is then
saved, together with the image.

This cube and image contain the residual gradients that need to be corrected in
science data but also small scale artifacts, especially strong variations at
the edges of the slices and slicer stacks.
The cube is therefore post-processed, to produce a correction that only
contains the large-scale gradient. To this end, a mask of the illuminated area
is created. If a vignetting mask was provided\footnote{The lower right corner
  of the MUSE WFM field was strongly vignetted before mid March 2017 in
  calibration exposures, causing higher values in the corner of flat-fielded
  twilight exposures.}, its area is not part of the illuminated region. The
illuminated area is then smoothed by a median filter ($5\times7$ pixels in size) to
remove very small-scale outliers, normalized to 1, and fit with a 2D polynomial
(by default, with order 2 in both directions), and normalized again. This is
repeated for all wavelengths in the cube.

If a vignetting mask was provided or NFM data is processed, a small area close
to the edge of the MUSE field is used to compute a two-dimensional correction
for the vignetted area: The original unsmoothed white-light image is corrected
for large scale gradients by dividing it with the smooth white image. The
residuals in the edge area (as defined by the vignetting mask or using the top
22 pixels of the field for NFM) are then smoothed using input parameters. By
default a $4\times4$ order polynomial is used for this, but Gaussian or median
filtering can be used instead. This smoothed vignetting correction is then
multiplied onto each plane of the smooth cube, before normalizing each
wavelength plane of the cube again.
The smoothed cube and white-light image are then saved to disk.

The pipeline module for this procedure is called {\tt muse\_twilight}. The
parameters that determine wavelength range and sampling are {\tt lambdamin},
{\tt lambdamax}, and {\tt dlambda}, the polynomial orders can be changed with
{\tt xorder} and {\tt yorder}, the vignetting model smoothing can be changed
with {\tt vignsmooth}, and the size of the built-in mask in case of NFM data
can be adjusted with {\tt vignnfmmask}.

\subsection{Line spread function measurement}\label{sec:lsfmeas}
Measuring the line spread function (LSF) is important within the MUSE pipeline
to get good sky subtraction and to model the laser-induced Raman lines, but can
also be used for scientific analysis on the reduced cube. Since the LSF is
determined on data recorded by the MUSE CCDs, it implicitly includes the slit
width (of the MUSE slices) and the bin width of the detector.

To derive the LSF, this module requires the same inputs as the wavelength
calibration (see Sect.~\ref{sec:wavecal}), as well as the final wavelength
solution derived from them. The LSF is improved and better sampled, if long
sequences of exposures of all three arc lamps are used, so that the arc lines
are detected with high S/N over the full MUSE wavelength range.
As before, the raw exposures are overscan-corrected, bias-subtracted, and
optionally, corrected for dark current (not by default) and flat-fielded.
Further processing depends on the type of LSF that is used. The MUSE pipeline
mainly uses an interpolated 2D image which represents the line shape as
a function of wavelength, for each IFU and slice of the instrument.
An alternative implementation uses a Gauss-Hermite function to model
the line shape as a function of wavelength. In this case, the coefficients are
tracked in a table structure instead of an image. Since using the LSF for
sky subtraction is computationally expensive and the former method is much
faster and more robust, it was chosen as default.
Both formats are further described in Sect.~\ref{sec:lsf}.

If the interpolated image approach is used, the image of all arc exposures are
converted into a special pixel table, that contains only the data around the
brightest and most isolated arc lines from the line list, within a certain
range from the known peak of the line. No combination of images is carried out
to avoid introducing biases at this stage.
This arc-line pixel table is then divided up into the individual slices to fit
the individual LSFs.
The computation is then carried out as a 2D regression, where the x-direction
is along the LSF, i.e.\ across the arc lines, and the y-direction the
wavelength, i.e.\ the central wavelength of the arc lines. For each step in LSF
direction, all pixels of all arc lines within a certain distance range from the
line peak are fit with a 2D polynomial, with order 2 in x- and 3 in
y-direction.  The resulting image is created by evaluating all these
polynomials at their nearest output image pixel.
The LSFs of all slices of one IFU are then stored together, as a datacube, and
saved to disk.

If, on the other hand, the Gauss-Hermite function is used, the individual
images are first combined, using given combination parameters.
The resulting image is converted into a pixel table, on which the actual fit is
run. The fit runs over all slices of the IFU for which the pixel table was
created.
In a first fit, the LSF width and the fluxes of all arc lines are minimized.
The 2nd step then keeps the fluxes constant, but fits all other Gauss-Hermite
coefficients.
A spectrum can then be simulated from the LSF parameters and the arc line list,
and subtracted from the data for debugging.
The Gauss-Hermite coefficients are finally stored into a table.

The module that implements this algorithm in the pipeline is called {\tt
muse\_lsf}. The parameter {\tt method} selects either the image-based
interpolation or the Hermite-based table algorithm.  One can adapt the
interpolation method by selecting the half-size of the wavelength window around
each arc line ({\tt lsf\_range}, 7.5\,\AA\ by default), the size of the
interpolated image in 2D ({\tt lsf\_size} for the LSF direction, by default 150
pixels, and in wavelength direction, {\tt lambda\_size}, defaulting to 30
pixels), and the size {\tt lsf\_regression\_window} in LSF direction used for
the interpolating fit.

\subsection{Sky subtraction}\label{sec:skysubmain}
As mentioned already in Sect.~\ref{sec:scipost}, the two ways to operate the sky
subtraction are
\case{a} determining and subtracting the sky in the science exposure itself and
\case{b} observe offset sky fields to characterize the sky background to then
subtract it from the actual science exposure.
The actual algorithm is the same for both, the only difference is how and when
the sky spectrum is created and decomposed.

In case \case{a}, the sky only needs to be determined and decomposed once. The
sky line fluxes have not changed and can be directly subtracted from the
science data. We describe the full procedure in Sect.~\ref{sec:skysub} below.

In case \case{b}, however, the sky lines are expected to have changed between
the exposure of the offset sky and the science field. The characterization
of the sky is the same, but needs to be stored in a way that it can be applied
to the science exposure. In the Sect.~\ref{sec:skydecom} we describe in detail
how this is done.

The sky subtraction algorithm used in the MUSE pipeline was already described
by \citet{SWB+11} and the basic ideas have not changed. However, the handling
of the line spread function (LSF) has evolved after on-sky data showed the
original implementation to be insufficient (see Sect.~\ref{sec:lsf}).
The MUSE pipeline tries to combine the ideas of \citet[][work on unresampled
data and use knowledge of sub-pixels information]{Kel03} and \citet[][employ
groups of sky-lines]{Dav07} to derive a good description of the sky background.
Since the sky continuum changes slowly (on timescales of tens of minutes) while
most of the telluric emission lines change rapidly (within minutes or less),
the sky spectrum gets decomposed. We model the sky emission lines (as described
below in Sect.~\ref{sec:skylinefit}) and adapt their fluxes to each exposure,
and propagate the sky continuum as the residuals of a spectrum of empty sky
regions.\footnote{Another way that only works on fields with a significant part
  of blank sky background, is to directly subtract the full sky spectrum instead
  of decomposing it into lines and continuum. This is implemented in the MUSE
  pipeline for the special case where flux calibration is not available.
  It can be activated in {\tt muse\_scipost} by switching {\tt skymethod}
  parameter to {\tt simple}.}

In the way that our algorithm decomposes the sky spectrum into emission
lines and continuum, it is very similar to the approach taken in the ESO {\sc
skycorr} tool \citep{skycorr}, developed at around the same time. However, {\sc
skycorr} is set up for application on (single) object spectra and hence not
well adapted to be applied on datacubes and cannot handle our unresampled pixel
tables.

\subsubsection{Fitting telluric emission lines}\label{sec:skylinefit}
The list of sky emission lines used as input to the procedure is of high
importance. For MUSE, we chose to divide all possible lines
\citep{2006JGRA..11112307C} in the wavelength range 3129\dots11000\,\AA\ into
52 groups. Most of the lines and groups are transitions from the OH molecule
(7800 lines with 5100 of them within the MUSE wavelength range, see
\citealt{2007JChPh.126k4314V}), and we form groups for lines that have the same
upper level, which approximately vary in flux together. Other line groups
originate from O$_2$ (4400 lines in the MUSE range), \oi (71 lines, including
the three brightest), \ion{H}{i} (the two Balmer lines), \ion{N}{i}\,5198,5200,
\ion{Na}{i}\,D\,5890,96, \ion{K}{i}\,7665,99, and \ion{He}{i}\,5015, where the
single lines and doublets are forming separate groups. Of the many lines in the
full list, those outside the wavelength range and those with a flux of less
than 1/10000th of the strongest lines are removed during the fitting process,
so that typically 4000 to 4500 lines in 40 groups get used.
The ESO {\sc skycorr} tool uses a slightly different grouping of emission lines.
The advantage of using groups of lines is that the fit to the line fluxes is
more robust, and that many of the lines are unresolved at MUSE resolution. A
drawback is that the flux calibration has to be accurate in a relative sense
over the wavelength of each group.  Since the sky line list is a FITS table
distributed to users, the file can be edited to optimize the sky subtraction
residuals, for example, by removing lines that are on top of features in object
spectra.

Parameters in the fitting process are flux factors for each line group as well
as a linear correction for the wavelengths of all sky lines.
The first guess uses the pixel value at the wavelength position of the
brightest line of each group.
The actual minimization then compares the modeled sky spectrum to the data in a
differential manner, where the intensity difference of neighboring pixels forms
the spectral residual $s(p)$:
\begin{equation*}
s^2(p) = \sum_i \dfrac{\left[\Delta I_m(\lambda_i, p) - \Delta I(\lambda_i)\right]^2}
                        {(\sigma_i^2 + \sigma_{i+1}^2) \Delta\lambda^2}
\end{equation*}
with $\Delta I_m(\lambda_i, p)$ as the modeled intensity difference between two
neighboring spectral bins $i$ and $i+1$, taking into account the fit parameters
$p$, $\Delta I(\lambda_i)$ is the measured intensity difference,
$\Delta\lambda$ the bin width of the spectrum, and $\sigma_i^2$ the estimated
variance at each spectral position.
When creating the simulated spectrum at each iteration, the LSF
(Sect.~\ref{sec:lsf}) and the fluxes of all relevant sky emission lines are
taken into account.

The fit is run using the global sky spectrum, averaged over all sky regions of
an exposure. To allow interpolation with higher precision for the final
subtraction, the sky spectrum is created with an oversampling factor of 4 to a
binsize of 0.3125\,\AA. The LSF for the fit is the
weighted average LSF of all spectra -- where the weighting is determined by how
many pixels contribute from which IFU and slice to the sky --, folded with the
rectangular function of the sampling.
After the sky line fit converged, the sky continuum is created by subtracting
the line fit, again using the same weighted average LSF, from the average sky
spectrum. By default, the continuum is created with the same oversampling as
the spectrum. This approach has the drawback that any residuals of the sky line
fit directly affect the continuum, but tests showed that any post-processing
of the continuum did not result in any significant improvement. Nevertheless,
the pipeline user can provide the sky continuum as input to the relevant
pipeline modules to override the computed continuum.

This emission line fit is run in the module {\tt muse\_create\_sky} and during
processing of the on-target science data in {\tt muse\_scipost}.

\subsubsection{Handling offset sky fields}\label{sec:skydecom}
To be able to subtract the sky background from exposures of large objects that
fully cover the MUSE field of view, observations of an offset sky field are
necessary (this was called case \case{b} above).
The main reason for this is to characterize the sky {\em continuum} which
changes on timescales of tens of minutes during the night and due to moon
illumination. The sky emission lines, however, change on shorter timescales.
These, therefore, usually need to be characterized or refined in the actual
science exposures.

Offset sky fields are processed in the same basic way as science data or standard
star exposures, so the module that handles the sky exposure starts by loading
and merging the pixel tables of the individual IFUs. It divides the data by
the smoothed average flat-field spectrum, and flux calibrates the data, using
the given response and extinction curves. The correction for telluric absorption
is carried out using the provided correction spectrum. Any kind of bad pixel is
then removed from the table and the remaining pixels are corrected for
differential atmospheric refraction (for WFM data).
The next step is to create the sky spectrum which is assumed to be constant
across the small MUSE field of view. To this end, a cube is reconstructed using
nominal instrument sampling, and a white-light image integrated over all
wavelengths is created from it. Thresholding is used to first remove bad pixels
(on the edge or in between image slicers in the MUSE field) to then select
the chosen lowest percentile of pixels. In most cases, the resulting mask
should cover a large fraction of the MUSE field, for example 75\%. This mask is
then used to select the corresponding pixels in the pixel table, so that the
further work is carried out on unresampled data.
The selected pixels are resampled into a 1D sky spectrum, normally using four
times the nominal sampling (\ie, 0.3125\,\AA\,pix$^{-1}$).  To remove any
remaining outliers, this is done iteratively, rejecting pixels deviating more
than $15\sigma$ on the second pass.

Next, the initial set of sky emission lines is read in from the specified
table, and the line spread function (LSF) for all slices is loaded as well.
The LSF is averaged using the relative contributions of the slices to the sky
background area and folded with the rectangular function to match the resampled
sky spectrum. This average LSF is then used to fit the fluxes of all sky
emission lines, using Levenberg-Marquardt minimization.
To make the fit robust and efficient, the underlying assumption is that most
sky emission lines are part of groups, which vary in flux together. The solution
allows for linear variation of the wavelengths against the wavelength
calibration present in the pixel table.
A telluric emission spectrum is simulated using this fit and the averaged LSF
and subsequently subtracted from the sky spectrum to compute the sky continuum.
The table of sky lines with the updated line fluxes is saved to disk as is
the resulting sky continuum. The auxiliary files (the white-light image, the
sky mask, and the sky spectrum) are stored as well.

Tests showed that only short exposures are necessary for the offset sky
field. Since the sky spectrum gets averaged over a large part of the MUSE
field, the S/N of the spectrum is extremely high after already 2\,min. We
estimate statistical $S/N\gtrsim250$ for the continuum and $S/N>1\,000$ for
even moderately bright emission lines (\eg, the \ion{N}{I}\,5197.92,5200.28
doublet) for a 240\,s sky exposure. When modeling the sky spectrum, systematic
errors like line blends and limiting accuracy of the line spread function
therefore start to dominate with sky exposures $>60$\,s.

The module that handles offset sky fields is {\tt muse\_create\_sky}.  The
parameter {\tt fraction} controls the area of the field of view to be regarded
as sky background and (default: 0.75, since we expect the offset sky field to
be almost empty), {\tt ignore} (default: 0.05) is the fraction of the field to
be ignored.  Further parameters control the wavelength range used by the
processing ({\tt lambdamin}, {\tt lambdamax}; default: all wavelengths) and set
the reference wavelength for the atmospheric refraction correction ({\tt
lambdaref}, default: 7000\,\AA). Expert users can also adapt the sampling used
for the sky spectrum and continuum; the relevant parameters are called
{\tt sampling} and {\tt csampling}.

\subsubsection{Science sky subtraction}\label{sec:skysub}
The sky \emph{subtraction} on the science data within the post-processing
module starts from a pixel table that is merged from all IFUs, corrected for
the lamp-flat spectrum and for atmospheric refraction. Optionally, the data are
corrected for Raman signatures (see Sect.~\ref{sec:raman}) and autocalibrated
(Sec.~\ref{sec:autocal}). The data also have to be flux-calibrated.
The user has to decide, if the science field was empty enough
to be used for the construction of a sky spectrum (\case{a} from above) and
provide only the list of sky emission lines. Or, if it was filled by an object
(\case{b}), and then a pre-fitted sky-line list and a sky continuum has to be
provided.

In both cases, a datacube is reconstructed from the pixel table and a
white-light image is created from it. The image is thresholded, assuming that
the darkest pixels (by default 5\% of the field of view) are artifacts to be
ignored, the faintest pixels (by default, the next 10\% but this can be adapted
for largely empty fields) are taken to be the actual sky background. All pixels
from the pixel table located at these sky positions are used to reconstruct a
spectrum of the sky background in the exposure.

In case \case{b}, the choice is usually to re-fit the emission lines
in the science exposure, since they are likely to have changed in flux since
taking the offset sky field. Should the sky spectrum reconstructed from the
science exposure be strongly affected by object emission lines, or if
processing time is an issue, the refit can be skipped. In this case, the
reconstruction of the cube and sky spectrum will not be done, either.

In case \case{a}, the input sky line list usually is the default list and an
emission line fit is done from the sky spectrum as described in
Sect.~\ref{sec:skylinefit}.

The continuum is linearly interpolated to the wavelength of each entry in the
pixel table, and subtracted. The oversampling mentioned above helps to reduce
artifacts when interpolating it onto each bin.
To subtract the sky emission lines, the spectrum is modeled and subtracted
separately from the data of each IFU and slice. The line fluxes are folded with
the corresponding LSF that was interpolated to the wavelength of the line and
the position in the MUSE field, and then subtracted from each entry in the
pixel table. This minimizes residuals due to changes of the LSF across the
field.

The sky subtraction is run as part of the {\tt muse\_scipost} module. The method
can be selected with the {\tt skymethod} parameter, with the possible values
{\tt model} (the default) and {\tt model-subtract} (no line refitting) for the
algorithm described here. {\tt skymodel\_ignore} allows to adjust the fraction
of darkest pixels to ignore as artifacts, {\tt skymodel\_fraction} defines the
fraction of the field to take as sky.
Expert users can adapt the sampling used for the sky spectrum and continuum.
The relevant parameters are called {\tt skymodel\_sampling} and {\tt
skymodel\_csampling}.

\subsection{Internal calibrations of science exposures}
Two calibration steps that can be applied to the science data itself are complex
enough to be described in more detail than possible in Sect.~\ref{sec:scipost}.
They are presented here.

\subsubsection{Correction of Raman-scattered laser light}\label{sec:raman}
After commissioning of the adaptive optics (AO) module with the MUSE instrument
in 2017, new emission lines of telluric origin were found in the spectra. These
turned out to be Raman-scattered light of the laser guide stars used to
stabilize the field \citep{2017PhRvX...7b1044V}. They appeared as features
originating from O$_2$ at around 6484\,\AA\ and N$_2$ at 6827\,\AA\ and consist
of bright peaks, unresolved at the spectral resolution of MUSE, with a band of
faint secondary peaks extending to about 50\,\AA\ from the main lines. Since
they vary spatially around the laser beams, they have to be modelled spatially
as described in the following when observing in the WFM AO modes. For NFM
and crowded WFM observations, objects in the field usually preclude such an
approach, and a constant is subtracted as part of the sky continuum.

For the modeling, it is assumed that the flux ratio of the peaks to the
secondary lines of each feature is constant. The relative fluxes of each single
Raman line are computed from molecular physics \citep[following the
prescription of][]{2019PhRvL.123f1101V}, and are given to the pipeline together
with the exact wavelength as input. To model the absolute fluxes across the
field, we extract the data of the affected wavelength ranges (by default,
$\pm10$\,\AA\ around the main peaks) and reconstruct an image for each range.
Sky regions are selected in these images, and the corresponding object pixels
as well as pixels marked as bad (cosmic rays or CCD effects) are removed from
the pixel tables. The remaining sky pixels are fit, using the estimations of
the LSF at this wavelength, with the Raman lines. The minimization adapts the
absolute fluxes of all Raman lines in a given feature, using a 2nd-order 2D
polynomial as model for the spatial domain.

This fit is run separately for the O$_2$ and N$_2$ features, and removed from
the science data. The corresponding spatial flux distribution can be output,
for both features, for further checks.

In the pipeline implementation,\footnote{This was first implemented in v2.4 of
  the pipeline, and updated with a new line list in v2.8.}
the Raman correction is done as part of the {\tt muse\_scipost} module. Besides
the input table of Raman lines, the only parameter is the extraction width,
this is called {\tt raman\_width}.

\subsubsection{Self-calibration of fluxes within each slice}\label{sec:autocal}
Flat-fielding removes spatial structure from MUSE exposures to about 1-1.5\%
accuracy. When integrating the datacube over many wavelength bins, for example,
when creating integrated images over broad-band filters or the whole wavelength
range, a pattern is left that contains the four stacks of slices within each
IFU and even to the level of single slices that have slightly different
illumination than neighboring pixels.
An autocalibration on the slice level can be activated when much of the MUSE
field is blank sky background to facilitate object detection in deep surveys.

This procedure uses the sky background signal to compute reference fluxes, and
is hence applied before sky subtraction. It is assumed that the background is
intrinsically flat across the MUSE field\footnote{For exposures taken in WFM
  with the AO laser system active, this needs the Raman correction described in
  Sect.~\ref{sec:raman} so that this is true also in the wavelengths around the
  O$_2$ and N$_2$ features. This was changed in v2.8.3 of the MUSE pipeline,
  previous versions ran the self-calibration first.} and that the science
exposure itself contains sky in at least a few ($\sim1/3$) spaxels (spectral
pixel, a spectrum in the datacube) of each slice, meaning that no large
objects are present.
By default, the MUSE pipeline then constructs a white-light image from the
pixel table, by resampling to a cube and then integrating it over the full MUSE
wavelength range.  Using this image, a sky mask is created
by thresholding it to $\pm15\sigma$ (in terms of median absolute deviation
around the median).  A morphological opening of the resulting mask with a
kernel of $3\times3$ pixels ensures that contiguous regions are marked as sky.
Then all pixels coinciding with the spatial position of the sky regions are
selected in the pixel table.
The actual algorithm to determine the flux correction factors for each slice
then divides the data into 20 wavelength ranges (19 for AO data where the NaD
region is masked). Correction factors are derived for all slices of all IFUs
and all wavelength ranges (``segments'' in short), so 23040 (21888) individual
factors are computed.\footnote{Similar functionality was available in the
  \href{https://mpdaf.readthedocs.io/}{MPDAF} package
  \citep{MPDAF_1710.03554} before it was ported to the MUSE pipeline v2.4 and
  subsequently phased out from MPDAF.}
These ranges are hardcoded and were chosen to end in between groups of skylines
to minimize the influence of the sampling that would be critical if a strong
telluric emission line would be on the edge.

Within each segment the reference level is computed first, if data of at least
50 pixels within 20 spaxels exist (\ie, are not marked as object).
The reference is determined in a multi-step process as the MAD-clipped mean and
deviation (where MAD-clipping refers to values computed after rejecting
outliers outside the $3\sigma$ median absolute deviation around the median) for
all spaxels, then again averaging the spaxels with $3\sigma$-MAD rejection on
both spatial halves for all IFUs, and then again averaging those with plain
mean and standard deviation rejection. In addition to the reference flux for a
particular wavelength range across the whole sky portion of the field of view,
reference values for each IFU and the spatial left and right half of each IFU
are tracked.
The correction factors in each segment are then computed as the ratio between
the reference flux and the flux in the segment.
Should the correction factor in one segment exceed the MAD-clipped ($15\sigma$)
mean value in the same half of the IFU, the correction factor is taken as the
mean correction for this half of the IFU instead.
The correction factors for the adjacent wavelength ranges are checked for large
deviations as well, so that if the deviation is larger than the maximum of 3\%
or $3\times$ the difference between these adjacent bins, the mean of the
corrections for those two ranges is taken instead.

The correction factors computed in this way are then applied to the pixel table
data, depending on the segment (slice and wavelength range) in question. The
corrections are applied in quadrature to the pixel variances.

Two special cases can be handled as well:
(i) If the positions of objects in the field are already known from ancillary
data, the user can provide an external sky mask. If it contains a world
coordinate system with sky coordinates, it is used to align the mask to the
MUSE data. The computationally expensive step of creating the white-light image
will then be skipped. Providing such an optimized mask can significantly
improve the results.
(ii) In some cases a large object is present in the field of view, but several
exposures of the same field were taken, so that different parts of the
instrument were illuminated by this object. In that case, the resulting tables
with correction factors could be combined with appropriate algorithms for
clipping, for better results in a second iteration of this reduction step. The
MUSE pipeline supports this by allowing the user to input a table with the
corrections.

This functionality was developed to improve final data quality of deep fields
observed with MUSE, in particular the HDF-S \citep{2015A&A...575A..75B} and UDF
\citep{2017A&A...608A...1B} fields, and the wavelength ranges, clipping methods
and $\sigma$-levels were adapted to give optimal results for these projects
after thorough testing. It is possible that for some other datasets individual
segments are assigned deviant correction factors. In this case, a discontinuity
in the spectra at the edge between the wavelength ranges would show
up.\footnote{This can be checked visually using the MPDAF function
  \hyperref{https://mpdaf.readthedocs.io/en/latest/api/mpdaf.drs.plot_autocal_factors.html}{mpdaf.drs}{plot_autocal_factors}{{\tt plot\_autocal\_factors()}}.}
Hand-optimising the sky mask is often a way to improve the results.

This functionality is available in the {\tt muse\_scipost} module, it is
switched on if the parameter {\tt autocalib} is set to {\tt deepfield}. The
user-provided table is read, if it is set to {\tt user} instead. By default,
this self-calibration is switched off (parameter set to {\tt none}).

\subsection{Standard star handling}\label{sec:stdstar}
In MUSE data processing, spectrophotometric standard stars are used to compute
the instrument throughput and to derive a spectrum of normalized telluric
absorption. The throughput is determined in the form of a response curve that
is used to flux-calibrate the science data.

Standard stars get the same basic processing applied as science data, so this
module also starts with merging the pixel tables, dividing by the smoothed
average flat-field spectrum and correcting for differential atmospheric
refraction (in WFM). A cube of the stellar field is then reconstructed, using
the nominal MUSE sampling of 1.25\,\AA\,pixel$^{-1}$. Object detection using
simple thresholding (with levels of 50-5 $\sigma$, until at least one object is
found) on the central wavelength plane of the cube determines initial positions
for all possible stars in the field.

Then the flux of all stars is integrated on each wavelength plane of the cube.
For WFM (both with and without AO), the point-spread function on the sky is
represented very well by a \citet{1969A&A.....3..455M} function
\citep{2016A&A...588A.148H,2018MNRAS.480.1689K}. When directly fitting this
function to extract the fluxes of the detected stars, changes due to noise can
result in unphysical differences even in adjacent wavelength planes. The
consequences would be increased noise in the output spectrum and wavy continua
in extracted spectra.
Hence, we used the idea of PampelMUSE \citep{KWR13} and first fit a free
elliptical Moffat in all wavelength planes in the reconstructed cube.  Then we
fit 2nd order polynomials to the central position and all Moffat parameters,
iteratively rejecting wavelengths where a parameter deviates by more than
$3\sigma$ from the polynomial.  Then we re-fit the Moffat at every wavelength
plane, fixing the parameters to the polynomial parameter solutions at every
plane, so that only the flux and the background level remain free parameters.
This ``smoothed Moffat'' results in high S/N spectra for all stars in the
field, and is used as the default extraction mechanism for WFM standard star
exposures.
For NFM, the profile is more complex and cannot be modeled fully
analytically.\footnote{The function described by \citet{2019A&A...628A..99F} in
  principle allows to fit the AO-correction NFM PSF in Fourier-space. However,
  tests show that it is not robust enough to be used in an automated pipeline
  environment, where small-scale artifacts might be present.}
So a simple circular aperture with a sky annulus is taken. Depending on the
mode, these are the automatically chosen defaults.\footnote{This automatic
  selection and the ``smoothed'' Moffat fit are new in v2.8, older versions
  always defaulted to non-iterative Moffat fits.}
The extraction window at each wavelength is determined from the spatial FWHM of
the exposure, given by metadata about ambient conditions -- the observatory
seeing -- or by measuring it on the central wavelength.  For the aperture
extraction, $4\times$FWHM is used, the (Moffat) fit is carried out over
$3\times$FWHM.
Once the total flux of each object over all wavelengths is known, the pipeline
selects the star to use (either the brightest one or the one closest to the
field center). For most standard star fields, only a single star is detected.

Then the measured fluxes of the selected star are compared to the interpolated
fluxes from the reference flux table of the target field, taking into account
the airmass of the target and the extinction curve that was provided by the
user.\footnote{This process was modeled following the widely used implementation
  in IRAF's {\sc onedspec} package, especially the {\sc sensfunc} and
  {\sc calibrate} tasks.}
The relation
\begin{equation}
s(\lambda) = 2.5 \log_{10} \left(\dfrac{d_\mathrm{ct}(\lambda)}
                                       {t_\mathrm{exp} \Delta\lambda
                                        f_\mathrm{ref}(\lambda)}\right)
           + f_\mathrm{ext}(\lambda) A
\end{equation}
describes the sensitivity $s$ computed at each wavelength $\lambda$, with the
recorded flux in counts (electrons) $d_\mathrm{ct}$, the
effective airmass $A$, the exposure time $t_\mathrm{exp}$, using the reference
flux $f_\mathrm{ref}$.
This curve is postprocessed as follows: wavelength ranges known to be affected
by telluric absorption are marked and interpolated across with 2nd order
polynomials. The fractional difference between the fit and the original data is
taken as the telluric absorption factor $f_\mathrm{tell}$; in between the
telluric regions it is set to 1.\footnote{This procedure assumes that the
  sensitivity is smooth within the telluric absorption bands. This is
  true for all approved spectrophotometric standard stars used by the MUSE
  calibration plan, but excludes cooler stars often used as telluric stars
  in other projects.}
The final response curve $f_\mathrm{resp}$ is obtained from $s(\lambda)$ by
extrapolation to the largest possible MUSE wavelength range and then smoothing it.
Smoothing can be done using a median filter, but piecewise cubic polynomials
followed by a sliding average is usually more effective to reduce noise and
reject outliers in single wavelengths.
Finally, using the known effective area of the telescope, a throughput
spectrum is computed from the smoothed response curve.

The whole procedure is repeated for each exposure given as input, but no
attempt is made to combine the resulting response curves or compute an improved
extinction curve. For each exposure, the response and the table of telluric
correction factors are saved to disk.

The pipeline recipe for this procedure is {\tt muse\_standard}. The parameter
that controls the method is called {\tt profile} and with {\tt select} the way
to choose among multiple detected sources can be changed. The smoothing behavior
can be influenced with the {\tt smooth} parameter.

\subsection{Astrometric calibration}\label{sec:astcal}
Since the geometric solution (Sect.~\ref{sec:geo}) only corrects per-slice
offsets, a calibration of the overall distortion and pixel scale of MUSE has to
be done on sky, using astrometric fields. These are located in Milky Way
Globular Clusters with existing Hubble Space Telescope imaging, so that
reference catalogs of a few hundred stars over a MUSE field exists.
Fields in the outer parts of the clusters were chosen for WFM, while positions
closer to the centers and with a bright star in the field were selected for
NFM.

Astrometric fields again get the same basic processing applied as all other
on-sky data, so this module again starts with merging the pixel tables,
dividing by the average flat-field spectrum and correcting for differential
atmospheric refraction (in WFM). Here, optionally, the data can be flux
calibrated, if the related input data (response, telluric, and extinction
curves) were given on input, but this is usually not necessary.
A cube of the medium-dense stellar field is then reconstructed. Since the
distortions are achromatic, a large sampling in wavelength is used
(50\,\AA\,pixel$^{-1}$) to improve S/N. The central three wavelength planes of
this cube are combined using the median to remove any last cosmic rays and
other artifacts. Thresholding at a given $\sigma$-limit is then applied to
detect the reference stars in the exposure, before their position is determined
more accurately using (Moffat) profile fits.
Two-dimensional pattern matching is used to identify the detected objects
against the sky positions from the reference catalog.
Here, we based our implementation on the {\tt kd-match} library of
\citet{2013MNRAS.433..935H}\footnote{\url{https://ubc-astrophysics.github.io/kd-match/}},
based on the search of quadrilaterals.\footnote{Originally, the triangle-based
  matching algorithm implemented in the FORS pipeline \citep{FORS_PipeMan} was
  used.  This turned out to not be robust enough on the scales of MUSE data and
  especially at NFM resolution. So with v2.8 this was changed to kd-match as
  default, but the older algorithm is still available. It also needed inputs
  regarding accuracy and matching radius, where the radius was automatically
  decreased $1.25\times$, until only unique matches were found.}
We improved on their code by optimizing it for the case of the actual MUSE
astrometric fields.
80\% of the detections have to match a catalog entry within the given radius
for a given transformation judged to be valid.
All matched objects (typically around 100) are then used to fit the astrometric
solution, using a six-parameter world coordinate representation (with zero point
position, two scales, rotation, and shear) in the gnomonic projection of the
tangent plane. The fit is typically iterated twice with a given
$\sigma$-clipping rejection. This reduces the effect of stars in the foreground
of the cluster, that have different proper motion, on the final solution.
Since the absolute astrometry of the astrometric field is not of interest, the
zero-point of the fit is ignored when the calibration is applied to science
data.

The corresponding pipeline module is called {\tt muse\_astrometry}. The most
important parameters are:
{\tt detsigma} the $\sigma$-level to use for object detection,
{\tt centroid} is the centroiding method (Gaussian or Moffat fits),
{\tt radius} is the matching radius,
{\tt niter} is the number of iterations of the final fit, and
{\tt rejsigma} sets the $\sigma$-level for the iterative rejection of the fit.

\subsection{Exposure offset calculation}\label{sec:offcalc}
When combining multiple MUSE exposures it is usually necessary to correct
for relative coordinate offsets. To apply an offset correction, a list of the
measured, relative offsets for each of the exposures has to be prepared, with
respect to a given reference.

The automatic calculation of offsets uses the reconstructed field-of-view
images of all exposures involved. It then measures the individual, relative
coordinate offsets with respect to the first exposure.  In a first step, the
algorithm creates a list of detected sources, one for each input image. An
implementation of the \textsc{daophot} \citep{1987PASP...99..191S} detection
algorithm is used to find sources in the MUSE field and thus targeted on the
detection of point sources. The parameters to fine-tune the \textsc{daophot}
star-finding algorithm, especially the FWHM of the Gaussian convolution filter,
and the roundness and the sharpness criteria can be adjusted.

For each of the input images the source detection is iterated adjusting the
detection threshold until the number of detected sources falls within
predefined limits given by the recipe options, or the maximum number of
iterations is exceeded. Starting with an initial detection threshold, the
detection threshold is adjusted in subsequent iterations by a given step size.
The initial detection threshold is either a given, absolute value, or it is
calculated from a given multiplier and an estimate of the background level
of the current input image and its uncertainty. In the latter case, the initial
detection threshold value is calculated as
uncertainty of the background above the background estimate, where the
background level and its uncertainty are estimated from a fraction of the pixel
values at the low end of the pixel value distribution of the image. Finally,
all sources located closer to a bad pixel than a given minimum distance are
discarded.

Once the source lists for all input images are created, the field offsets are
computed iteratively for a sequence of decreasing search radii. For a given
search radius and for every possible, pairwise combination of the fields the
relative offsets between the objects which were detected in each of the two
exposures are calculated. The offsets in right ascension and declination are
computed for every possible combination of the detected objects of the two
fields.

The offsets measured for each pair of objects are then used to calculate the
relative offset of two fields. The initial estimate of the relative field
offset, in other words the estimate for the largest search radius, is determined as the
mode of the 2D histogram of the relative offsets of the object pairs. Object
pairs whose distance are larger than the search radius are excluded.  In
subsequent iterations, smaller search radii are used to refine the estimates of
the field offsets. Here, the object pairs used to calculate the field offset
are again selected by applying the previously mentioned distance criterion,
taking into account the field offset calculated in the previous iteration step.
The median of the relative offsets of the selected object pairs is then used as
the measured relative offset of two fields.

Once the field offsets for all pairwise combinations of fields have been
measured, the final field offsets for a given search radius are determined by a
least-squares fit of all the measured field offsets. If weighting is enabled,
the weights for the fit are initially the peak value of the histogram, and, in
subsequent iterations, the variance of the relative offsets of the selected
object pairs.

The relative field offsets which are eventually written to the offset list are
the field offsets computed for the smallest search radius. If one of the input
exposures is found to not overlap with any of the other exposures, its relative
field offsets are set to zero. During the combination of the exposures these
relative offsets are used to correct the initial field center given by the FITS
header keywords \texttt{RA} and \texttt{DEC} of the exposures. For input
exposures which do not overlap with any other input exposure this means that no
correction is applied, and the exposure is combined with the other input
exposures using the header entries as they are.

Together with the offset list an approximate preview image of the combined
field of view is generated. This allows the visual assessment of the computed
offsets. It is also possible to, optionally, create an exposure map for the
field of view.

The pipeline module that implements this algorithm is
{\tt muse\_exp\_align}\footnote{First introduced into the MUSE pipeline v1.2.}.
The parameter that set roundness and sharpness are {\tt roundmin}/{\tt
roundmax} and {\tt sharpmin}/{\tt sharpmax}, the FWHM of the convolution is
{\tt fwhm}, the number of sources is given with {\tt srcmin}/{\tt srcmax} and
the maximum number of iterations are controlled by {\tt iterations}. One can
modify the initial threshold and step-size using {\tt threshold} and {\tt step}.
The initial threshold is taken as an absolute pixel value if {\tt threshold} is
larger than zero, and otherwise its absolute value is used as multiplier when
calculating the initial threshold from the background estimate. The pixel
values used for the background estimate can be chosen using {\tt bkgignore}
and {\tt bkgfraction}, to exclude and select a fraction of the pixel values.
The minimum distance between a detected source and the closest bad pixel is
given using {\tt bpixdistance}. The search radii can be adjusted using
{\tt rsearch} and weights are taken into account for the final fit when
{\tt weight} is set. The number of histogram bins is {\tt nbins}.

\section{Common algorithms}\label{sec:algo}
Some processing steps are used in several of the modules of the MUSE pipeline
or represent central design choices. These procedures or algorithms are
described in detail in this section.

\subsection{Error propagation}\label{sec:errorprop}
The initial estimate of the noise of each pixel is done from the data itself.
For this we assume that the errors of the individual pixels are independent.
The variance of pixels in an exposed frame then consists of the photon noise,
the read-out noise, and the error of the read-out noise estimate:
\begin{equation}
\qquad
\sigma^2_\mathrm{initial} = \dfrac{v - b}{g}
                          + \left(1 + \dfrac{1}{n_\mathrm{b}}\right) \sigma^2_\mathrm{b}
\label{eq:initvar}
\end{equation}
\citep[following][]{GR02} where $\sigma^2_\mathrm{b}$ is the variance of a bias
frame (derived from the RON, see Sect.~\ref{sec:bias}), in units of adu$^2$,
$n_\mathrm{b}$ is the number of pixels used to determine the RON, $v$ is the
values of a pixel in adu, $b$ is the bias level of the exposure in adu, $g$ is
the detector gain or conversion factor in electrons per adu. Computed values of
$\sigma^2_\mathrm{initial} < 0$ (in regions of low signal) are set to zero.
This is a noisy estimator, as discussed by \citet{2017A&A...608A...1B}, and if
used directly (\ie, for inverse variance weighting), can lead to wrong results,
if a source is detected as low S/N.

For any subsequent operation on a pixel, Gaussian error propagation ensures
that the variance after the operation is computed correctly. For any function
$f$ that affects two images a and b, the variance is therefore computed as
follows:
\begin{equation*}
\qquad
\sigma^2_{f(a,b)} = \left(\dfrac{\partial f}{\partial a}\right)^2 \sigma^2_a
                  + \left(\dfrac{\partial f}{\partial b}\right)^2 \sigma^2_b
\end{equation*}
A simple, non-weighted average of a sequence of n images then takes the form
\begin{equation}\label{eqn:avsigma}
\qquad
\sigma^2_\mathrm{average} = \dfrac{1}{n^2} \sum_{i=1}^n \sigma^2_i
\end{equation}
It should be pointed out that the treatment of the science data in the MUSE
pipeline does not depend on the variance estimate which is only propagated through
the multiple processing steps.

\subsection{Bad pixel processing}\label{sec:badpix}
The MUSE pipeline tries to suppress CCD defects, cosmic rays, and other
artifacts so that they do not contaminate the science data in the final
datacube. Such artifacts are found in a multi-stage process. Some hot columns
can be detected on bias images, some as dark columns on flat-fields, etc. These
are marked, first on the image level in the \DQ extension of the master
calibration products, then in the data quality column of the MUSE pixel table.
To encode the nature of the artifact, we use the data quality convention
invented for the Euro3D format \citep{2004AN....325..159K}.
The possible quality values are documented as bitwise flags in \citet[][their
Sect.~4.6.1]{Euro3D}. From those flags we use only a subset in the MUSE
pipeline which we list in Tab.~\ref{tab:euro3d}.

\begin{table*}
\caption{Bad pixel flags used in the MUSE pipeline}\label{tab:euro3d}
\begin{tabular}{r r l}
Flag value & bit-shift & Data quality condition\\
\hline
       0 &         0 & good pixel -- no flaw detected\\
       1 &         1 & affected by telluric feature (corrected)$^a$\\
       2 &  1 $<<$ 1 & affected by telluric feature (uncorrected)$^a$\\
      32 &  1 $<<$ 5 & cosmic ray (unremoved)$^b$\\
      64 &  1 $<<$ 6 & low QE pixel ($< 20\%$ of the average sensitivity;
                       \eg, defective CCD coating, vignetting...)\\
     256 &  1 $<<$ 8 & hot pixel ($>5\sigma$ median dark)\\
     512 &  1 $<<$ 9 & dark pixel (permanent CCD charge trap)\\
    4096 &  1 $<<$12 & A/D converter saturation (signal irrecoverable,
                       but known to exceed the A/D full scale signal)\\
    8192 &  1 $<<$13 & permanent camera defect (such as blocked columns,
                       dead pixels)\\
   16384 &  1 $<<$14 & bad pixel not fitting into any other category$^c$\\
2$^{30}$ &  1 $<<$30 & missing data (pixel was lost)\\
2$^{31}$ &  1 $<<$31 & outside data range (outside of spectral range,
                       inactive detector area, mosaic gap, ...)\\
\hline
\end{tabular}\\
$^a$ In the MUSE pipeline, telluric features are only marked in the standard
     star processing, but never propagated to science data.\\
$^b$ The corresponding ``removed'' cosmic ray from the Euro3D specifications is
     not used.\\
$^c$ In the MUSE pipeline this is used for non-positive pixels in flat-field
     images.
\end{table*}

Any process that uses bad pixels (those with flags greater than zero) propagates
the bits of the flag on to subsequent processing, using bitwise logical OR.
Since not all bad pixels can be found using automated means, an extra table can
be given to all pipeline modules that load raw data. Pixel positions and flags
stored in that table are then propagated to the data in the \DQ extension and
the pixel table in the same manner.

In the final cube (see Sect.~\ref{sec:cuberec}) written by the pipeline, the
voxels (volume pixel, one element of the data cube)
flagged in the \DQ extension are replaced by {\tt NAN} values in both
the \DATA and the \STAT extension to save 1/3 of the data volume. Since the
cube is resampled using 3D information, most of the bad pixels in the middle of
the data have been interpolated over from neighboring good pixels, especially
if the cube was reconstructed from multiple overlapping exposures, and only
voxels on the edges of the cube are usually left as flagged.

\subsection{CCD overscans and trimming}\label{sec:ovsc}
The CCD overscan regions\footnote{In the FITS images of the MUSE raw data, the
  ``overscans'' are located in a cross in the inside, between the CCD
  quadrants.  They result from just continuing to read the CCD beyond the
  physically available pixels. The ``prescan'' regions which represent actual
  pixels on the outskirts of the CCD which do not get illuminated are present as
  well, but they do not well represent the bias level in the data section. Both
  are visible in Fig.~\ref{fig:raw}.}
play an important role in the MUSE instrument. Since the bias level in the data
section\footnote{The ``data section'' corresponds to the illuminated part of the
  CCDs.}
of the CCDs has gradients that change with time, these regions can be analyzed
to correct for this.

Pixels with (strongly) deviant values sometimes appear in the analysis of the
overscans. These are caused either by cosmic ray hits or by hot columns or
pixels whose read-out spills into the overscan regions. When computing
statistics on the overscans, these should therefore be removed. In the MUSE
pipeline this can be done using the DCR cosmic ray rejection routine
\citep{2004PASP..116..148P} that was tuned to do optimal rejection in the
overscan regions. Other possible options are only useful for testing, and
iteratively fit a constant to the whole overscan region or to not do any value
rejection.
When computing statistics of the overscans, it can also be useful to ignore a
few pixels that are located next to the illuminated data section, so that flux
that may spill over due to less than optimal charge transfer efficiency of
the CCD electronics does not get included.  Typically, discarding bands of
three pixels in width is good enough to get clean statistics, even if an
exposure is illuminated close to saturation.

The first possibility is to ignore the overscan region. In this case the user
is still warned, if the overscans are found to be very different from the bias
level in the data section (if this can be determined, \ie, for bias images).
The $\sigma$-level of this warning can then be tuned.
The second possibility is to compute the mean value in the overscans belonging
to each quadrant. When combining multiple exposures or subtracting the master
bias from other images, the offset between the mean values of the images in
question is taken out.
The last possibility and the one that is typically used because it gives the
best bias correction for MUSE data, is to model the slope of the vertical
overscan with a polynomial. This polynomial is computed iteratively, by default
with a $30\sigma$ clipping. The order of the polynomial is increased until a
good match is found, depending on the root-mean square (RMS) of the residuals
and the $\chi^2$ of the fit. Testing against deep dark frames (see
Sect.~\ref{sec:biassub}) showed that most CCD overscans can be well modeled
with a 5th order polynomial, but 1-2 quadrants even need up to 15th order for a
good fit. To not use such high polynomial orders for every vertical overscan,
the next higher-order polynomial is only chosen, if it decreases the RMS by
more than a factor of 1.00001 and the fit quality $\chi^2$ decreases by more
than $1.00001$.\footnote{Until v2.6, a maximum 5th order polynomial
  was used, with a 1.01 RMS decrease and $\chi^2<1.04$.}
In practice, 83\% of the CCD overscans are fit with polynomial order of 5 or
lower.  The polynomial is then subtracted from the data of the whole quadrant,
before combining it with other exposures or before removing a master bias which
was already treated in the same way.

Once the abovementioned overscan-based processing was executed, only the data
section of the raw data remains relevant, and the regions of overscan and
prescan are cut off. Depending on the processing stage, statistics of the
overscans are kept internally to facilitate combination with other exposures
etc.

\subsection{Image combination methods}\label{sec:imcomb}
Several calibration processes require a sequence of exposures to be combined on
the CCD level. This is done using image combination. While this is widely
available in data processing packages (\eg, IRAF\footnote{IRAF was written and
  supported until 2013 by the National Optical Astronomy Observatories (NOAO) in
  Tucson, Arizona, see \citet{1993ASPC...52..173T}. A
  \href{https://iraf-community.github.io/}{community edition is now available}
  for legacy applications.}),
the specialty in the MUSE pipeline is that these procedure are aware of bad
pixels and pixel variances.  In the simplest case, the {\em average}, this
means that any pixels flagged as bad at the same pixel position are discarded
when computing the mean value. The propagated variance is then given by
Eq.~\ref{eqn:avsigma}. In case all pixels at one position are flagged as bad,
the pixel with the least severe flaw and its variance are taken, and its flag
is propagated.

In case of an image {\em sum}, any flagged pixels are discarded and the total
sum is then computed by scaling the partial sum back to the total sum of the
number of images involved. The variance is computed the same way, with the
factor in quadrature.

An image {\em median} is computed by sorting the input values and taking either
the middle value (for an odd number of input images) or the average of the two
central values (for even inputs). Then the variance corresponding to either the
middle value (again, for odd numbers) or the propagated average of the two
middle values (for even numbers) is taken as output variance. This again uses
only unflagged input pixels and only falls back to the least severe flaw in case
all inputs were flagged.

Since these simple methods are either affected by noise peaks or cosmic rays or
do not deliver the optimal S/N, two more options using value rejection are
implemented for MUSE. These sort the values at each position, and then discard
the outliers, either using simple {\em minmax} or {\em sigclip} rejection. The
first uses a fixed, user-selected, number of outliers at each end of the pixel
distribution, while the latter computes median and median
deviation\footnote{This might be better described by average absolute deviation
  against the median.}
and discards values outside boundaries given by this deviation and two factors.
Pixel selection for the initial histogram only uses unflagged pixels. The
average of the remaining data is then computed with variance propagation as
explained above.

Those MUSE pipeline modules which carry out such image combinations from raw
data (\eg, {\tt muse\_bias}, {\tt muse\_wavecal}), the default is always the
{\tt sigclip} option. In all cases, the sigma clipping factor was optimized for
the data typically handled by the module, but can be adjusted (parameters {\tt
lsigma} and {\tt hsigma}). For special cases, other combination methods can be
chosen by the user, with the parameter {\tt combine}.

\subsection{Cube reconstruction}\label{sec:cuberec}
Since two of the high-level goals in the design of the MUSE pipeline were to
propagate the variance from the raw data to the final product and to create
optimal data quality, only a single resampling step is used in the science
reduction. This is the last step, the reconstruction of the output cube.
As elsewhere in this paper, we call the elements entering this process
{\em pixels}\footnote{These pixels are internally represented as rows in the
  pixel table(s), and still in a one-to-one relation to the original CCD pixels
  of the one or several exposures involved in this process.}
while the output elements of the cube are called {\em voxels}.

To make resampling computationally tractable, the input pixels are sorted into
three-dimensional grid cells, each representing one output voxel. Each grid
cell can contain no pixel, one pixel, or many pixels, and the pixels are
assigned to them in a nearest-neighbor fashion depending on their 3D
coordinates with respect to the coordinates of the grid cells. To then compute
the data value of the output voxel, each input pixel is assigned a weight. This
weight depends on the resampling method chosen. Flagged pixels are ignored.

The cube reconstruction works in the same way for a single and multiple
exposures.  In the latter case, however, multiple pixel tables enter the
process, and so multiple pixels may be assigned to each grid cell, where
exposures overlap. The process assumes that all exposures were taken under
similar conditions. Strong deviations in spatial FWHM (seeing) and transparency
(cloud cover) may lead to artifacts.

Note that this resampling process causes variance to be transferred into
covariances. Since we cannot store those covariances due to the size of the
data involved, they are lost after reconstructing the cube. However, as the
MUSE pipeline only resamples the data once, this means that the variances
stored in the cube are accurate {\em per voxel} to the level of which we can
derive them from the data by the procedure given in Sect.~\ref{sec:errorprop}.
Just when further analyzing the data by integrating voxels in any of the three
dimensions, the variance of the integrated data will be underestimated.

The actual method to compute the output value of each voxel using the pixel
grid are discussed below. In the pipeline modules that allow this, the method
can be chosen by setting the {\tt resample} parameter.

\subsubsection{Cosmic ray rejection}\label{sec:cosmics}
Cosmic ray detection is an issue for all astronomical observations which record
the data on CCDs, MUSE is no exception. Most of the time, the solution is to do
statistics of pixels in small apertures \citep[\eg,][]{2004PASP..116..148P} in
2D or to filter the CCD image \citep[\eg,][]{2001PASP..113.1420V,2012A&A...545A.137H}.
But with 3D data, there are three advantages: ({\it i}) Each pixel has 26
instead of 8 neighbors, so the statistics can detect outliers to much lower
levels. ({\it ii}) Additional power comes from the fact that data contributing
to voxels that are adjacent in the 3D grid partly originate from regions in the
raw CCD-based images that are very far from each other. This reduces the number
of pixels in the statistics to about 18, but makes it highly unlikely that the
reference pixels are contaminated by a cosmic ray hit as well. And ({\it iii}),
if there are multiple ($n$) exposures that overlap on the sky and hence at
least partially in the 3D grid, one can compare the statistics of
$18 \times n$ (or $26 \times n$) pixels, and even more efficiently detect
cosmic rays.

The implementation in the MUSE pipeline follows the approach of running the
cosmic ray rejection once the grid-cells (that define the output voxels, as
mentioned above) are set up and linked with all pixels whose centers are
located within them in 3D.
Basically, the pipeline then loops through all grid cells, and then through
all pixels assigned to the grid cell itself and the directly surrounding grid cells
and computes statistics of all pixel values, and their variances.
By default, the variances are ignored, and only the median of all surrounding
pixels as well as the median of the absolute median deviation are computed.
All pixels in the central grid cell that are above a given $\sigma$-level
with respect to the median are then marked as cosmic rays.
Alternative statistics can be selected instead, then either mean and standard
deviation are used to compute the lower limit for cosmic rays. Or an estimator
using the estimated pixel variances can be used, which was inspired by the {\sc
avsigclip} rejection algorithm available in several IRAF tasks.  In the latter
case, the mean is compared to the local average noise.
Both alternatives are faster to compute the statistics, but much less effective
at similar $\sigma$ levels, and much more likely to clip real peaks with
correspondingly tighter constraints. They are therefore only useful for special
purposes.

If the angle between the input data and the output grid is a multiple of
90\degr (or within 5\degr), a simple operation can determine which pixels
originate from the same slice on the CCD. If this is the case, the pipeline
will ignore those close neighbors when computing the statistics and make cosmic
ray rejection more effective.

In the pipeline modules that resample data into a cube, the parameter {\tt
crtype} can be changed to set the type of statistics to use while {\tt crsigma}
allows to change the $\sigma$-level for the rejection. Tests have shown that
for single exposures a $15\sigma$ cutoff works well with median statistics,
while for multiple exposures a tighter $10\sigma$ cutoff is more effective
without destroying data. These are set as defaults.
In the typical range of exposure times used with MUSE, the pipeline detects
cosmic ray hits which affect between about 0.06\% (for 600\,s) and 0.11\% (for
1\,500\,s) of the pixels of a MUSE exposure.

\subsubsection{Drizzle-like resampling}
The primary method for this step uses a technique inspired by the Drizzle
algorithm used for HST imaging data \citep{Drizzle09}. In this case, the weight
for each pixel is computed using the geometrical overlap between pixel and
voxel.

The calibrations applied to the data ensure that the coordinates of the center
of each pixel within the output grid are known to high accuracy. However, they
do not form a contiguous image as in the case of the original drizzle
algorithm, but are irregularly spaced within the output grid. Due to the size
of the data, the orientation and individual input size of the pixels cannot be
tracked. Hence, we have to assume that all pixels have the same size and that
the rotation is negligible.
Then, the formula to compute the weights is
\begin{equation*}
\qquad t =
  \begin{cases}
\max\left[t_\mathrm{out}, 0\right],
         & \mathrm{if}\ t_\mathrm{out} + \mathrm{d}t \le t_\mathrm{in}\\
\max\left[(t_\mathrm{in} + t_\mathrm{out}) / 2 - \mathrm{d}t, 0\right],
         & \mathrm{otherwise}
  \end{cases}
\end{equation*}
where $t$ is one of the three coordinates ($x$, $y$, $z$), $t_\mathrm{in}$ is
the input pixel size, $t_\mathrm{out}$ is the output voxel size, and d$t$ is
the distance between the centers of input pixel and output voxel.  The total
drizzled weight $w$ for an output pixel can then be computed as
\begin{equation*}
\qquad w = \prod_{x,y,z} \dfrac{\max[t, 0]}{t_\mathrm{in}}
\end{equation*}

This algorithm is conserving the flux of the objects within the field of
view, it was therefore chosen to be the default method in the MUSE pipeline.
It can be set by selecting the {\tt drizzle} option of the {\tt resample}
parameter. Scaling of the input pixel size to give $t_\mathrm{in}$ is set using
{\tt pixfrac} which can be different in all three dimensions.

\subsubsection{Nearest neighbor}
Another, much simpler approach to compute the output voxel values, is to just
use the value of the pixel lying closest to the center of each grid cell. This
is the fastest method, but if one is resampling more than one exposure into a
cube, the improvement in S/N is lost. It is therefore only useful to provide a
quick look.

This method can be set by selecting the {\tt nearest} option of the {\tt
resample} parameter.

\subsubsection{Other weighted resampling types}
The pipeline allows to use several other weighted resampling schemes. Common to
all is that for each input pixel a weight is computed that is derived from the
distance between the center of the grid cell defining each voxel and the 3D
location of the center of the pixel. It should be pointed out that these
methods do not conserve the flux as well as the Drizzle method.

In case of {\tt linear}, the weight is computed as the inverse linear distance,
while for {\tt quadratic} the squared inverse distance is taken.  The option
{\tt renka} is following the approach of \citet{Ren88} and uses the relation
\begin{equation*}
\qquad w = \left[\dfrac{r_\mathrm{c} - r} {r_\mathrm{c} r}\right]^2
\end{equation*}
where $r_\mathrm{c}$ is the critical distance after which the weight is set to
zero.

The normalized sinus cardinal function $\sinc(r) = \sin(\pi r) / (\pi r)$ is
the only function that represents the Fourier-space box filter in real space
and therefore does not add additional correlated noise to resampled data
\citep[see, \eg,][]{Dev00}. Many algorithms have therefore been developed to find
a computationally reasonable implementation of this function which does not
have to be integrated over the whole dataset. The Lanczos functions, that cut
off for distances larger than $k$, have been used in many applications:
\begin{equation*}
\qquad L_k(r) =
  \begin{cases}
    \sinc(r) \sinc(r/k), & \mathrm{if\ } |r|<k\\
                      0, & \mathrm{otherwise}
  \end{cases}
\end{equation*}
This can be selected with the {\tt lanczos} option, and $k$ can be set with
{\tt ld}.

\subsection{Variances of the resampled datacubes}\label{sec:variance}
\setlength{\unitlength}{1mm}
\begin{figure*}
\hspace*{0mm}
\begin{picture}(180,150)
\put(0,0){\includegraphics[width=\columnwidth]{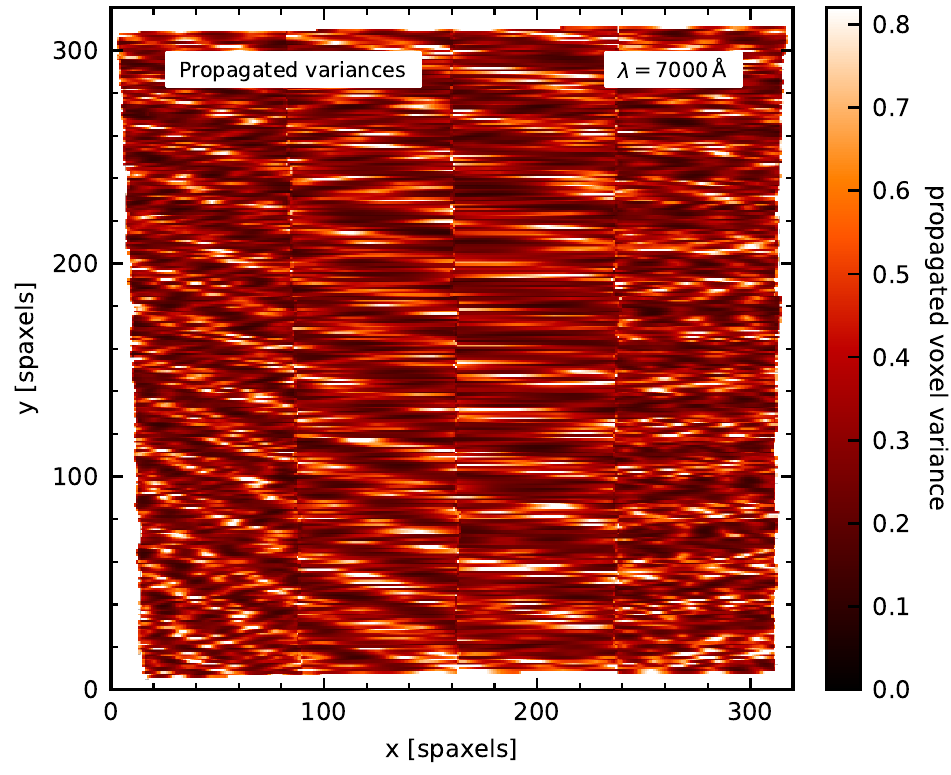}}
\put(0,74){\includegraphics[width=\columnwidth]{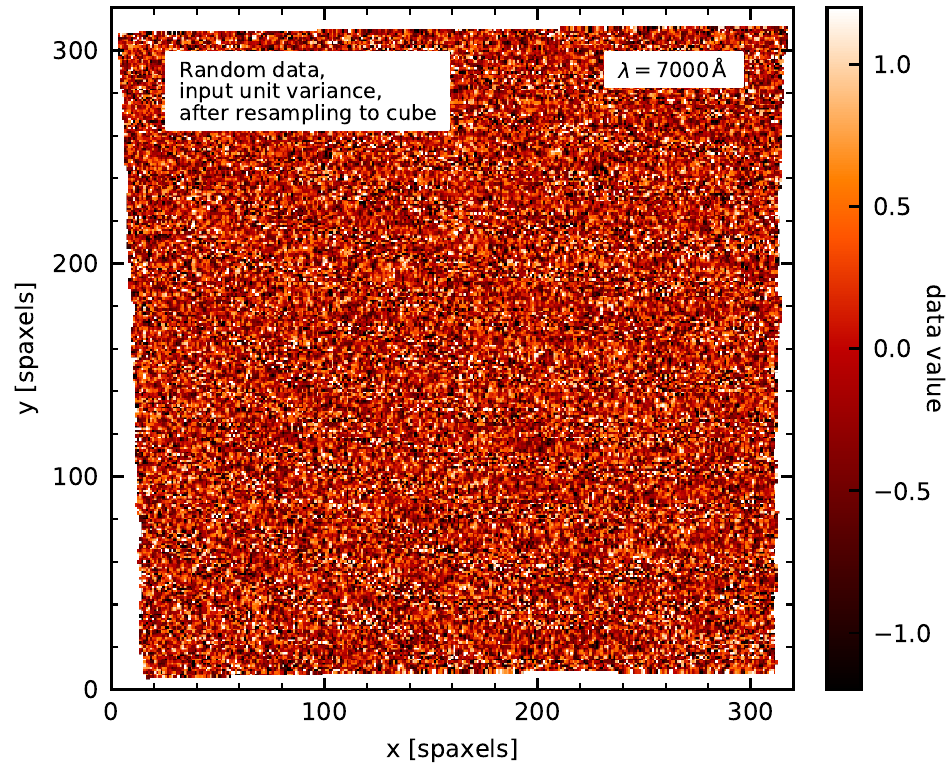}}
\put(93,76){\includegraphics[width=0.98\columnwidth]{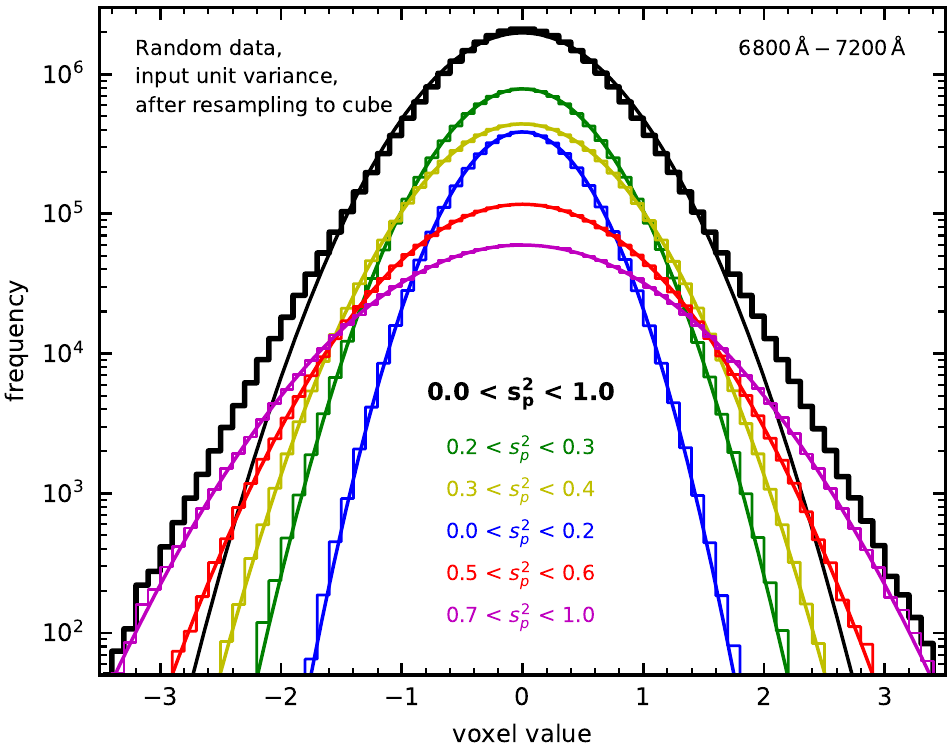}}
\put(94,2){\includegraphics[width=0.97\columnwidth]{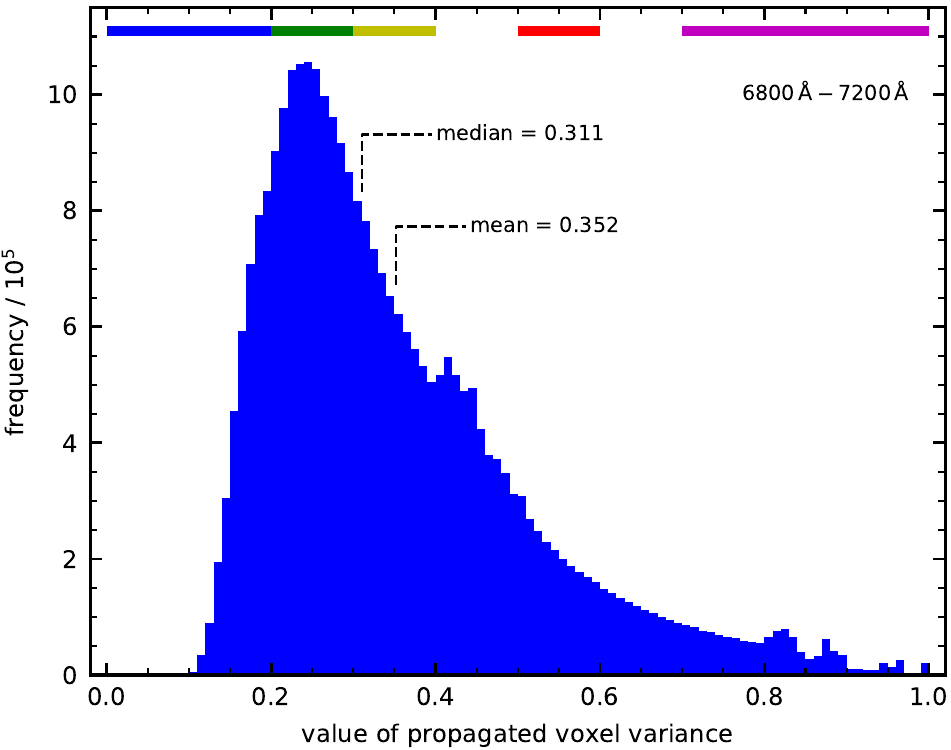}}
\end{picture}
\caption{Behaviour of the propagated variances \spp after resampling a pixel
         table with pure random data of unit variance into a datacube.
         The upper left panel displays a single wavelength plane of this cube;
         the same plane of the corresponding variance cube is shown below.
         The upper right panel presents histograms of the resampled random data
         evaluated over a subcube of 320 planes (400\,\AA) bandwidth, together
         with Gaussian fits to these histograms. Each of the coloured curves
         represent a different subset of voxels selected by a specific range of
         the propagated variances in these voxels, as indicated by the labels.
         The lower right panel shows the histogram of the propagated variances,
         for the same subcube. The coloured horizontal bar in the top indicates
         the selected variance ranges.
}
\label{fig:variances}
\end{figure*}

We already mentioned in Sect.~\ref{sec:cuberec} that the resampling step invariably
introduces cross-talk between adjacent voxels (with the details depending on
the actually adopted resampling algorithm). While the MUSE pipeline formally
propagates the individual pixel errors through the resampling process, it is
unavoidable that the real uncertainties of the data get partly diffused into
off-diagonal covariance terms which are not recorded by the pipeline. This has
three principal consequences: (i) The propagated variances \spp are
systematically lower than the true variances $\sigma^2$. (ii) The propagated
variances show substantial but unphysical spatial variations. (iii) The
noise distribution of voxels becomes apparently non-Gaussian. In the following
we discuss these quantitatively.

To isolate the effects on the variances caused by the resampling of a pixel
table into a datacube we performed the following simple experiment: We created
a MUSE pixel table filled with pure random numbers of zero mean and unit
variance. We converted this pixel table into a datacube using the drizzle-like
resampling option with default parameters (pixfrac of 0.8 in all directions).
The top-left panel of Fig.~\ref{fig:variances} shows one arbitrarily
selected wavelength plane of this cube. Already a quick visual inspection of
this image reveals a pattern: In some regions the pixels appear to be spatially
smoothed, in others they are largely unaffected. This variable smoothing
pattern is a direct consequence of the nontrivial geometric mapping of the
detector plane into a regularly gridded datacube. Voxels with projected
locations falling between detector pixels inherit similar amounts of flux from
multiple pixels implying a large covariance term, whereas voxels nearly
congruent with pixels suffer from little such cross-talk.
Fig.~\ref{fig:variances} also shows that there is spatial coherence in these
patterns, with strongly and weakly smoothed horizontal bands alternating in
vertical direction.

The same structures are in fact evident in the propagated variances \spp
displayed in the bottom left panel of Fig.~\ref{fig:variances}. Lower values of
\spp correspond to heavier local smoothing of the voxels. Around each of the
three vertical slicer stack transitions the patterns changes abruptly.
But there are also drastic changes between slices within one stack, and even
across  a single slice -- typically resulting from a nonzero tilt between
detector rows and the datacube grid. For this reason the spatial coherence
length is much shorter in the two outer slicer stacks which are more tilted.
Additional variations are affected by the resampling in wavelength together
with the nonlinear dispersion relation. The amplitude of these variations are
substantial, up to a factor $\sim 6$ in the value of \spp even between adjacent
voxels -- recall that all input data to this datacube have a variance of
exactly $\sigma^2 = 1$. But a similar underlying pattern will be present in all
real science data, only less visible because of the noisy nature of the
variance estimates when using Eq.~\ref{eq:initvar}. These variations are purely
geometric in origin and obviously imply no changes of the level of
trustworthyness of the data (\eg, as a S/N ratio). Because of the
horizontal striping in \spp one cannot even define a small aperture over which
the variations get averaged out.

The bottom right panel of Fig.~\ref{fig:variances} presents the actual
histogram of propagated variances, revealing a very skewed distribution with a
most frequent value of 0.24, implying an artificial increase of the apparent
S/N ratio by a factor 2; median and mean are correspondingly
larger. Only a tiny fraction of the voxels has \spp values close to 1. The
smallest values of \spp correspond to the voxels suffering from the heaviest
cross-talk. Since for resampling with the given setup each
voxel can be inheriting flux from maximally eight pixels, the propagated
variance can be reduced by up to that number, $s_\mathrm{p}^2 \ge
0.125\times\sigma^2$, corresponding precisely to the lower bound in the
histogram.

One further consequence of the variable scale of spatial (and spectral)
smoothing is that the histogram of actual voxel values at any given wavelength
deviates from a Gaussian even if the input data follow a perfect normal
distribution, as can be seen in the upper right panel of
Fig.~\ref{fig:variances}. Not only is the histogram (thick black line; standard
deviation of $s_\mathrm{p} = 0.593$) much narrower than the input distribution,
it also shows extended wings that in real data might lead the user to suspect
significant non-Gaussian errors. However, Fig.~\ref{fig:variances} demonstrates
that these wings can be decomposed into a superposition of essentially perfect
Gaussians when performing the histogram analysis on subsets of voxels within a
narrow range of \spp values; each of these histograms corresponds to a
different degree of local smoothing and therefore reduced variance.

We emphasize that this does not mean to say the propagated variances contain no
useful information. In fact they describe the actual noise level \emph{per
voxel} accurately (apart from the limitation of Eq.~\ref{eq:initvar}).
In other words, any repeat experiment with identical resampling
geometry would reveal random fluctuations in each voxel in accordance with the
corresponding value of \spp. But as the variations in adjacent voxels are
correlated, the amplitude of any aperture-based quantity would be larger than
formally expected if interpreting \spp as white noise variance. For this reason
it also would not help to self-calibrate the variance level by measuring the
actual voxel-to-voxel fluctuations in an observed datacube; these fluctuations
will always be significantly lower than the real noise.

A relatively simple recipe to obtain realistic variances for background-limited
observations would involve three steps:
\begin{enumerate}
\item Run a similar simulation as described in this subsection but specifically
      for the chosen resampling setup.  This needs to be done only once per
      setup.
\item At each wavelength, obtain the histogram of propagated variances in the
      observed target datacube and estimate, for example, its mean or median.
\item Scale that value by the inverse of the corresponding mean or median of
      \spp in the random numbers datacube, and replace the fluctuating variance
      plane by the resulting number as a constant.
\end{enumerate}
This approach implicitly also solves the problem of the initially noisy
variances.
It is clearly limited to cases where the objects of interest are faint enough
that their photons can be neglected as contribution to the shot noise per
pixel. An object model with defined simulated noise would have to be used for
the case of a field where objects contribute significantly.

\subsection{Image reconstruction}\label{sec:imagerec}
Once a datacube was reconstructed, it is possible to integrate it over the
wavelength direction to create an image of the field of view. By default, a
white-light image is created, where each voxel of the cube is weighted equally,
and only data outside the wavelength range 4650\dots9300\,\AA\ is discarded.
If a filter function was given, e.\,g.\ for the Johnson $V$ filter, then each
voxel in the cube is weighted according to the filter transmission $w_\lambda$
at its wavelength:
\begin{equation}
\qquad f_\mathrm{pixel} = \dfrac{\sum_\lambda w_\lambda \Delta\lambda f_\lambda}
                                {\sum_\lambda w_\lambda \Delta\lambda}
\end{equation}
For cubes with constant wavelength sampling this is simply
\begin{equation}
\qquad f_\mathrm{pixel} = \dfrac{\sum_\lambda w_\lambda f_\lambda}
                                {\sum_\lambda w_\lambda}
\end{equation}

\subsection{Correction of atmospheric refraction}\label{sec:dar}
The correction of atmospheric refraction is important for all spectroscopic
data, especially for an instrument with a wavelength range as long as the one
of MUSE.
In WFM, the red and blue ends are shifted significantly (more than a spatial
resolution element). In NFM, the instrument has a built-in atmospheric
dispersion corrector (ADC), and in most cases no software correction appears to
be necessary.
However, the ADC was built to allow for up to 2 pixels residual shift for
zenith distances of 50\degr, so an empirical correction may be necessary (see
below).

The algorithm in the MUSE pipeline was originally developed by
\citet{2008A&A...486..545S} and implemented in the {\sc p3d} software
\citep{SBR+10}. Using known atmospheric parameters for humidity, temperature,
and pressure at the observatory, the refractive index of air is computed, for a
reference wavelength and all other wavelengths of the MUSE data. The relative
shift between the data in both spatial directions is then computed depending on
parallactic and position angle of the exposure. This shift is then applied to
the coordinates of all pixels in the pixel table.

Computation of the refractive index uses the formulae from
\citet{1982PASP...94..715F}
by default. Following the refinement of \citet{SWTV+12}, the MUSE pipeline also
implemented different, more accurate methods to calculate the refractive index
of air based on the abovementioned atmospheric conditions at the time of the
observation. These were taken from
\citet{1967ApOpt...6...51O},
\citet{1966Metro...2...71E},
\citet{1993Metro..30..155B},
and \citet{1996ApOpt..35.1566C}.
However, since it was shown that the \citeauthor{1982PASP...94..715F} approach
gives satisfactory results with MUSE data \citep[see][]{WeilbacherM42}, it
remains the default.
For the typical case of an observation at airmass 1.5, shifts of about
1\farcs2 occur at the blue end of the MUSE wavelength range. All four methods
correct this shift equally well to below 0\farcs04, far smaller than the MUSE
sampling on the sky.

An additional step is implemented in the MUSE pipeline (parameter {\tt
darcheck}).
It reconstructs a cube with nominal spatial but long (10\,\AA) wavelength
pixels to improve the S/N per wavelength plane. It then runs a simple
threshold-based object detection on the central plane of the cube. All objects
detected are then centroided in each plane. A polynomial is then fitted to
these measurements. If the cube was already corrected for atmospheric
refraction effects, a quadratic polynomial is used, otherwise a 4th-order fit
is done. This fit can be used (setting {\tt darcheck=correct}) to either
correct the data for atmospheric refraction effects if none of the above
analytical methods were applied before, or to correct residual shifts after an
analytical correction. Alternatively, one can use it (setting {\tt
darcheck=check}) to quantify any residual effects in the dataset, in terms of
maximum deviation from the reference wavelength.
This empirical procedure is not activated by default, since it takes a
significant amount of extra time, and the accuracy of its operation depends on
the science field in question. In the case of a single bright source in the
field, correction to sub-pixel accuracy should be possible. This is the typical
case for NFM observations, and can be used to correct for residuals caused by
the ADC. In a more complex case or in fields without any directly visible
continuum sources, a separate characterization outside the pipeline will be
necessary.

\subsection{Flux calibration}\label{sec:fluxcal}
When applying the response curve and the telluric corrections that were
computed from the standard star (see Sect.~\ref{sec:stdstar}) to science data,
the calibration data (the response $f_\mathrm{resp}$ and the telluric
correction $f_\mathrm{tell}$, as well as the atmospheric extinction
$f_\mathrm{ext}$) are linearly interpolated to the wavelength $\lambda$ of each
pixel and applied to the data in counts $d_\mathrm{ct}$ using
\begin{equation}
d_\mathrm{cal}(\lambda) = \dfrac{10^{-0.4 f_\mathrm{resp}(\lambda)}
                                 10^{0.4 f_\mathrm{ext}(\lambda) A}
                                 f_\mathrm{tell}(\lambda)}
                                {t_\mathrm{exp} \Delta\lambda}
                          \ d_\mathrm{ct}(\lambda)
\end{equation}
to yield the calibrated flux $d_\mathrm{cal}$ in each pixel in units of
$10^{-20}$\cgsflux, where $A$ is the effective airmass of the given exposure.

This procedure is used for science exposures, and implemented in the {\tt
muse\_scipost} recipe. It is also used in the same way for offset sky exposures
(in {\tt muse\_create\_sky}) and can be applied to astrometric exposures as
well (from {\tt muse\_astrometry}).

\subsection{Line Spread Function}\label{sec:lsf}
The Line Spread Function (LSF) is used in the MUSE pipeline for the description
of sky emission lines and their subtraction (see Sect.~\ref{sec:skysub}), as
well as in the modeling of the laser-induced Raman lines (Sect.~\ref{sec:raman}).
Sect.~\ref{sec:lsfmeas} described how it is determined, here we provide a
description of the two possible representations that are used in the various
pipeline modules.

\begin{figure}
\includegraphics[width=\linewidth]{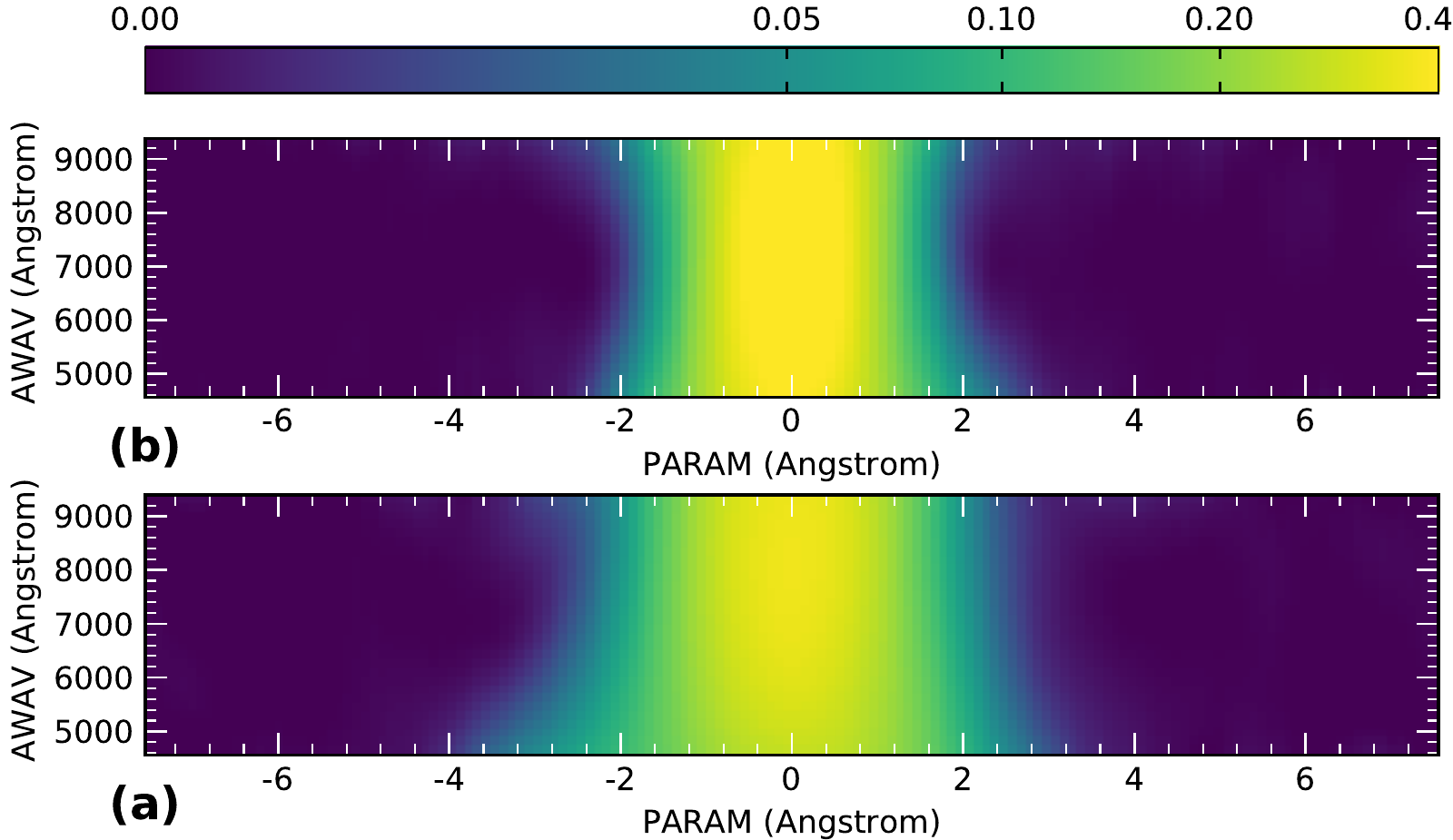}
\caption{The interpolated image of the LSF for two MUSE slices, computed from
         the arc exposure set started on 2016-05-12T11:40:31.
         {\bf (a)} Shows the LSF for slice 1 of IFU 1 while
         {\bf (b)} shows the result for slice 22 in IFU 12.
         The value range is the same and uses arcsinh scaling, the horizontal
         axis shows the LSF direction whereas the vertical axis is the MUSE
         wavelength range.
}
\label{fig:lsfinterpol}
\end{figure}

The interpolated LSF is saved in image form, where each image stores the LSF
for a given slice in a given IFU of MUSE. The line profile includes the slit
width of the instrument (determined by the height of the image slicer in the
spectrographs) and the bin width (the CCD pixel size). The LSF from each IFU is
stored in one file or FITS extension, thereby using the FITS cube format, with
48 pixels in the third axis. In Fig.~\ref{fig:lsfinterpol} we show two
representative slices, one with a narrow LSF (a slice in the middle of the
field of view, also located in the middle of the CCD), and a wider LSF (a slice
at the left edge of the CCD, and near the bottom of the MUSE field).
Since this interpolated, empirical description provides a superior and much
faster sky subtraction, it is used by default.

The {\em alternative} LSF description is a parametrization using damped
Gauss-Hermite function of the following form:\footnote{Our definition was
   inspired by \citet{1996MNRAS.282.1223Z}, who used a modified Gauss-Hermite
   polynomial to parametrize line of sight velocity distributions.}
\begin{equation*}
\qquad L(x) = e^{\frac{x^2}{2w^2}} + e^{\frac{x^2}{w^2}}
            \sum_{i=3}^6 k_i H_i\left(\frac{x}{w}\right)
\end{equation*}
with the coefficients $k_i$ of the Hermitian polynomials $H_i(x)$ and the LSF
width $w$ as slice and wavelength dependent fit parameters ($x = \lambda -
\lambda_0$ with $\lambda_0$ as the wavelength of the emission line).  This LSF
is then analytically convolved with rectangular functions representing the slit
width and the CCD binning. The LSF width $w$ is parametrized with a quadratic
dependency of the wavelength, the Hermitean coefficients $k_i$ with linear
function.
This alternative LSF description was mainly used before MUSE was first tested
on sky. Commissioning data then showed this method not to be robust enough and
leave too strong residuals of the telluric emission, so the interpolating LSF
above was developed, and since then used by default.

Within the MUSE pipeline, the LSF is only used for background modeling (sky and
Raman lines). However, it can be used later in other tools for analysis, if the
sampling of the output cube is taken into account. For example, \citet{WMV+17}
use the FWHM of the averaged LSF as determined by the pipeline in their pPXF
fits, using the log-spaced pixel sizes in wavelength direction to convolve the
unbinned LSF.
It should also be noted, that the LSF is very similar between the four wide-field
modes of MUSE, but has different shape and width for the narrow-field mode.

\section{Implementation}\label{sec:impl}
\begin{figure*}
\includegraphics[width=\linewidth]{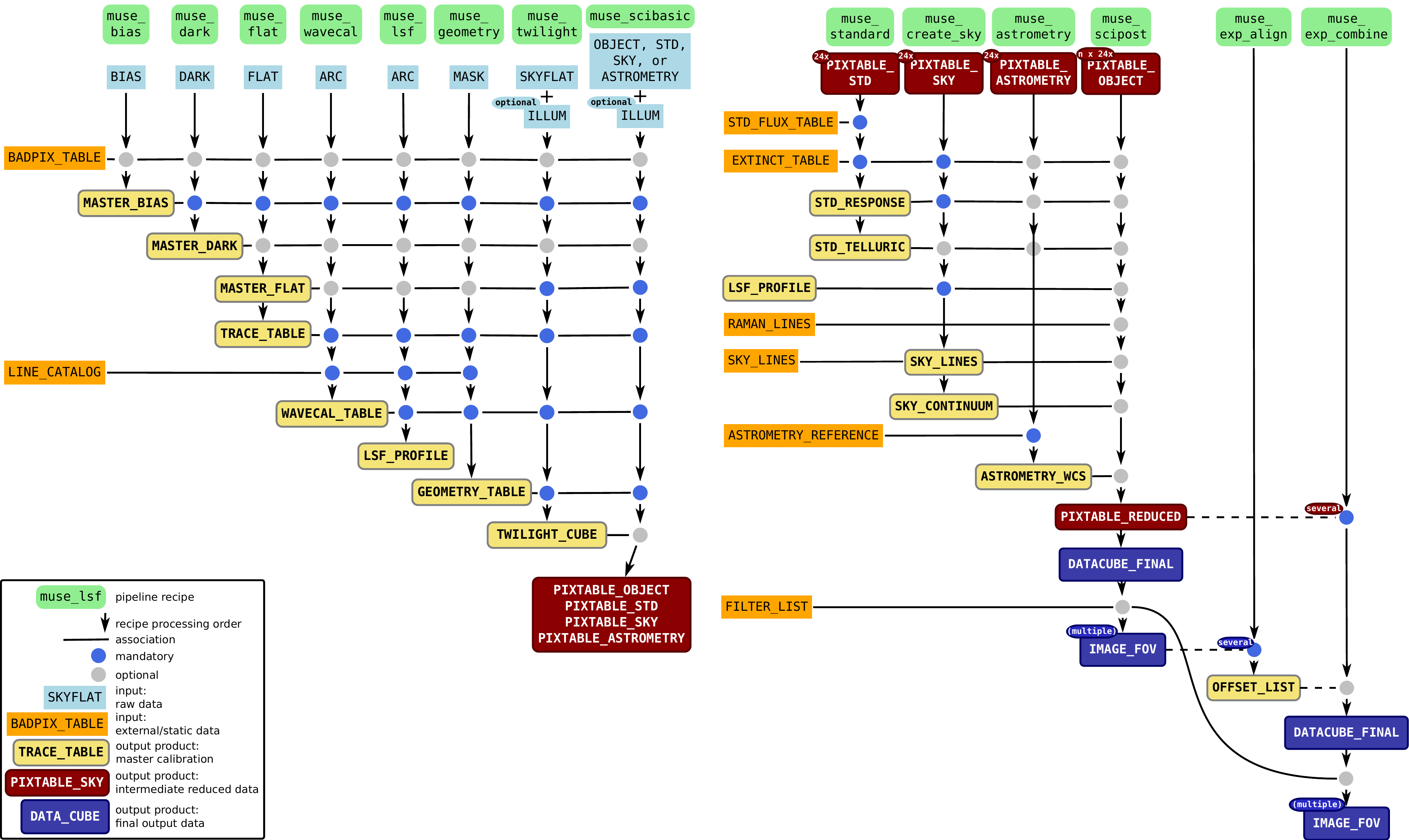}
\caption{{\bf Left}: {\em Basic processing} recipes and products of the MUSE pipeline.
         {\bf Right}: {\em Post-processing} steps in the MUSE pipeline.
         Compared to Fig.~\ref{fig:sciproc} this shows all input and output
         files and includes all calibration modules.}
\label{fig:cascades}
\end{figure*}

Since the pipeline discussed in the present paper is written to serve as data
reduction environment for an ESO instrument, it is implemented in the ESO
software framework. It is written in pure C and uses the ESO Common Pipeline
Library \citep[CPL;][]{BBI+04,2014ascl.soft02010E} as its base library for all
internal data structures. CPL in turn uses CFITSIO \citep{2010ascl.soft10001P}
and wcslib \citep{2011ascl.soft08003C} as well as FFTW \citep{FFTW05} for parts
of its functionality.

This means that the data reduction modules are available as shared libraries
that can be used as plugins with esorex \citep{2015ascl.soft04003E} alone or in
the Reflex environment \citep{2013A&A...559A..96F}. While the latter makes MUSE
reduction very easy and hides a lot of the complexity and data association
behind a graphical user interface, the former forces the user to think about
the relation between science data and calibrations, and allows scripting. A
Python interface \citep[\texttt{python-cpl},][]{SW12} can be used to start the
modules as well.\footnote{It actually works for all ESO pipelines built around
  the plugin concept of the CPL.}

Each individual reduction module (also called {\em recipe}) calls into the
shared library to carry out its task. The user-facing side of the recipes and
its basic C code on the other hand is generated from XML descriptions, that
provide the possibility to create documentation as well.  The XML interface was
finally also used to integrate the MUSE pipeline into the MUSE-WISE data
processing system \citep{PSV12,2015scop.confE..28V} that is based on the
AstroWISE concept \citep{VMS+07}. This is used to process part of the data
collected by the MUSE collaboration in its guaranteed observing time.

To allow the MUSE pipeline to efficiently handle the data processing on modern
hardware, it was planned to be parallelized from the start. The basic
processing part operates in parallel for the data from the 24 IFUs, and can
therefore only make use of 24 cores.  This can be run either as internal
parallelization using OpenMP or parallelized externally using scripting to run
multiple processes. The post-processing steps are always parallelized
internally, using OpenMP loops, and can employ all cores that the computing
hardware offers. However, testing showed that not all steps benefit from
parallel operation, so that operations that read data from disk or save files
are serial. The same is true for operations involving concatenation of large
data buffers in memory, for example when merging multiple exposures.

The pipeline code is Open Source (GPL v2) and distributed through the ESO
pipeline website.\footnote{\url{https://www.eso.org/sci/software/pipelines/muse/muse-pipe-recipes.html}}
The releases of the MUSE consortium are code-identical but up to v2.6.2
used a slightly different packaging.\footnote{\url{https://data.aip.de/projects/musepipeline.html}}

The full data reduction cascade of the implementation, complete with the tags
of the input and output files, and all relevant data reduction recipes
described in this paper is shown in Fig.~\ref{fig:cascades}.

\section{Testing of the data quality}\label{sec:qual}
Testing the quality of the data reduction was central throughout the development
of the MUSE pipeline. Before on-sky data became available in 2014, testing
relied heavily on the comprehensive simulation of raw data provided by the
Instrument Numerical Model \citep[INM,][]{Jarno+10,JBP+12}. Especially
developing complex procedures like the geometrical calibration and testing
cosmic ray rejection or the combination of multiple exposures with offsets
benefitted strongly from the availability of such data. A multitude of
tests, too many to be reported here, were run for each data processing module
using this simulated data, to ensure readiness for on-sky operation. Since the
data format, especially the metadata recorded in the FITS headers of the raw
exposures that the MUSE pipeline now requires to be present, has changed since
then, we cannot use these data any more for verification of the current
software version.

Instead, we here show a few cases that are reproducible with current, publicly
available, datasets, to show the performance of the pipeline processing.

\subsection{Bias subtraction accuracy}\label{sec:biassub}
\begin{figure}
\includegraphics[width=\linewidth]{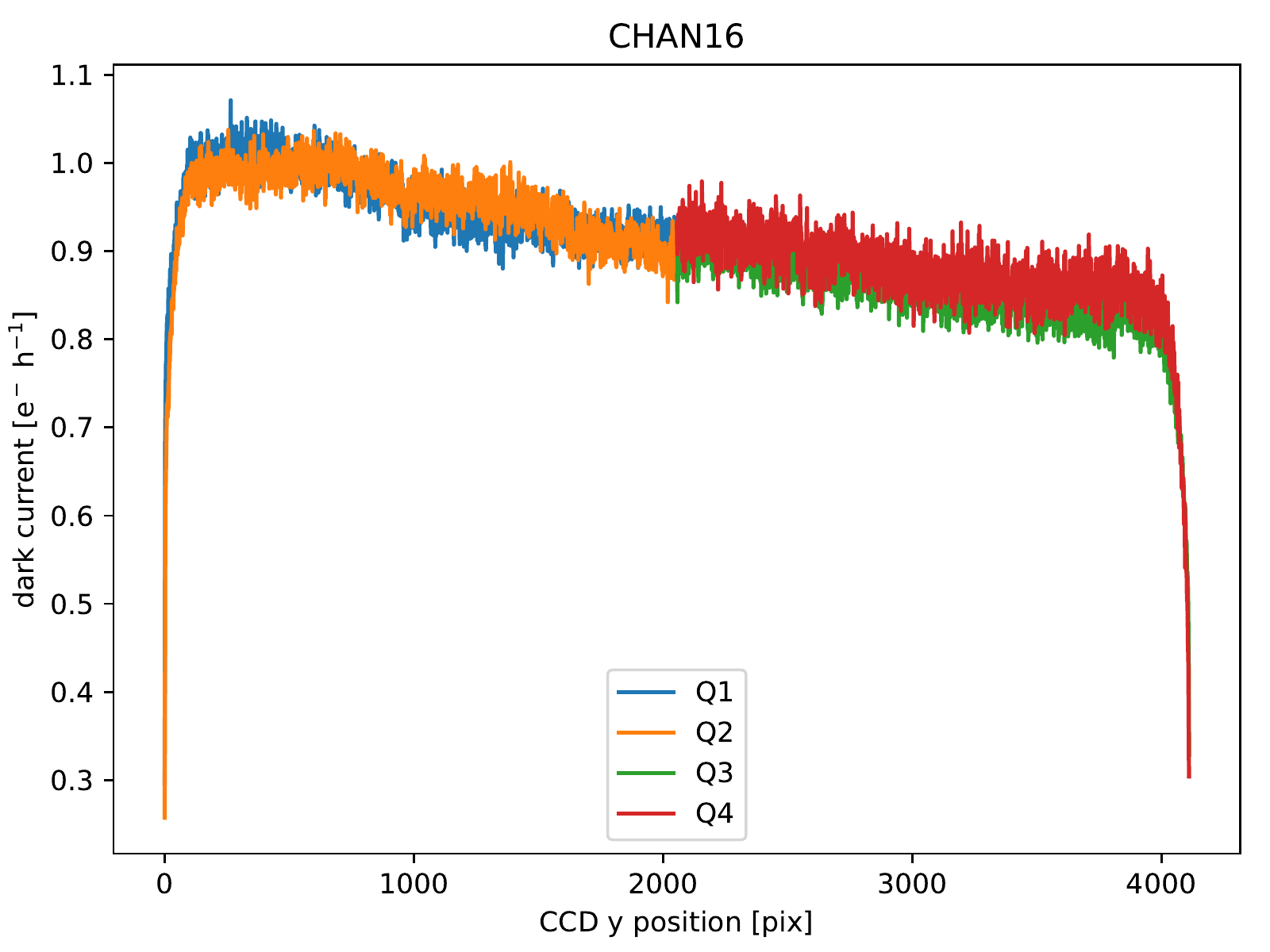}
\caption{Dark current distribution in CCD `pallas' of MUSE channel 16, measured
         after bias subtraction, averaged over 1000 CCD columns.}
\label{fig:darkres}
\end{figure}
The quality of the bias subtraction, including overscan modeling, can be best
judged by checking the flatness of darks. To be able to see differences in
small fractions of an ADU, one needs to check the combination of many dark
frames, meaning the combined master dark of a long series.
Such a series was taken in 2018, between June 25 and Aug.~21, comprising 149
darks of 30\,min exposure time each. All bias frames which were taken on the
same days as the dark exposures (418 in total) were used to create master bias
images for bias subtraction. The process used the default parameters of
pipeline v2.8.
A vertical cut through a typical combined master dark image is shown in
Fig.~\ref{fig:darkres}. The vertical cut was computed as the average over 1000
CCD columns in the dark image, ignoring flagged pixels. The vertical gradient
of the dark current (about 0.15\,e$^-$\,h$^{-1}$ over 4000 pixels) seems to be
a property of the e2v CCDs as used in the MUSE instrument together with the ESO
NGC electronics.  The strong gradients in the first and last few
pixels are edge effects that appear differently in bias and dark frames; these
are not of relevance to science data, since these extreme CCD pixels are at
wavelengths that are not used for normal analyses. In the case of the displayed
dark of channel 16, a small jump (about 0.05\,e$^-$\,h$^{-1}$) between the four
CCD quadrants is visible. While the displayed jump is typical, for some of the
other 23 CCDs it is not visible at all and for a few others the discontinuity
is stronger.
However, this discontinuity is undetectable without large-scale binning and it
could also be caused by a horizontal gradient in the affected CCDs, an error in
the determination of the CCD gain in a quadrant, or undetected bad pixels that
deviate from the surrounding pixels in a subtle manner.

Any residual features due to bias subtraction seem to be
$\lesssim$0.1\,e$^-$\,h$^{-1}$, so are below $1.5\times10^{-21}$\cgsflux for
typical atmospheric transparency.

\subsection{Wavelength solution}\label{sec:waveacc}
To check the wavelength solution, we use a sequence of interleaved arc
sequences taken in the different instrument modes on 2018-04-21, between
18:06:05 and 20:37:01 UTC. Five exposures in each mode were taken after each
other before changing the arc lamp, so that each full sequence of 15 exposures
(three lamps, with five exposures per lamp, in five modes) was as close in time
as possible. This was done to exclude any changes in flexure in the
spectrographs between the modes, that might other\-wise happen due to
temperature changes.

For the test we {\em derive} the wavelength solution using the 15 arc exposures
taken in non-AO extended mode (WFM-NOAO-E).
We then use the arcs taken in AO extended mode (WFM-AO-E)
as {\em comparison}.  We average them at the CCD level (to save processing
time) but reduce them as if they were science exposures, and let the pipeline
create a combined pixel table and also reconstruct a cube.
We ensure that corrections for atmospheric refraction, sky subtraction,
astrometric calibration, and radial velocity are not carried out.
We then measure the centers of the brightest and most isolated eleven arc
emission lines using Gaussian fits, with spectra reconstructed directly from
the pixel table with 0.1\,\AA\ bins for each slice of the instrument and for
the whole field of view, as well as for each spaxel in the cube.

\begin{table}
\caption{Arc line wavelengths recovered from the pixel table.}\label{tab:wavept}
\begin{tabular}{r@{.}l c r@{.}l r@{.}l r@{.}l}
\multicolumn{2}{c}{$\lambda$} & ion & \multicolumn{2}{c}{mean} & \multicolumn{2}{c}{median} & \multicolumn{2}{c}{std.~dev.}\\
\multicolumn{2}{c}{\AA}       &     & \multicolumn{2}{c}{\AA}  & \multicolumn{2}{c}{\AA}    & \multicolumn{2}{c}{\AA}\\
\hline
4799&912  & \ion{Cd}{i} &  0&0094 &  0&0083 & 0&0095\\
5085&822  & \ion{Cd}{i} &  0&0127 &  0&0086 & 0&0104\\
5460&750  & \ion{Hg}{i} & -0&0095 & -0&0110 & 0&0117\\
6438&470  & \ion{Cd}{i} & -0&0233 & -0&0211 & 0&0183\\
6506&528  & \ion{Ne}{i} &  0&0074 &  0&0033 & 0&0125\\
6929&467  & \ion{Ne}{i} &  0&0005 & -0&0013 & 0&0103\\
7173&938  & \ion{Ne}{i} & -0&0043 & -0&0037 & 0&0094\\
7245&167  & \ion{Ne}{i} &  0&0022 & -0&0011 & 0&0105\\
8231&634  & \ion{Xe}{i} & -0&0045 & -0&0082 & 0&0110\\
8819&410  & \ion{Xe}{i} & -0&0047 & -0&0062 & 0&0082\\
9045&446  & \ion{Xe}{i} & -0&0130 & -0&0131 & 0&0117\\
\hline
\end{tabular}
\end{table}

\begin{figure}
\includegraphics[width=\linewidth]{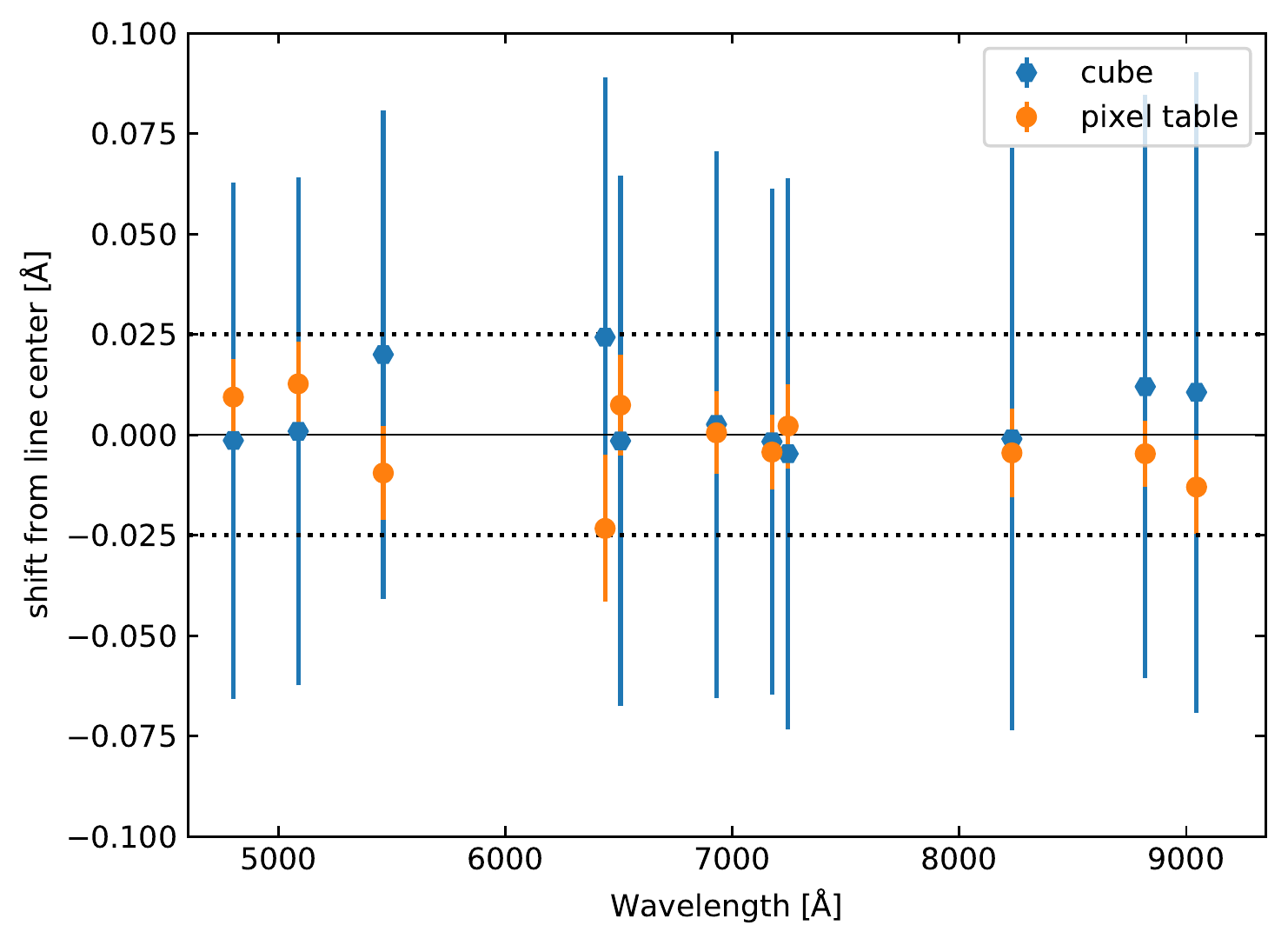}
\caption{Mean arc line positions recovered from the pixel table (orange) and
         data cube (blue). The error bars represent $1\sigma$ standard
         deviations. The dotted horizontal lines mark the range of $1/100$th of
         the instrumental resolution. This is the visual representation of
         Tables \ref{tab:wavept} and \ref{tab:wavecube}.}
\label{fig:wavelambda}
\end{figure}

The measured line positions for the pixel table (see Tab.~\ref{tab:wavept})
represent the intrinsic wavelength calibration accuracy that can be achieved by
the pipeline process.  This is not degraded by the resampling to (typically)
1.25\,\AA\ bins of the cube. Similar levels of accuracy can be recovered during
data analysis when using full-spectrum fitting to overcome the sampling problem
of individual lines. The table lists the arc line wavelength in air as given in
the NIST database \citep{NIST_ASD_2014} and used as wavelength reference in the
MUSE pipeline, the ion it originated from and the offset from the reference in
the mean across the whole field, in the median across all slices, and the
standard deviation measured over all 1152 MUSE slices. Given the width of the
MUSE instrumental profile of about 2.5\,\AA, both the absolute and the relative
accuracy reached is $100\times$ better than the instrumental resolution for all
arc lines, namely below 0.024\,\AA. This corresponds to a velocity accuracy of
$\sim1$\,\kms. In the central part of the MUSE wavelength range
($6500\dots8500$\,\AA) where the density of high-S/N arc lines is higher and
hence the polynomial fit is better constrained, even 0.01\,\AA\ (equivalent to
$\sim0.4$\,\kms) are reached.
This is in line with the velocity accuracy that was reached by fitting high-S/N
stellar spectra with solar metallicity, as was demonstrated by
\citet{2018MNRAS.473.5591K}.
The orange data points in Fig.~\ref{fig:wavelambda} represent these pixel-table
based measurements.

\begin{table}
\caption{Arc line wavelengths recovered from the datacube.}\label{tab:wavecube}
\begin{tabular}{r@{.}l c r@{.}l r@{.}l r@{.}l}
\multicolumn{2}{c}{$\lambda$} & ion & \multicolumn{2}{c}{mean} & \multicolumn{2}{c}{median} & \multicolumn{2}{c}{std.~dev.}\\
\multicolumn{2}{c}{\AA}       &     & \multicolumn{2}{c}{\AA}  & \multicolumn{2}{c}{\AA}    & \multicolumn{2}{c}{\AA}\\
\hline
4799&912  & \ion{Cd}{i} & -0&0014 & -0&0032 & 0&0643\\
5085&822  & \ion{Cd}{i} &  0&0009 & -0&0005 & 0&0633\\
5460&750  & \ion{Hg}{i} &  0&0200 &  0&0195 & 0&0608\\
6438&470  & \ion{Cd}{i} &  0&0243 &  0&0243 & 0&0646\\
6506&528  & \ion{Ne}{i} & -0&0015 &  0&0006 & 0&0661\\
6929&467  & \ion{Ne}{i} &  0&0026 &  0&0034 & 0&0681\\
7173&938  & \ion{Ne}{i} & -0&0017 & -0&0001 & 0&0629\\
7245&167  & \ion{Ne}{i} & -0&0047 &  0&0019 & 0&0686\\
8231&634  & \ion{Xe}{i} & -0&0010 &  0&0025 & 0&0725\\
8819&410  & \ion{Xe}{i} &  0&0120 &  0&0164 & 0&0726\\
9045&446  & \ion{Xe}{i} &  0&0106 &  0&0131 & 0&0798\\
\hline
\end{tabular}
\end{table}

\begin{figure}
\includegraphics[width=\linewidth]{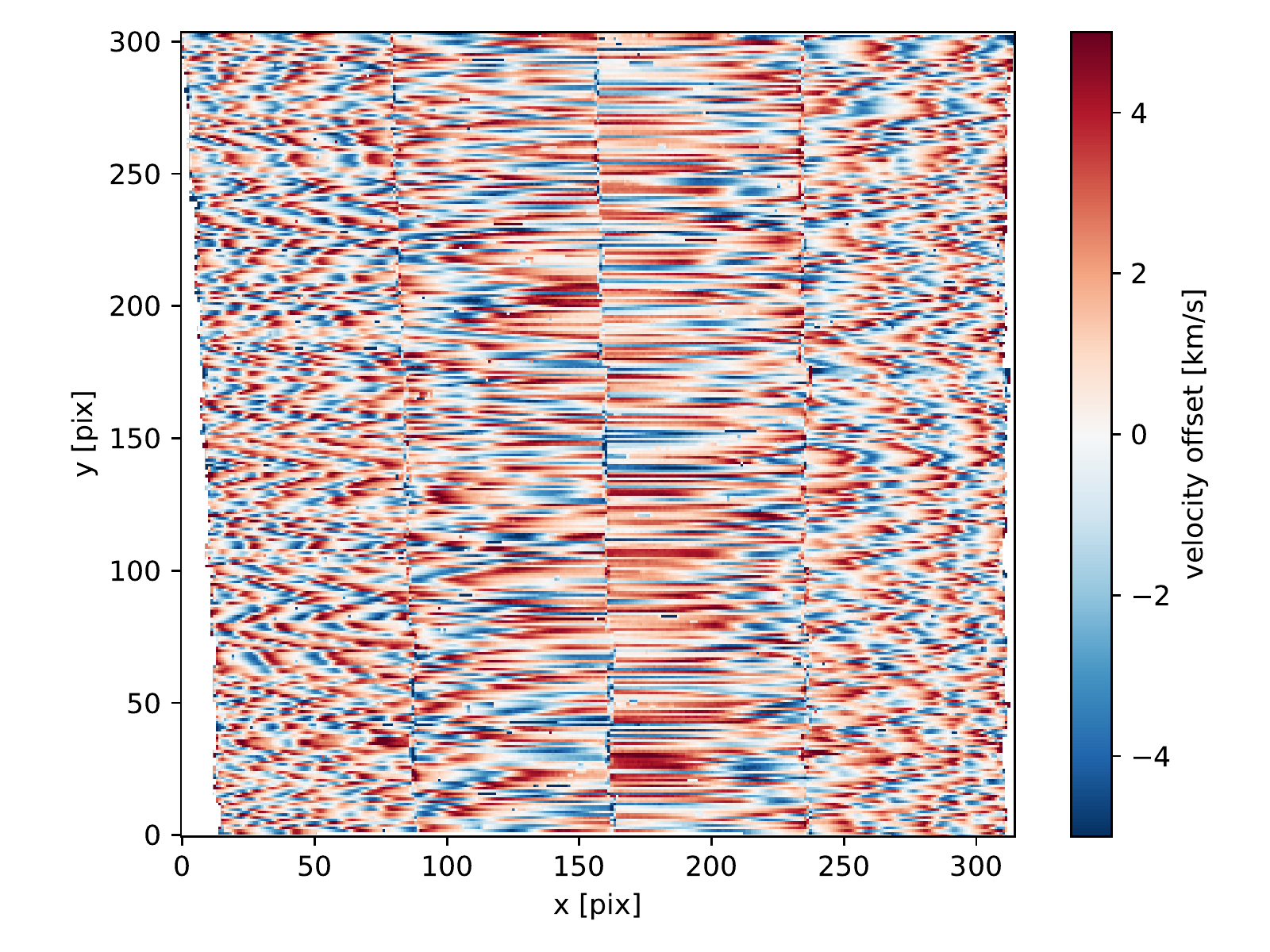}
\caption{Velocity offsets across the MUSE field of view as measured from the arc
         line \ion{Xe}{i}\,8819 on a reconstructed cube.}
\label{fig:wavecube8819}
\end{figure}

The results for the fits of individual lines at the cube level (see
Tab.~\ref{tab:wavecube}) are what a naive user can recover using fits to single
(emission) lines. Since this folds the intrinsic wavelength calibration with
the binning of the cube, the recovered line center depends on the sampling of
the original pixel on the CCD with respect to the final sampling of the
line-spread function in the cube. The values show that the average and median
line centers can be measured with the same precision as for the pixel table.
This is visually presented as the blue data points in Fig.~\ref{fig:wavelambda}.
The relative offsets for all pixel positions (the standard deviation) are
affected by the resampling, and only allow to reconstruct the per-spaxel
wavelength to $0.06\dots0.08$\,\AA\ accuracy, corresponding to a $1\sigma$
velocity precision of $2.5\dots4.0$\kms. The spatial distribution of this
pattern is shown in Fig.~\ref{fig:wavecube8819} for the example of the Xe\,I
line at 8819.410\,\AA. This effect can be mitigated in the analysis of
astronomical objects, if the measurement of multiple lines of similar S/N can
be combined.

\subsection{Sky subtraction residuals}\label{sec:skyacc}
\begin{figure*}
\includegraphics[width=\linewidth]{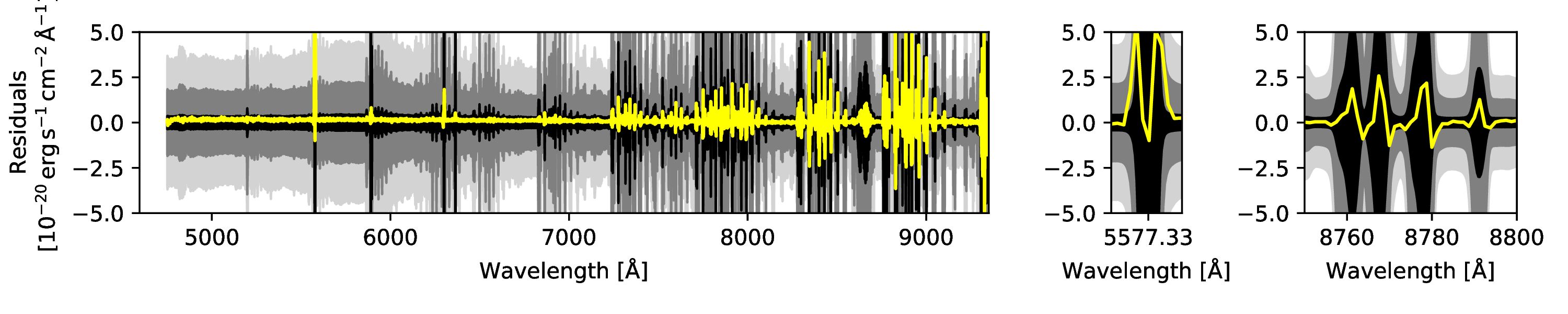}
\includegraphics[width=\linewidth]{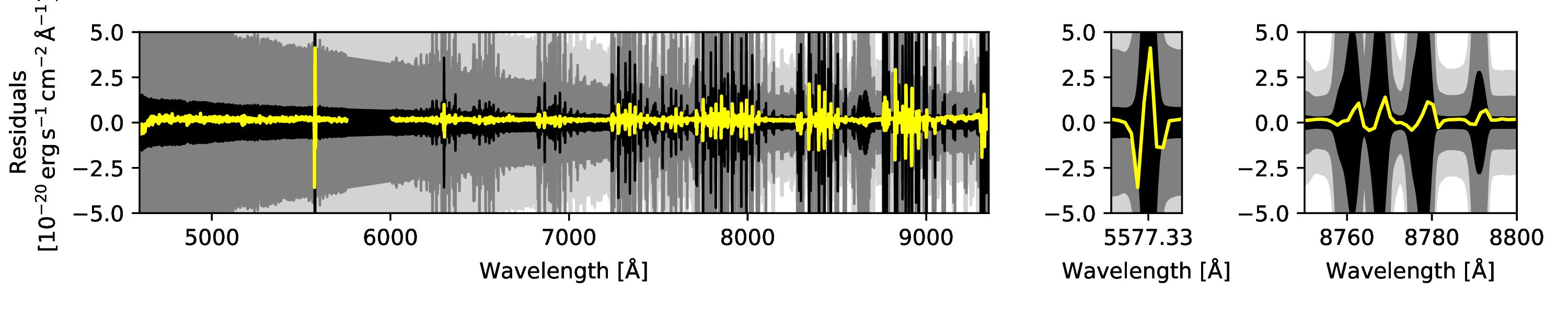}
\includegraphics[width=\linewidth]{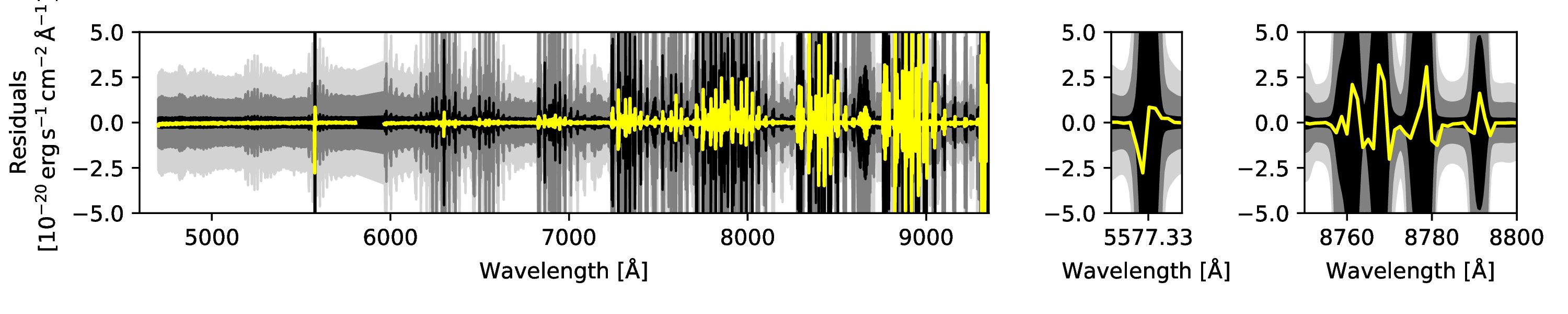}
\caption{Sky background percentages (light grey: 10\%, grey: 5\%, black: 1\%)
         and median residuals (yellow) in the three MUSE exposures:
         {\bf (1)} (top): The galaxy AM\,1353-272 in WFM-NOAO-N mode on
                          2014-04-29T04:22:00 UTC,
         {\bf (2)} (middle): the tidal dwarf NGC\,7252\,NW in WFM-AO-E mode on
                             2017-07-16T09:13:11 UTC,
         {\bf (3)} (bottom): a QSO in WFM-AO-N mode on
                             2018-02-15T05:08:06 UTC.}
\label{fig:skyresiduals}
\end{figure*}

To assess the quality of the sky subtraction performed by the MUSE pipeline we
take publicly available MUSE data of three different instrument modes as
examples, and reduce them in the standard way.  The science post-processing
module outputs the cube, a white-light image, and files to characterize the sky
subtraction. We use the high-S/N sky spectrum of the exposure (averaged over
the sky regions of the exposure) to assess the original sky level at each
wavelength. The residuals still present in the corresponding exposure are taken
as the integrated spectrum of the cube. To exclude any object features, we used
the median over the field of view (for targets that cover a small fraction of
the field), or the median after masking the brightest 75\% of the white-light
image (for nearby galaxies).

We plot the median residuals of the cube as yellow lines in
Fig.~\ref{fig:skyresiduals}, where $\pm$10\%, $\pm$5\%, and $\pm$1\% level
deviation from the original sky background are highlighted as areas shaded in
light grey, dark grey, and black.  We show the whole MUSE wavelength range in
the left panels, a zoom on the small region around the strong line
[\ion{O}{i}]5577 in the middle panel, and an enlarged display of the range
around the blue end of the OH\,7-3 band.
The MUSE pipeline was designed to reach a sky subtraction accuracy of at least
5\% with a goal of 2\%, outside wavelengths affected by strong sky emission
lines. These goals were clearly reached, the yellow lines are always within the
dark grey 5\% region, and in the continuum within the 1\% range as well. It
only happens for a few lines, that the line-spread function was not determined
with enough precision. In those cases, the wings of the line residuals extend
beyond to about 2\% of the original sky level. This is visible in the
[\ion{O}{i}]5577 in the middle panels in the two top rows in
Fig.~\ref{fig:skyresiduals}. While residuals for other lines are apparent, they
do not extend beyond the 1\% range.

\subsection{Flux calibration accuracy}\label{sec:fluxacc}
\begin{figure}
\includegraphics[width=\linewidth]{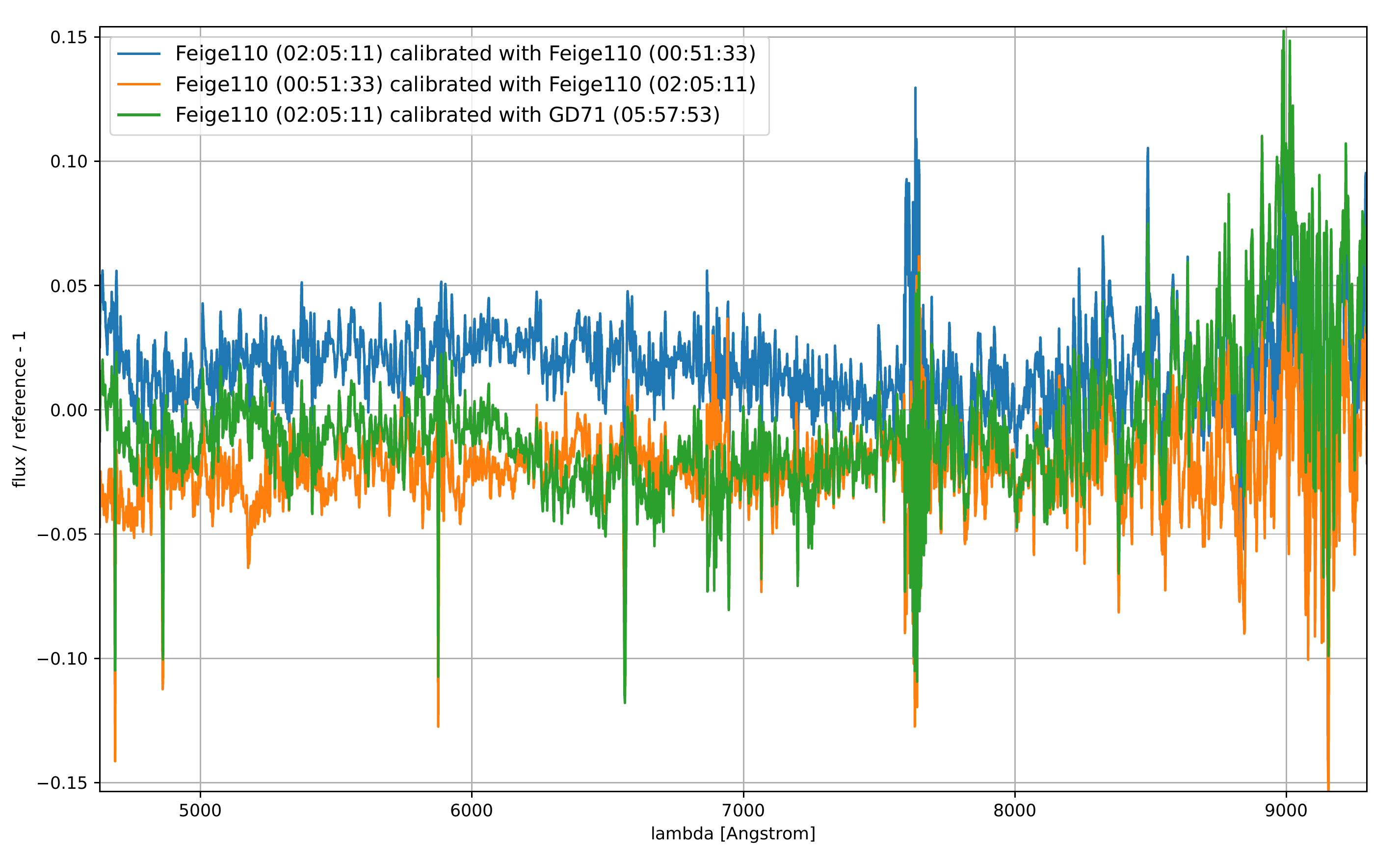}
\caption{Residuals in the flux calibration accuracy as a function of wavelength.
         Three different exposures of the standard star Feige\,110 observed
         on 2018-11-22 are shown. The calibrations were done with another
         exposure of the same star and with GD\,71 at different time of the same
         night. The UTC times of the individual exposures are given as labels.}
\label{fig:fluxacc}
\end{figure}

The flux calibration accuracy of the pipeline reduction was already shown to
be within 4-7\% of fluxes measured by other instruments, depending on the
emission line in question, and to vary about 5\% across the field of view of
the instrument \citep{WeilbacherM42}. Here, we investigate how the accuracy of
the flux calibration depends on the wavelength.

We reduced standard stars taken during the night of 2018-11-22 with the
flux-calibration module ({\tt muse\_standard}), resulting in a response curve
and the telluric correction spectrum. Then we treated these exposures, as if
they were science data, using the science module ({\tt muse\_scipost}),
including the response and telluric correction derived from another standard
star exposure. Since the night was classified photometric between 00:58 and
06:55 UTC, we did not expect strong atmospheric changes during the observing
sequence.
Then we used the same extraction method (the smoothed Moffat profile described
in Sect.~\ref{sec:stdstar}) to measure the resulting fluxes again, and computed
the ratio of the resulting spectrum with respect to the reference spectrum.
The residuals can be seen in Fig.~\ref{fig:fluxacc} for the two exposures of
the star Feige\,110 calibrated with the other exposure of the same star and an
exposure of GD\,71 observed during the same night. In these datasets the
residuals are $\lesssim$2\% on average, with a standard deviation of 2.4 to
3.7\%.

The instrument mode used for these exposures was the extended wide-field mode
(WFM-NOAO-E) that incurs a 2nd order overlap in the red part. Since the white
dwarfs used for flux calibration are very blue, this effect is particularly
strong. Therefore the deviations at wavelengths beyond $\sim8300$\,\AA\ can be
larger. It is also obvious, that during the night the atmospheric absorption
in the telluric A- and B-bands changed, so that strong outliers and enhanced
noise are visible around 6900 and 7650\,\AA.

To summarize, the flux calibration of MUSE data by the pipeline should be
accurate to within the 3-5\% across the wavelength range of the instrument.

\subsection{Astrometric precision}\label{sec:astprec}
The combination of geometrical (see Sect.~\ref{sec:geo}) and astrometric
(Sect.~\ref{sec:astcal}) calibration is not meant to give an absolute world
coordinate system (WCS) but should give a high-precision relative coordinate
solution within the spatial extent of a MUSE cube. The astrometric calibration,
derived from typically 50-100 stars within the MUSE field and repeated about
once a month, provides a measure of the median residuals of the final WCS
solution in both axes.
The average over 36 such calibrations is 0\farcs048$\pm$0\farcs018 in horizontal
direction of a cube (in right ascension), and 0\farcs027$\pm$0\farcs0.006 vertically
(in declination). The calibration is therefore thought to be better than 1/4 of a
MUSE sampling element in WFM.
In the high-resolution NFM that has started operating only in 2019, the five
existing calibrations show average residuals of 9\,mas, so about 1/3 pixel, in
both directions.

\begin{figure}
\centering
\includegraphics[trim={9 11 55 36},clip,width=\linewidth]{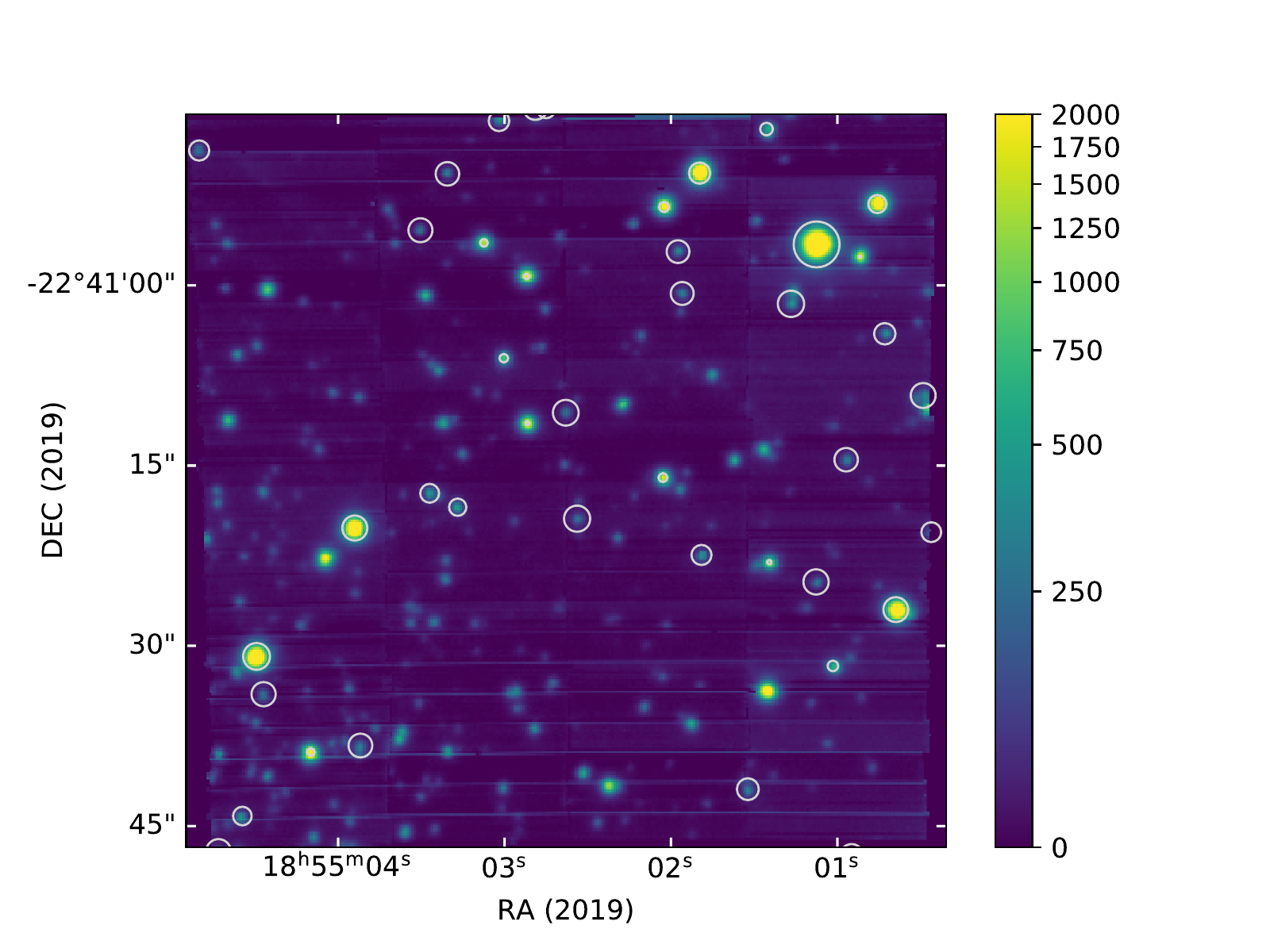}
\caption{Image of a stellar field in the globular cluster NGC\,6717. Positions
         from the Gaia DR2 catalog are shown as circles.}
\label{fig:astimage}
\end{figure}

For the WFM fields, enough stars are available in the Gaia DR2 catalog
\citep{GaiaDR2_astrometry} to carry out an independent check. We used the
astrometric calibration created using a field in the outskirts of the globular
cluster NGC\,3201 (observed at 2019-10-29T08:38:37.911 UTC) together with the
geometry sequence created from calibrations taken on 2019-10-26T09:20:36. We
then reduced the data of a globular cluster field in NGC\,6717 (observed on
2019-10-25T23:40:52.126 UTC) as a science exposure, using this calibration.
The resulting image, integrated over a wavelength range with little sky
contribution ($7700\dots8100$\,\AA), is shown in Fig.~\ref{fig:astimage}.
There are 15 stars from the Gaia DR2 catalog within the MUSE field with
magnitudes $G<17.65$\,mag that have high-quality proper motions, and that can
therefore be projected to the epoch of observations (2019.81640).
We match these positions with those measured as Gaussian centroids using the
ESO {\tt skycat} tool.
When correcting for the zero-point offset, the 15 stars show a separation of
$0\farcs036\pm0\farcs018$ ($\lesssim1/5$ MUSE pixels) with standard deviations
of 0\farcs031 in RA and 0\farcs024 in DEC, between the epoch-corrected Gaia
positions and the their centers in the MUSE data.
Unfortunately, Gaia DR2 does not provide stars with enough density to do
the same check for small fields taken in the NFM.

This suggests that the quality of the astrometric solution computed by the
pipeline gives a realistic representation of the real quality on sky, which
corresponds 1/4th of a spatial element of MUSE or better.

\subsection{S/N behavior}\label{sec:sncheck}
\begin{figure*}[!ht]
\includegraphics[width=0.40\linewidth]{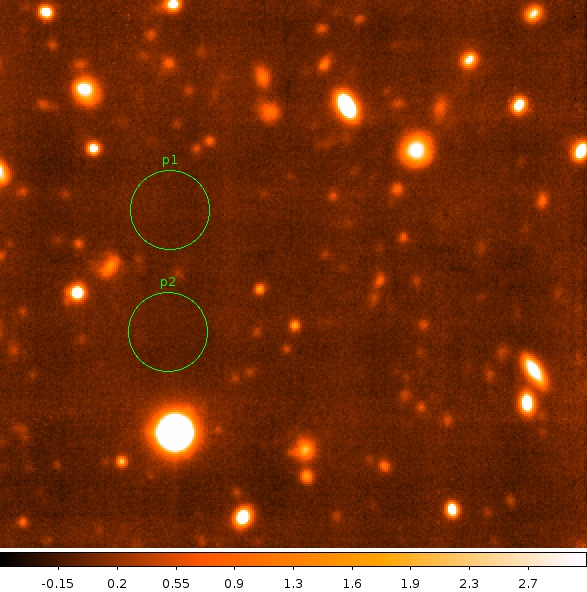}
\hspace{0.5cm}
\includegraphics[width=0.55\linewidth]{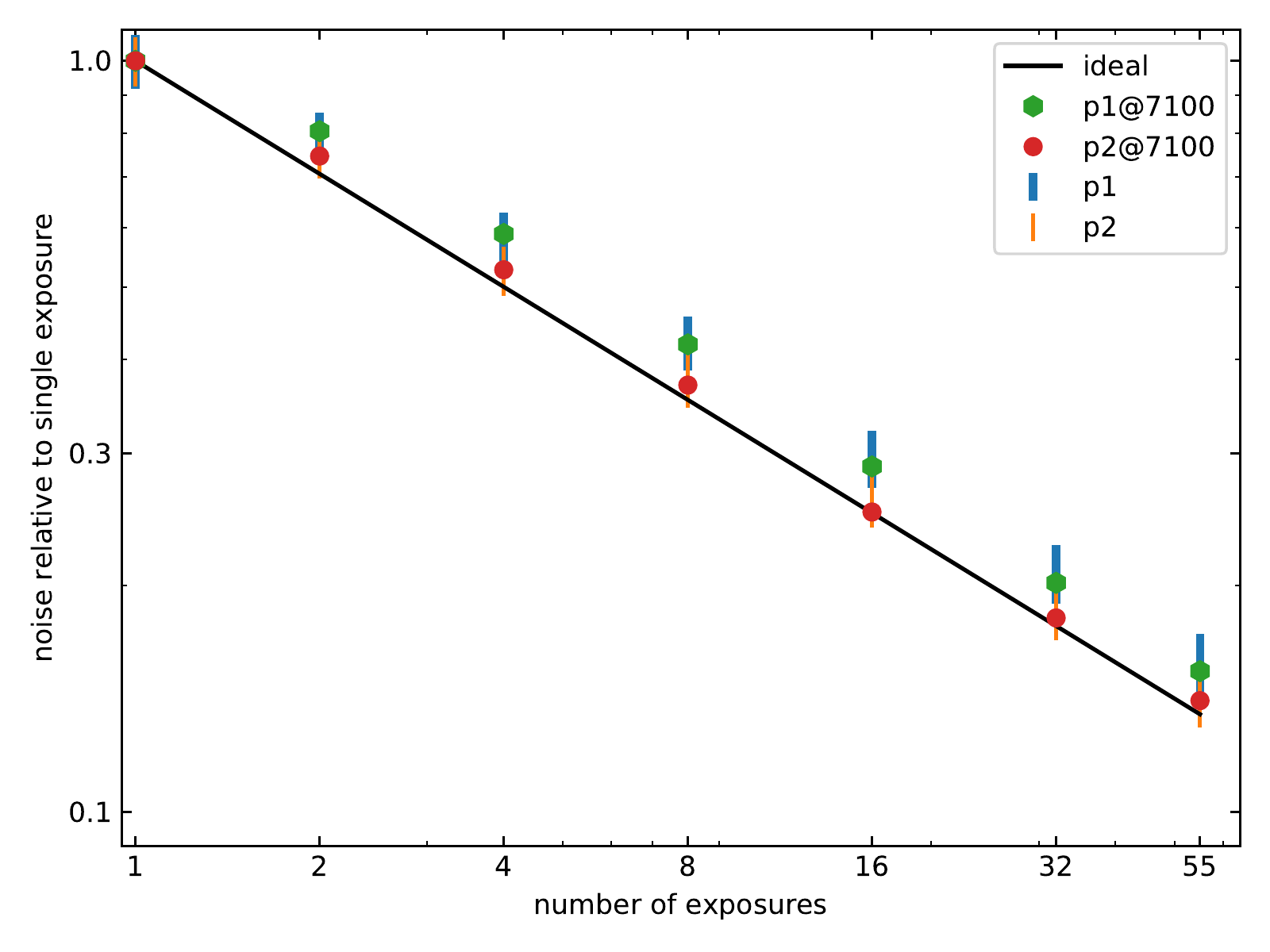}
\caption{{\bf Left}: MUSE image of a region in the Hubble Deep Field South
         integrated over the wavelength range 7000-7200\,\AA.
         The two circular apertures that were used for measurements of the
         noise are marked as green circles. The displayed version shows the
         full depth, 55 exposures corresponding to 27.4\,h. We selected the
         apertures to be located in a region without any detectable objects.
         {\bf Right}: Improvement of the noise measured in the final cube as a
         function of the number of exposures, measured as the standard
         deviation across two circular areas without any objects. See text for
         more details.}\label{fig:sn}
\end{figure*}

In the ideal case, a data reduction system should remove any systematics from
the data so that the combination of a number of $n$ exposures results in a S/N
improvement of $\sqrt n$. For a deep MUSE dataset this was already tested in
\citet{2015A&A...575A..75B} to show that the combination of deep datasets
reaches this ideal case to within a factor of 1.2. In that work, however, the
authors used heavy post-processing of the individual cubes, and the exposures
were not combined using the MUSE pipeline but externally at the cube-level.

We therefore re-run that experiment with the MUSE dataset of the Hubble Deep
Field South, using only the pipeline recipes and combining 1, 2, 4, 8, 16, 32,
and 55 exposures at the pixel table level, without any exposure weighting
using the {\tt muse\_exp\_combine} module of the pipeline.
Since we do not intend to use the data for science, we restrict the experiment
on the wavelength range 7000 to 7200 Angstrom, where a number of strong sky
residuals and a clean wavelength range are present. We do not use the slice
autocalibration (Sect.~\ref{sec:autocal}).

To check the S/N behavior with the number of exposures, we measure the noise in
two circular apertures of 20 pixel radius (as displayed in
Fig.~\ref{fig:sn} left), in regions where the white-light image
\citep[of][]{2015A&A...575A..75B} does not show any object to be present. We
measure the standard deviation over these areas for all seven cubes and for
each wavelength plane of the cubes.
The mean and standard deviations of all wavelengths for a given cube are
plotted in Fig.~\ref{fig:sn} (right) for both regions. The results for the
wavelength plane at 7100\,\AA\ is plotted without error bars to show a
particularly good case, since this wavelength plane does not contain any
significant residual of telluric emission.

Overall, the results show that the S/N improves almost as expected
from the ideal curve, but in region p1 the systematics are slightly stronger.
The final datapoint, for the cube that was combined using all 55 exposures of
the field, is only a factor of
1.05 (1.17) higher than the theoretical expectation for region p2 (p1).

\section{Conclusions \& Outlook}\label{sec:concl}
At the time of writing, more than 200 publications have produced new science
results using the MUSE instrument.
All of them used the MUSE pipeline to process the raw data, or directly used
datacubes made available by ESO through their Phase 3
process\footnote{\url{https://www.eso.org/sci/observing/phase3.html}}.
While most publications used the pipeline as described above
\citep{2015MNRAS.452....2K,2016A&A...591A.143G,2016A&A...588A.148H,2018A&A...618A...3R},
others try to improve sky subtraction or background uniformity using different
ways \citep[\eg,][]{2017A&A...608A...1B,2016ApJ...831...39B,2019A&A...624A.141U,2019MNRAS.490.1451F},
often using the external ZAP-tool \citep[Zurich Atmosphere
Purge,][]{2016MNRAS.458.3210S} to improve sky residuals. This tool was
integrated into the ESO MUSE Reflex workflow to improve accessibility for
users.

In hindsight we can say that there was no single key to the success, but
several ingredients helped to write the pipeline as described in this paper:
{\bf (i)} We started prototyping early, shortly after the final properties of
the instrument were fixed.
{\bf (ii)} In parallel to the reduction software a simulation of the raw data
was created as well, which was finally used for a complete ``dry-run''.
{\bf (iii)} The pipeline was actively used to process test data during the
hardware tests, by multiple people at several institutes.
{\bf (iv)} The main pipeline developer had a vested interest in making the
instrument work, even after commissioning, as a member of the team exploiting
the data collected in the guaranteed time.
{\bf (v)} We used multiple communication channels during different stages of
the project, to collect ideas for features and algorithms, to keep track of the
implementation, and track bugs.
{\bf (vi)} The software was published completely, without holding back
consortium-internal parts, once testing of new features was completed.

While the pipeline hence works well to produce a plethora of science results on
many different topics, some ideas exist that could improve its performance.
Prototype code to compute and correct nonlinearities of the data exist, and
implementation in the public pipeline releases is planned and have the
potential to improve data accuracy for low and high illumination levels.
To improve removal of telluric absorption, coupling with the {\sc molecfit}
tool \citep{molecfit} could be possible. Likewise, sky emission lines could
be treated by taking line groups from {\sc skycorr} \citep{skycorr} instead
of those distributed with the MUSE pipeline.
To improve the accuracy with which the subsequent data analysis can be carried
out, propagation of the line-spread function and limited propagation of
covariances are further possibilities.

Some of the algorithms described in this paper would also be applicable for
future integral-field instruments, especially for the ESO project HARMONI for
the Extremely Large Telescope \citep{2016SPIE.9911E..1ZP} and the possible
BlueMUSE spectrograph for the VLT \citep{BlueMUSE_1906.01657}.

\begin{acknowledgements}
The authors thank Joris Gerssen for contributions to the pipeline code at early
stages of the development.
We also thank all members of the MUSE collaboration for feedback on the
pipeline during all stages of development, and especially Sebastian Kamann
for preparing the astrometric reference tables.
ESO staff, most notably Danuta Dobrzycka and Lodovico Coccato, helped to
improve usage, capabilities, and algorithms since commissioning.
We are also grateful for the anonymous referee who provided a very helpful
and timely report on this long paper in difficult times.\\
PMW, OS, and TU received support through BMBF Verbundforschung (project
MUSE-AO, grant 05A14BAC) to develop the pipeline, PMW and OS were further
supported by the MUSE-NFM project (grant 05A17BAA).
We thank the developers of the Image Reduction and Analysis Facility
\citep[IRAF,][]{1986SPIE..627..733T}, as ideas from some of its tasks were used
in the design of the MUSE pipeline.
\end{acknowledgements}

\bibliographystyle{aa}
\bibliography{pipeline.bib}

\begin{appendix}

\section{Instrument modes}\label{sec:appmodes}
\begin{table*}
\caption{MUSE instrument modes and nominal properties.}\label{tab:modes}
\begin{tabular}{l c l c l l}
Mode$^a$  & FOV                    &sampling &wavelength$^b$&AO& filter$^b$\\
\hline
WFM-NOAO-N&$1\arcmin\times1\arcmin$&0\farcs2  &4750\dots9350\,\AA& no   & 2nd order ($<4750$\,\AA)\\
WFM-NOAO-E&$1\arcmin\times1\arcmin$&0\farcs2  &4600\dots9350\,\AA& no   & none\\
WFM-AO-N  &$1\arcmin\times1\arcmin$&0\farcs2  &4750\dots9350\,\AA& GLAO & blocking ($5800\dots5970$\,\AA), 2nd order ($<4750$\,\AA)\\
WFM-AO-E  &$1\arcmin\times1\arcmin$&0\farcs2  &4600\dots9350\,\AA& GLAO & blocking ($5750\dots6010$\,\AA)\\
NFM-AO-N  &$7\farcs5\times7\farcs5$&0\farcs025&4750\dots9350\,\AA& LTAO & blocking$^b$ ($5780\dots6050$\,\AA), 2nd order ($<4750$\,\AA)\\
\hline
\end{tabular}\\
$^a$ -N stands for {\em nominal}, -E for {\em extended} wavelength range.\\
$^b$ NaD blocking filter in the GALACSI AO system.
\end{table*}

Table \ref{tab:modes} lists all five modes of the MUSE instrument and their
properties.

\section{Instrument layout}\label{sec:appfov}
\begin{figure*}
\includegraphics[width=\linewidth]{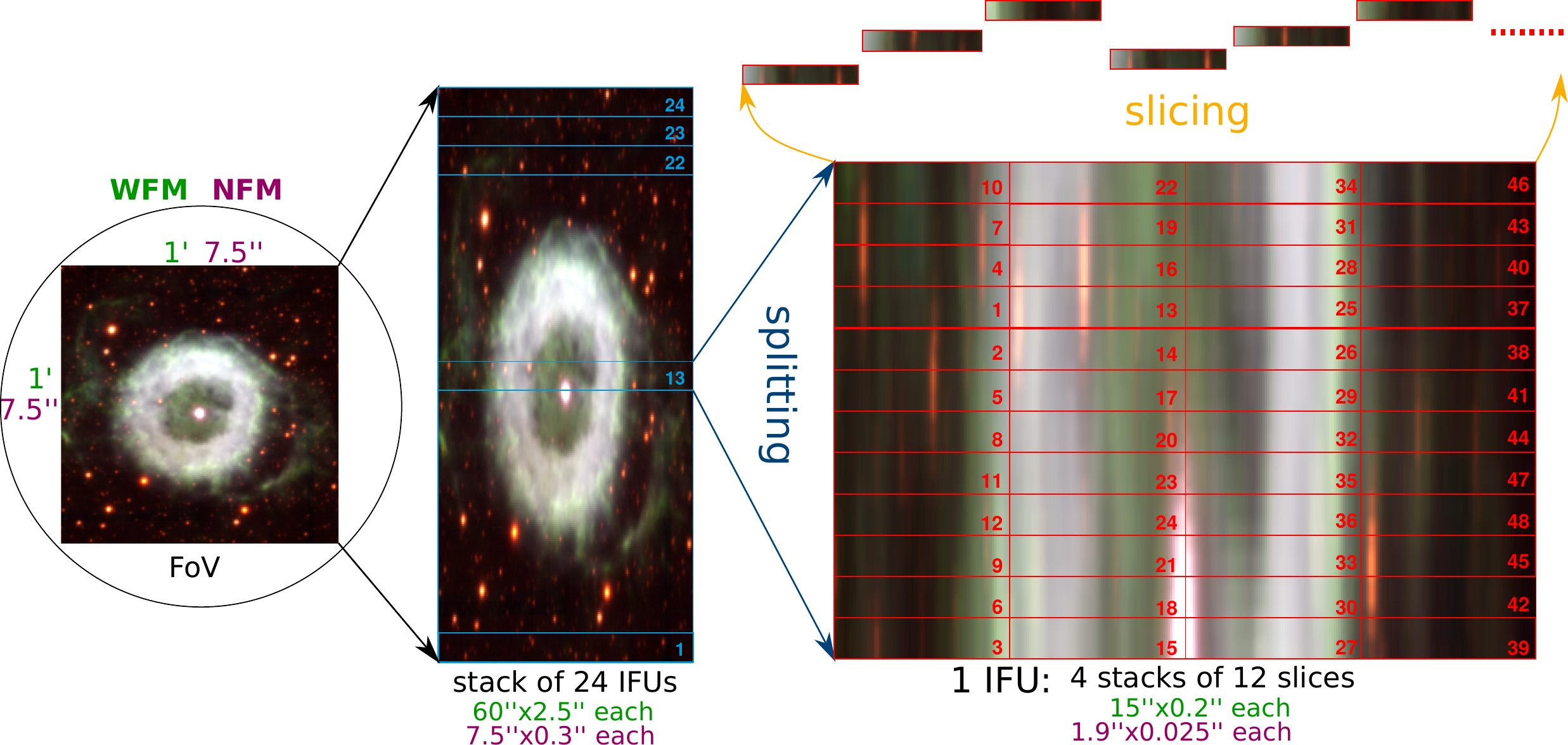}
\caption{The layout of the MUSE instrument, from the focal plane of the VLT
        (left) to the individual slices (top right). The light from the slices
        is subsequently dispersed and recorded on the CCD, and there forms the
        step-pattern visible in Fig.~\ref{fig:raw}.
        The sizes of the different elements are given as nominal values (green:
        for WFM, violet: for NFM), the actual sizes are slightly different and
        change across the field.
        Overlaid on the images are the channel numbers (on the IFU stack) and
        the numbers of the slices as counted left-to-right on the raw data
        images (on the slicer stack).
        The example image is a color picture of the planetary nebula NGC\,6369
        as observed with MUSE in WFM-AO-N on 2017-07-15.}
\label{fig:instlayout}
\end{figure*}

Fig.~\ref{fig:instlayout} shows a sketch of the instrument layout, and how the
field is distributed over the 24 IFU and the 48 slices per IFU.

\end{appendix}

\end{document}